\documentclass[preprint,12pt,authoryear]{elsarticle}
\usepackage[round]{natbib} 

\usepackage[margin=2.5cm]{geometry}
\usepackage{enumitem}
\newlist{inlinelist}{enumerate*}{1}
\setlist*[inlinelist,1]{%
  label=(\roman*),
}
\usepackage{soul}
\usepackage{hhline}
\usepackage{subfig}
\usepackage{multirow}
\usepackage{xcolor}
\usepackage{graphicx}
\usepackage{upgreek}
\usepackage{amssymb}
\usepackage{amsfonts,amsthm,bm,amsmath} 
\usepackage{appendix}
\usepackage{times}
\usepackage{caption}
\usepackage{subfig}
\newcommand{\psubref}[1]{\protect\subref{#1}}
\usepackage{multirow}
\usepackage{xcolor}
\usepackage[linesnumbered,ruled,vlined]{algorithm2e}
\usepackage{tablefootnote}

\SetCommentSty{mycommfont}

\SetKwInput{KwInput}{Input}                
\SetKwInput{KwOutput}{Output}              

\usepackage[colorlinks]{hyperref} 
\hypersetup{ 
    colorlinks=true,       
    linkcolor=red,          
    citecolor=blue,        
    filecolor=magenta,      
    urlcolor=cyan           
}

\newcommand{\fref}[1]{Fig.~\ref{#1}}

\newcommand{\eref}[1]{Eq.~(\ref{#1})}

\newcommand{\sref}[1]{Section~\ref{#1}}
\newcommand{\tref}[1]{Table~\ref{#1}}

\journal{arXiv}
\begin{document}

\begin{frontmatter}

\title{A deep learning energy-based method for classical elastoplasticity}
\author[]{Junyan He$^1$}
\author[]{Diab Abueidda$^2$\corref{mycorrespondingauthor1}}
\ead{abueidd2@illinois.edu}
\author[]{Rashid Abu Al-Rub$^3$}
\author[]{Seid Koric$^{1,2}$}
\author[]{Iwona Jasiuk$^1$\corref{mycorrespondingauthor2}}
\address{$^1$ Department of Mechanical Science and Engineering, University of Illinois at Urbana-Champaign, Champaign, IL, USA \\
$^2$ National Center for Supercomputing Applications, University of Illinois at Urbana-Champaign, Champaign, IL, USA \\
$^3$ Advanced Digital \& Additive Manufacturing Center, Khalifa University, Abu Dhabi, UAE \\}
\cortext[mycorrespondingauthor1]{Corresponding author 1}
\cortext[mycorrespondingauthor2]{Corresponding author 2}
\ead{ijasiuk@illinois.edu}
\begin{abstract}

The deep energy method (DEM) has been used to solve the elastic deformation of structures with linear elasticity, hyperelasticity, and strain-gradient elasticity material models based on the principle of minimum potential energy. In this work, we extend DEM to elastoplasticity problems involving path dependence and irreversibility. A loss function inspired by the discrete variational formulation of plasticity is proposed. The radial return algorithm is coupled with DEM to update the plastic internal state variables without violating the Kuhn-Tucker consistency conditions. Finite element shape functions and their gradients are used to approximate the spatial gradients of the DEM-predicted displacements, and Gauss quadrature is used to integrate the loss function. Four numerical examples are presented to demonstrate the use of the framework, such as generating stress-strain curves in cyclic loading, material heterogeneity, performance comparison with other physics-informed methods, and simulation/inference on unstructured meshes. In all cases, the DEM solution shows decent accuracy compared to the reference solution obtained from the finite element method. The current DEM model marks the first time that energy-based physics-informed neural networks are extended to plasticity, and offers promising potential to effectively solve elastoplasticity problems from scratch using deep neural networks.

\end{abstract}

\begin{keyword}
Deep energy method \sep Plasticity \sep Variational formulation \sep Radial return \sep Cyclic loading 
\end{keyword}

\end{frontmatter}

\section{Introduction}
\label{sec:intro}
The deep energy method (DEM) is a recently developed physics-informed neural network model that can be used to find continuous solutions to partial differential equations from scratch \citep{ew2018deep,nguyen2020deep,samaniego2020energy}. The method has gained popularity due to its ease of implementation and the ability to solve partial differential equations in a mesh-free manner \citep{fuhg2022mixed}. It has been applied in the Poisson's equation \citep{ew2018deep}, linear elasticity \citep{samaniego2020energy,he2022use}, hyperelasticity \citep{nguyen2020deep,he2022use}, viscoelasticity \citep{abueidda2022deep}, piezoelectricity \citep{samaniego2020energy}, fracture mechanics \citep{samaniego2020energy}, strain gradient elasticity \citep{nguyen2021parametric}, and topology optimization \citep{zehnder2021ntopo,he2022deep}. Central to DEM is the definition of a loss function, whose value is reduced during the training process through the use of an optimizer. In solid mechanics, the principle of minimum potential energy (PMPE) states that the potential energy of a system attains a local minimum at its equilibrium displacement field. This principle coincides nicely with the minimization structure of the DEM model. Thus, DEM employs the total potential energy as the loss function and approaches equilibrium displacements by minimizing the system potential. This means that extending the DEM method to nonlinear problems such as hyperelasticity is straightforward, as one only needs to implement the corresponding strain energy expression, and is much simpler than extending to nonlinearity in the finite element method (FEM). One of the key advantages of DEM is that the loss function, being an energy form, only requires the first-order gradient of the displacement fields (except for strain gradient models as in \cite{nguyen2021parametric}). Prior to DEM, a notable physics-informed neural network to solve partial differential equations is the deep collocation method (DCM) \citep{raissi2018deep,abueidda2021meshless,guo2021deep,haghighat2021physics}. It is based on the strong form of governing equation, and the loss function consists of residuals in the strong form evaluated at the nodes (collocation points). This means that for the elasticity problem, the second-order spatial gradients of the displacement field are required, thus it is more computationally expensive (to evaluate second-order and first-order gradients) compared to the DEM. Additionally, researchers have employed a mixed formulation, incorporating both the strong form and energy method, to find the mechanical response in regions with stress concentrations and high solution gradients \citep{fuhg2022mixed, abueidda2022enhanced, rezaei2022mixed}.

Since all of the above-mentioned DEM models are based on PMPE, restrictions applied to PMPE also apply to those models. This means that the energy-based method can only be applied to the elastic deformation of structures under static equilibrium. However, plastic deformations often occur in engineering structures when a relatively large load is applied, and the non-conservative nature of plasticity renders PMPE no longer valid. Plasticity problems have been widely studied by neural networks and other data-driven approaches. For example, machine learning has been used to approximate elastic and plastic constitutive laws in simulations \citep{ali2019application,huang2020machine,fuhg2021model,zhang2020using,zhang2022predicting,pi2021data}, to learn constitutive laws directly from displacement field or other experimental measurements \citep{yang2020learning,liu2020learning,muhammad2021machine}, to act as a constitutive model in FEM code \citep{jang2021machine,qu2021towards,tancogne2021recurrent,ibragimova2021new,al2006prediction}, and to directly predict the stress-strain curve and energy absorption during plastic deformation \citep{he2022exploring,abueidda2021deep,mozaffar2019deep}. However, the models introduced in those works require training data, typically generated from the many FEM solutions to different elastoplasticity problems. They do not, however, solve the partial differential equation directly. \cite{abueidda2021meshless} leveraged the DCM to solve the plastic bending of a cantilever beam using the residual formulation of the strong form of the governing equations. Close agreement with FEM is observed, although a very refined grid with 1.2 $\times 10^{11}$ nodes was used in the training process. Considering the ease of implementation and the advantage of only requiring lower-order spatial derivatives, it is of high research interest to extend the present DEM framework to treat plasticity problems. Therefore, the objective of this work is to propose a definition of the loss function that is suitable for elastoplasticity problems, which is inspired by the discrete variational formulation of plasticity \citep{simo2006computational}. In addition, we propose to combine the classical radial return algorithm with DEM to update the plastic internal state variables. 

This paper is organized as follows: \sref{sec:methods} presents an overview of neural networks, the deep energy method, and a summary of $J_2$ plasticity and radial return method. \sref{sec:results} presents and discusses the results of five numerical examples. \sref{sec:conc} summarizes the outcomes and highlights possible future works.

\section{Methods}
\label{sec:methods}
\subsection{Neural networks}
\label{NN}
\begin{figure}[h!] 
    \centering
         \includegraphics[width=0.6\textwidth]{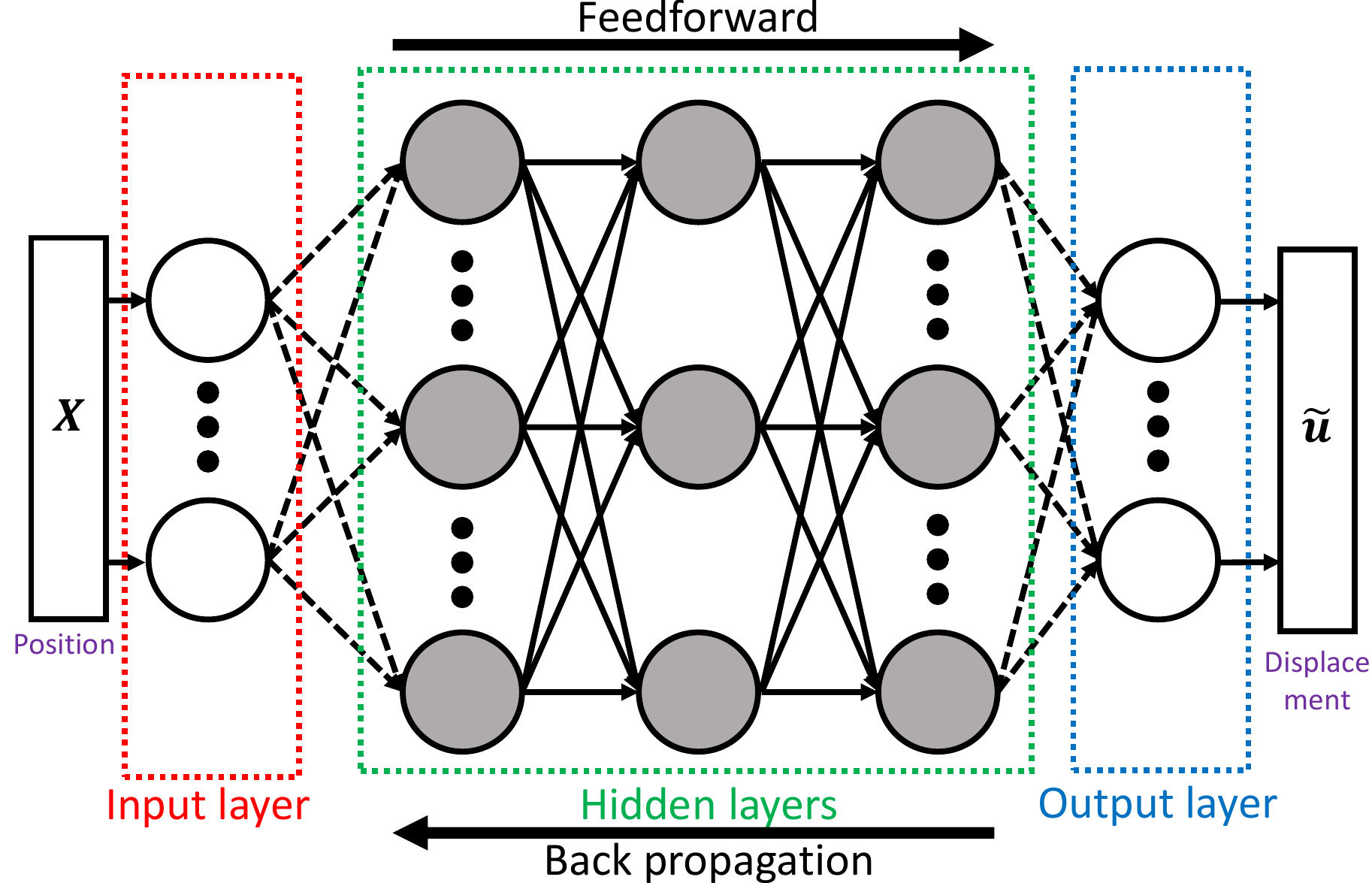}
    \caption{Schematic of a fully connected neural network that maps position vectors $\bm{X}$ to displacement vectors $ \Tilde{\bm{u}}$.}
    \label{DNN}
\end{figure}

Neural networks seek to approximate a continuous function of an unknown functional form. A feed-forward, fully connected neural network consists of multiple layers of interconnected neurons, and the layers themselves are also connected. \fref{DNN} shows an example of a fully connected neural network model, which maps position vectors $\bm{X}$ into displacements $ \Tilde{\bm{u}}$. 
The neurons of consecutive layers are connected by a set of weights $\bm{W}$ and biases $\bm{b}$. The output $\bm{y}^i$ of layer $i$ can be computed as:
\begin{equation}
    \bm{y}^i = f_{\rm{act}}^i( \bm{W}^i \bm{y}^{i-1} + \bm{b}^i ),
\end{equation}
where $f_{\rm{act}}^i$ denotes the activation function. The trainable weights and biases are updated in a process known as training, where a loss function $\mathcal{L}$ is minimized through optimizers such as stochastic gradient descent \citep{bottou2010large}.

\subsection{Deep energy method in solid mechanics}
\label{DEM}
The universal approximation capability of deep neural networks \citep{hornik1989multilayer} has been widely leveraged in the field of computational solid mechanics to approximate complex displacement fields. In particular, the PMPE for elastic structures under static equilibrium offers a physics-informed way to define the loss function of a neural network, namely the potential energy of the system. The resulting method is known as the deep energy method. For linear elastic material in the absence of any body and inertial forces, the potential reads:
\begin{equation}
    \mathcal{L}(\bm{u}) = \frac{1}{2} \int_{\Omega} \bm{\sigma} : \bm{\epsilon} \, dV - \int_{\partial \Omega_t} \Bar{\bm{t}} \cdot \bm{u} \, dA.
    \label{PE}
\end{equation}
where $\bm{\sigma}$, $\bm{\epsilon}$ and $\Bar{\bm{t}}$ denote the stress, strain, and applied boundary traction, respectively. Neumann boundary conditions (BCs) naturally appear in \eref{PE}, while Dirichlet BCs can be enforced exactly by modifying the displacements predicted by DEM:
\begin{equation}
   \bm{u}(\bm{X})  =  \Tilde{\bm{u}}(\bm{X}) \circ \bm{m}(\bm{X}) + \bm{u}_0(\bm{X}),
\end{equation}
where $\bm{u}$ denotes a modified displacement, $\circ$ denotes the Hadamard product between two vectors, and the vectors $\bm{m}$ and $\bm{u}_0$ are chosen such that:
\begin{equation}
    \bm{m}(\bm{X}) = \bm{0} \,\, {\rm{and}} \,\, \bm{u}_0 (\bm{X}) = \Bar{\bm{u}}, \forall \bm{X} \in \partial \Omega_u
\end{equation}
By design, $\bm{u}$ satisfies Dirichlet BCs for all arbitrary $ \Tilde{\bm{u}}$ produced by DEM. The solution to the elasticity problem, as given by the DEM model, is defined according to PMPE as:
\begin{equation}
    \bm{u}^* = {\rm{arg}}\min_{\bm{u} } \, \mathcal{L}( \bm{u} ).
\end{equation}
However, since the above DEM model is based on PMPE, it is only applicable to elastic structures in static equilibrium.

\subsection{$J_2$ plasticity and the return mapping algorithm}
\label{Return}
In the absence of any body and inertial forces, the equilibrium equations and BCs (Dirichlet and Neumann) under small deformation can be stated in terms of the Cauchy stress tensor $\bm{\sigma}$ as:
\begin{equation}
\begin{aligned}
    \nabla \cdot \bm{\sigma} = \bm{0}, \;\; \forall \bm{X} \in \Omega,\\
    \bm{ u } = \Bar{\bm{u}}, \;\; \forall \bm{X} \in \partial \Omega_u,\\
    \bm{\sigma} \cdot \bm{ n } = \Bar{\bm{t}} , \;\; \forall \bm{X} \in \partial \Omega_t,
    \label{strong}
\end{aligned}
\end{equation}
where $\bm{n}$, $\Bar{\bm{u}}$, and $\Bar{\bm{t}}$ denote the outward boundary normal, prescribed displacement, and prescribed traction, respectively. In the small deformation setting, the total strain tensor is given by:
\begin{equation}
    \bm{\epsilon} = \frac{1}{2} ( \nabla \bm{u} + \nabla \bm{u}^T ),
    \label{strain}
\end{equation}
which is additively decomposed into its elastic part $\bm{\epsilon}^e$ and plastic part $\bm{\epsilon}^p$:
\begin{equation}
    \bm{\epsilon} = \bm{\epsilon}^e + \bm{\epsilon}^p.
    \label{decomposition}
\end{equation}
For linear elastic and isotropic material, the constitutive equation is:
\begin{equation}
    \bm{\sigma} = 2\mu (\bm{\epsilon}^e)' + \kappa {\rm{tr}}(\bm{\epsilon}^e) \bm{I},
    \label{stress}
\end{equation}
where $\mu$ and $\kappa$ are the shear and bulk moduli, $(\bm{\epsilon}^e)'$ is the deviatoric part of the elastic strain tensor\footnote{$\bm{A}' = \bm{A} - \frac{1}{3} tr(\bm{A}) \bm{I}$, where $\bm{I}$ is the second-order identity tensor.}.

In this work, we focus on $J_2$ plasticity with an associative flow rule, which assumes incompressible plastic strains. The following von Mises yield function $f$ is used, which accounts for both isotropic and kinematic hardening:
\begin{equation}
    f = \sqrt{ \bm{\eta} : \bm{\eta} } - \sqrt{\frac{2}{3}} \sigma_{y}( \Bar{\epsilon}^p )
\end{equation}
where $\bm{\eta} = \bm{\sigma}' - \bm{q}'$ is the shifted stress. $\bm{\sigma}'$ and $\bm{q}'$ denote the deviatoric parts of the stress tensor $\bm{\sigma}$ and the back stress tensor $\bm{q}$. $\sigma_{y}$ denotes the yield stress of the material, which can be a function of the equivalent plastic strain $\Bar{\epsilon}^p$ (an internal state variable). To complete the definition of plasticity, we need the Kuhn-Tucker conditions on consistency, which read:
\begin{equation}
\begin{aligned}
    f( \bm{\sigma}' , \bm{q} , \Bar{\epsilon}^p ) \le 0,\\
    \gamma \ge 0,\\
    \gamma f( \bm{\sigma}' , \bm{q} , \Bar{\epsilon}^p ) = 0,
\end{aligned}
\label{consistency}
\end{equation}
where $\gamma$ is the consistency parameter. 

The evolution laws for $\bm{\epsilon}^p$ and $\bm{q}$ are given by:
\begin{equation}
\begin{aligned}
    \Dot{ \bm{\epsilon}^p } = \Dot{ \gamma } \bm{N},\\
    \Dot{ \bm{q} } = \frac{2}{3} \Dot{ \gamma } C \bm{Z},
\end{aligned}
\end{equation}
where 
\begin{equation}
\begin{aligned}
    \bm{N} = \frac{ \bm{\eta} }{  \sqrt{ \bm{\eta} : \bm{\eta} } }  ,\\
    \bm{Z} = \frac{ \bm{\sigma} - \bm{q} }{ \sqrt{ ( \bm{\sigma} - \bm{q} ) : ( \bm{\sigma} - \bm{q} ) } }.
\end{aligned}
\end{equation}
Here, $\bm{N}$ is the normal to the yield surface $f=0$ (associative flow rule), and $\bm{Z}$ connects the center of the yield surface ($\bm{q}$) to the current stress point ($\bm{\sigma}$). $C$ is the (constant) kinematic hardening modulus in the linear Ziegler kinematic hardening model \citep{ziegler1959modification}.

Elastoplastic problems in computational mechanics are typically solved iteratively, where at each step $i+1$, we are given a known current strain (or in the form of a total strain increment). The updated values of the state variables ($\bm{\sigma}$, $\bm{\epsilon}^p$, $\Bar{\epsilon}^p$, and $\bm{q}$) can be obtained by integrating the constitutive laws with known initial conditions. For $J_2$ plasticity with associative flow rule, the radial return algorithm \citep{wilkins1964methods} is commonly used.

Given the current state vector (at step $i$) $[ \bm{\sigma}_i , \bm{\epsilon}^p_i , \Bar{\epsilon}^p_i , \bm{q}_i ]$ and a known strain increment $\Delta \bm{\epsilon}$, the radial return algorithm begins with a fully elastic trial state:
\begin{equation}
\begin{aligned}
    \bm{\sigma}'_{trial} = \bm{\sigma}'_i + 2\mu \Delta \bm{\epsilon}' ,\\
    \bm{\eta}_{trial} = \bm{\sigma}'_{trial} - \bm{q}'_i,\\
    f_{trial} = \sqrt{ \bm{\eta}_{trial} : \bm{\eta}_{trial} } - \sqrt{\frac{2}{3}} \sigma_{y}( \Bar{\epsilon}^p_i ),
\end{aligned}
\end{equation}
where $\Delta \bm{\epsilon}'$ denotes the deviatoric part of the strain increment. If $f_{trial} \le 0$, the trial state is accepted, the updated stress is given by $\bm{\sigma}_{i+1} = \bm{\sigma}'_{trial} + \kappa tr(\Delta \bm{\epsilon}) \bm{I}$, and no updates to the plastic internal state variables are necessary. If $f_{trial} > 0$, this indicates active yielding has occurred, and the internal state variables are updated as follows:
\begin{equation}
\begin{aligned}
    \bm{N}_{i+1} = \frac{ \bm{\eta}_{trial} }{ \sqrt{ \bm{\eta}_{trial} : \bm{\eta}_{trial} } }, \\
    \bm{\epsilon}^p_{i+1} = \bm{\epsilon}^p_i + \Delta \gamma \bm{N}_{i+1} , \\
    \Bar{\epsilon}^p_{i+1} = \Bar{\epsilon}^p_i + \sqrt{\frac{2}{3}} \Delta \gamma , \\
    \bm{\sigma}_{i+1} = \bm{\sigma}'_{trial} + \kappa tr(\Delta \bm{\epsilon}) \bm{I} - 2\mu \Delta \gamma \bm{N}_{i+1},\\
    \bm{Z}_{i+1} = \frac{ \bm{\sigma}_{i+1} - \bm{q}_{i} }{ \sqrt{ ( \bm{\sigma}_{i+1} - \bm{q}_{i} ) : ( \bm{\sigma}_{i+1} - \bm{q}_{i} ) } },\\
    \bm{q}_{i+1} = \bm{q}_i + \frac{2}{3} \Delta \gamma C \bm{Z}_{i+1}.
\end{aligned}
\end{equation}
where $\Delta \gamma$ is the consistency parameter increment to be determined. In this work, we focus on linear isotropic hardening:
\begin{equation}
    \sigma_{y}( \Bar{\epsilon}^p ) = \sigma_{y,0} + H \Bar{\epsilon}^p,
\end{equation}
where $\sigma_{y,0}$ and $H$ denote the initial yield stress and the constant isotropic hardening modulus, respectively. In this case, $\Delta \gamma$ is given directly as:
\begin{equation}
    \Delta \gamma = \frac{ f_{trial} }{ 2 ( \mu + \frac{H+C}{3} ) }.
    \label{linear_return}
\end{equation}

The radial return algorithm can be computationally expensive to evaluate for complex material models with highly nonlinear hardening behaviors. Therefore, surrogate models such as a neural network for plastic state update have been developed by \cite{zhang2020using}, \cite{masi2021thermodynamics}, \cite{,yu2022elastoplastic}, and \cite{bonatti2022importance} and used in elastoplastic FE simulations to replace the radial return algorithm.

\subsection{Deep energy method based on the variational formulation of plasticity}
\label{var_plastic}
In this section, we introduce a loss function for DEM that is suitable for $J_2$ plasticity, to replace \eref{PE} that only applies to elastic deformation. Inspired by the discrete variational formulation of plasticity, which is used to develop finite element formulations for elastoplasticity, we define the total free energy functional $\mathcal{P}_{i+1}$ \citep{simo2006computational}:
\begin{multline}
    \mathcal{P}_{i+1} =
    \int_{\Omega} \Bigl( W_{i+1} + \frac{1}{2} \bm{v}_{i+1} \bm{D}^{-1} \bm{v}_{i+1} - \Delta \gamma f_{i+1} + \\
    ( \bm{\epsilon}^p_{i+1} - \bm{\epsilon}^p_{i} ) : \bm{\sigma}_{i+1}
    - \bm{v}_{i+1} \bm{D}^{-1} ( \bm{v}_{i+1} - \bm{v}_i ) \Bigr) dV + \mathcal{P}_{ext},
    \label{freeE}
\end{multline}
where $ W_{i+1} = \frac{1}{2} \bm{\sigma}_{i+1} : \bm{\epsilon}^e_{i+1}$ is the elastic strain energy, $\bm{v}_{i+1}$ denotes the collection of internal state variables defined as:
\begin{equation}
    \bm{v}_{i+1} = 
    \begin{bmatrix}
        H \Bar{\epsilon}^p_{i+1} \\
        \bm{q}_{i+1}
    \end{bmatrix}.
\end{equation}
$\bm{D}$ is the matrix of hardening moduli:
\begin{equation}
    \bm{D} = 
    \begin{bmatrix}
        H & \bm{0}_{[1\times3]} \\
        \bm{0}_{[3\times1]} & C \bm{I}
    \end{bmatrix}.
\end{equation}
With the absence of body forces, the potential energy of the external loading at the current time, $\mathcal{P}_{ext}$, is given by:
\begin{equation}
    \mathcal{P}_{ext} = - \int_{\partial \Omega_t} \Bar{\bm{t}} \cdot \bm{u} \, dA
    \label{pext}
\end{equation}

In this work, we use the radial return algorithm to update the plasticity variables, this leads to the satisfaction of the discrete counterpart of \eref{consistency} at step $i+1$:
\begin{equation}
    \Delta \gamma f_{i+1} = 0.
    \label{discrete_consistent}
\end{equation}
Similar to how a trained neural network can replace radial return in elastoplastic FE simulations, the discrete consistency condition can also be satisfied by updating the plastic state variables through a neural network, whether data-driven or physics-informed. Additional knowledge on the material behavior, such as plastic strain direction and pressure dependence of plastic strain, can be incorporated by these separately trained surrogate models. The rest of the DEM framework is designed to be agnostic to the plastic state update kernel. In this work, the radial return algorithm is used to demonstrate the rest of the framework and is chosen since it avoids the additional errors accompanying any surrogate neural networks used for state update. 

Substituting \eref{discrete_consistent} into \eref{freeE}, and specializing the equation to isotropic hardening only ($H_{i+1} \ne 0, C = 0$), we have:
\begin{equation}
    \mathcal{P}_{i+1}^{iso} = \int_{\Omega} \Bigl( W_{i+1} + \frac{1}{2} H ( \Bar{\epsilon}^p_{i+1} )^2 + ( \bm{\epsilon}^p_{i+1} - \bm{\epsilon}^p_{i} ) : \bm{\sigma}_{i+1} - H \Bar{\epsilon}^p_{i+1} ( \Bar{\epsilon}^p_{i+1} - \Bar{\epsilon}^p_{i} ) \Bigr) dV + \mathcal{P}_{ext}.
    \label{freeE_iso}
\end{equation}
Similarly, when specializing to kinematic hardening only ($H_{i+1} = 0, C \ne 0$), we have:
\begin{equation}
    \mathcal{P}_{i+1}^{kin} = \int_{\Omega} \Bigl( W_{i+1} + \frac{1}{2 C } \bm{q}_{i+1} : \bm{q}_{i+1} + ( \bm{\epsilon}^p_{i+1} - \bm{\epsilon}^p_{i} ) : \bm{\sigma}_{i+1} - \frac{1}{C} \bm{q}_{i+1} : ( \bm{q}_{i+1} - \bm{q}_{i} ) \Bigr) dV + \mathcal{P}_{ext}.
    \label{freeE_kine}
\end{equation}

Since the Euler-Lagrange equations of the discrete functional $\mathcal{P}$ yield the equilibrium equations, flow rule, and hardening law \citep{simo2006computational}, it follows from variational calculus that $\mathcal{P}$ attains a local minimum at the equilibrium state, similar to PMPE applied to elastic deformation. Therefore, we define the loss function of DEM as the total free energy of the system $\mathcal{L} = \mathcal{P}$, and minimize its value during the training of DEM.

Automatic differentiation of the neural network displacement output is commonly used \citep{nguyen2020deep} to compute the spatial gradient of the displacement field. However, \cite{he2022use} demonstrated that when the applied load magnitude is large, automatic differentiation can fail to penalize unrealistic strain localizations that occur in between the nodes, thus leading to an incorrect solution. \cite{he2022use} further showed that a shape-function-based spatial gradient calculation method correctly penalizes any unrealistic strain localizations that may occur and yields accurate solutions even at large deformations. Therefore, the shape-function-based approach is adapted here for its robustness. In this case, the displacement gradient is given by:
\begin{equation}
    \frac{ \partial \bm{u} }{ \partial \bm{X} } \approx \frac{\partial \bm{\phi} }{\partial \bm{\xi}} \bm{J}^{-1} \cdot \bm{u},
\end{equation}
where $\bm{\phi}$ , $\bm{\xi}$, and $\bm{u}$ denote the finite element shape functions, natural coordinates, and displacement vector, respectively. $\bm{J} = \frac{\partial \bm{X} }{\partial \bm{\xi}}$ denotes the Jacobian matrix of the isoparametric mapping of the finite element. The need to discretize the domain with isoparametric finite elements renders this version of DEM no longer a meshless method, which is distinct from other meshless DEM implementations \citep{nguyen2020deep,nguyen2021parametric}. After the gradients are calculated at the quadrature points, Gauss quadrature is used to evaluate the volume and surface integrals in \eref{freeE}. Since the constitutive update is performed at the quadrature points of the element, to save computation time, reduced integration with one integration point per element is used.

\begin{figure}[h!] 
    \centering
         \includegraphics[trim={0cm 0cm 0cm 0cm},clip,width=1\textwidth]{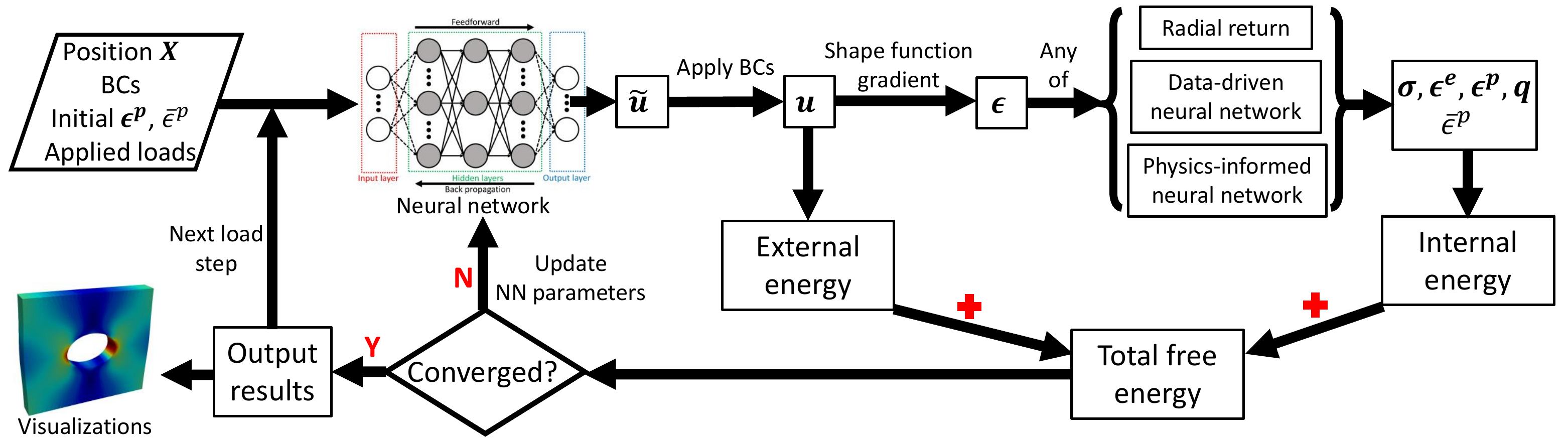}
    \caption{Complete workflow of the plasticity DEM framework. Note that the framework is modular in the sense that different state update kernels can be plugged into the model to update the plastic state variables.}
    \label{flowchart}
\end{figure}

When the solution is completed in multiple load steps, we do not reinitialize the weights and biases of the DEM model after the first step. Instead, we used the trained weights and biases from the previous load step as an initial guess for the parameter values in the next load step, which is a form of transfer learning. Since the loss function, $\mathcal{L}$, is defined as the total free energy of the system, whose value at equilibrium is not known a priori, absolute convergence criterion like $| \mathcal{L} - \mathcal{L}_{eq} | < \epsilon_{tol}$ cannot be used to determine convergence of the DEM model. Therefore, a relative convergence criterion that monitors the relative change in the loss function is used. Convergence is said to be achieved if:
\begin{equation}
    \left | \frac{ {\rm{mean}}([ \mathcal{L}_{j-2N+1} , \mathcal{L}_{j-2N+2} \cdots \mathcal{L}_{j-N} ]) - {\rm{mean}}([ \mathcal{L}_{j-N+1} , \mathcal{L}_{j-N+2} \cdots \mathcal{L}_{j} ]) }{ {\rm{mean}}([ \mathcal{L}_{j-N+1} , \mathcal{L}_{j-N+2} \cdots \mathcal{L}_{j} ]) } \right | \le \epsilon_{tol},
\end{equation}
where $[ \mathcal{L}_{j-2N+1} , \mathcal{L}_{j-2N+2} \cdots \mathcal{L}_{j} ]$ denotes an array of $2N$ loss values generated by the optimizer, where $N$ is a user-defined patience parameter that defines the number of loss values to be included in the check. Since loss values from the last $2N$ iterations are needed, convergence is not checked until the optimizer has gone through at least $2N$ iterations. In essence, this criterion computes the mean of the loss function in the last $N$ iterations (involving indices $j-N+1, j-N+2 \cdots j$) and compares it to the mean in the last $N$-$2N$ iterations (involving indices $j-2N+1, j-2N+2 \cdots j-N$). If the two means are close within a relative tolerance of $\epsilon_{tol}$, then the optimizer is no longer efficiently reducing the loss and any further iterations are unlikely to improve the results significantly, hence the solution is considered converged, and the training is terminated. $N=10$ was used in this work. Note that, by definition, this criterion merely measures the relative change of the loss value (which reflects the effectiveness of the optimizer) and does \emph{not} necessarily imply the correctness of the converged solution. Finally, we use algorithm \ref{DEM_$J_2$} to summarize the DEM model for $J_2$ plasticity. The flowchart for the plasticity DEM framework is presented in \fref{flowchart}.
\begin{algorithm}[!ht]
\DontPrintSemicolon
    \KwInput{ Network architecture, domain size, grid size, material properties, boundary conditions, external loading, number of load steps}
    
    \KwOutput{Displacement, stress, and plastic internal state variables at each step }

    \tcc{Initialization}
    Initialize weights and biases of the DEM model $\mathcal{M}$ \\
    Initialize all plastic internal state variables to 0 \\
    $ i \gets 0 $ \\

    \tcc{Begin load stepping}
    \While{ $i < $ max load step }{

    \tcc{Begin DEM training}
    \While{ relative loss change $> \epsilon_{tol}$ }{
        Obtain $\Tilde{\bm{u}}$ from $\mathcal{M}$, apply Dirichlet BCs to get $\bm{u}$ \\
        Use shape function gradient operator to compute $\bm{\epsilon}$ \\
        Update stress and plastic state variables using user-defined kernel\\
        Compute external work potential \\
        Compute loss \\
        Update the weights and biases of $\mathcal{M}$ through back-propagation
    }

    Store converged values of plastic internal state variables \\

    \tcc{Output results}
    Write output to vtk file, save trained model parameters for this step
    
    $ i \gets i + 1 $
    }
\caption{A deep energy method model for $J_2$ plasticity}
\label{DEM_$J_2$}
\end{algorithm}

\section{Results and discussion}
\label{sec:results}
In this section, we present four numerical examples to show the application of the new DEM model in elastoplasticity problems. All examples share identical elastic material properties, they are: $\mu = 384.62$ MPa and $\kappa = 833.33$ MPa, and the initial yield stress is 50 MPa unless otherwise specified. The DEM model used in this work was implemented in PyTorch (version 1.13.0) \citep{NEURIPS2019_9015}. The underlying neural network is a multi-layer perceptron (MLP) model, which has 6 layers, including input and output. The number of neurons in each layer is 3, 100, 200, 400, 200, 100, and 3, respectively, resulting in a total of 3009 trainable parameters. The hyperbolic tangent function was used as the activation function for all layers except the output, which has linear activation. The L-BFGS optimizer \citep{zhu1997algorithm} with a fixed learning rate of 0.5 was used to train the models. All DEM model training and inference were conducted using a single Nvidia A100 GPU card on Delta, an HPC cluster hosted at the National Center for Supercomputing Applications (NCSA). Abaqus/Standard \citep{Abaqus2021} was used to solve all example problems using identical meshes to obtain the reference FE solutions. All FE simulations were conducted using 40 high-end AMD EPYC 7763 Milan CPU cores. The absolute difference between the DEM and FEM solutions for a field variable $\theta$ is computed as:
\begin{equation}
    AD_\theta = |\theta_{DEM} - \theta_{FE}|.
\end{equation}
The $L^2$ error norm for the displacement vector is defined as \citep{nguyen2021parametric}:
\begin{equation}
    ||\bm{e}||_{L^2} = \frac{ || \bm{u}_{DEM} - \bm{u}_{FE} ||_{L^2}  }{ || \bm{u}_{FE} ||_{L^2} } \times 100\%.
\end{equation}

In all examples presented in this work, strains as large as 20$\%$ were applied to the structures, even when the small-strain formulation was used. This is intended to demonstrate the ability to generate a convergence solution even at relatively large applied strains. As a consistent comparison, all reference FE simulations were also conducted using the small strain formulation even when the applied strain is large, and the error induced by using the small-strain theory in finite deformation was ignored in this work.

\subsection{DEM-simulated stress-strain response}
\label{SS}
\begin{figure}[h!] 
    \centering
     \subfloat[]{
         \includegraphics[trim={0cm 0cm 0cm 0cm},clip,width=0.32\textwidth]{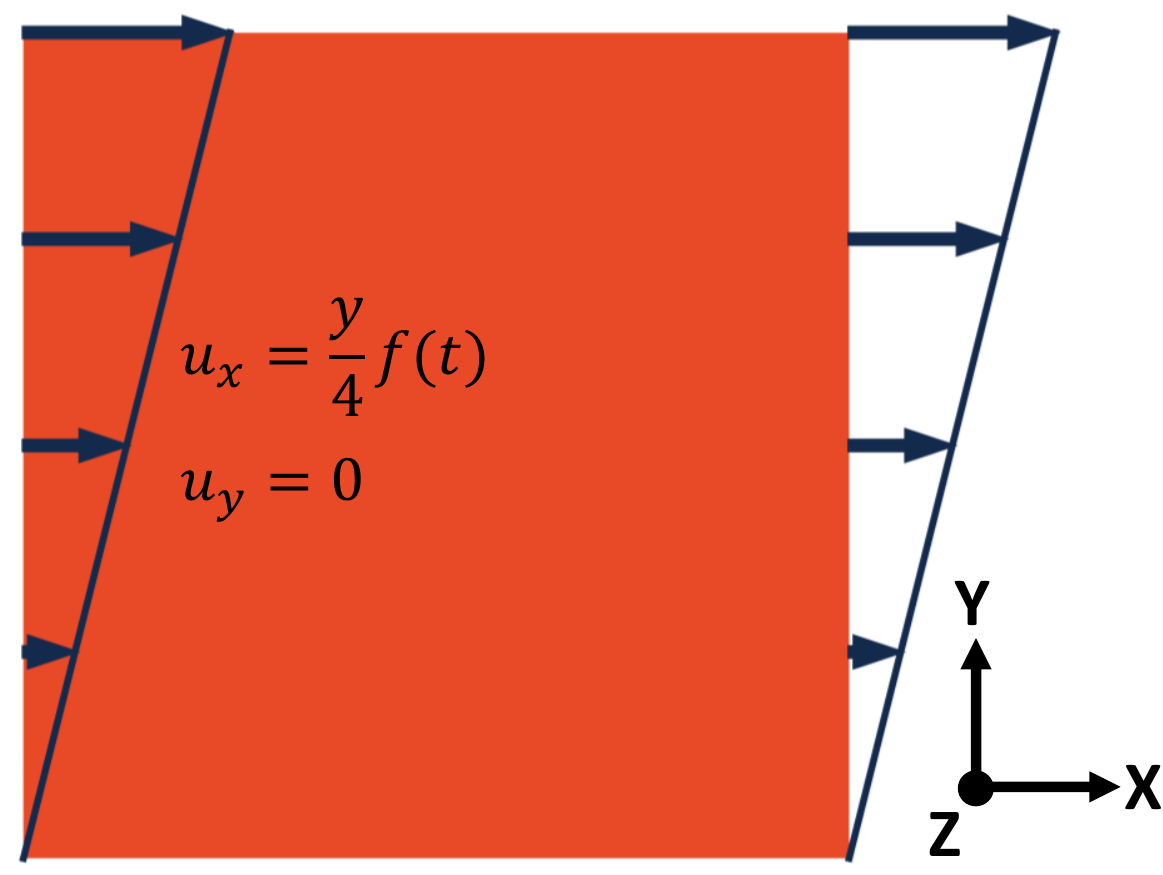}
         \label{uniform_shear}
     }
     \subfloat[]{
         \includegraphics[trim={0cm 0.45cm 0cm 0cm},clip,width=0.34\textwidth]{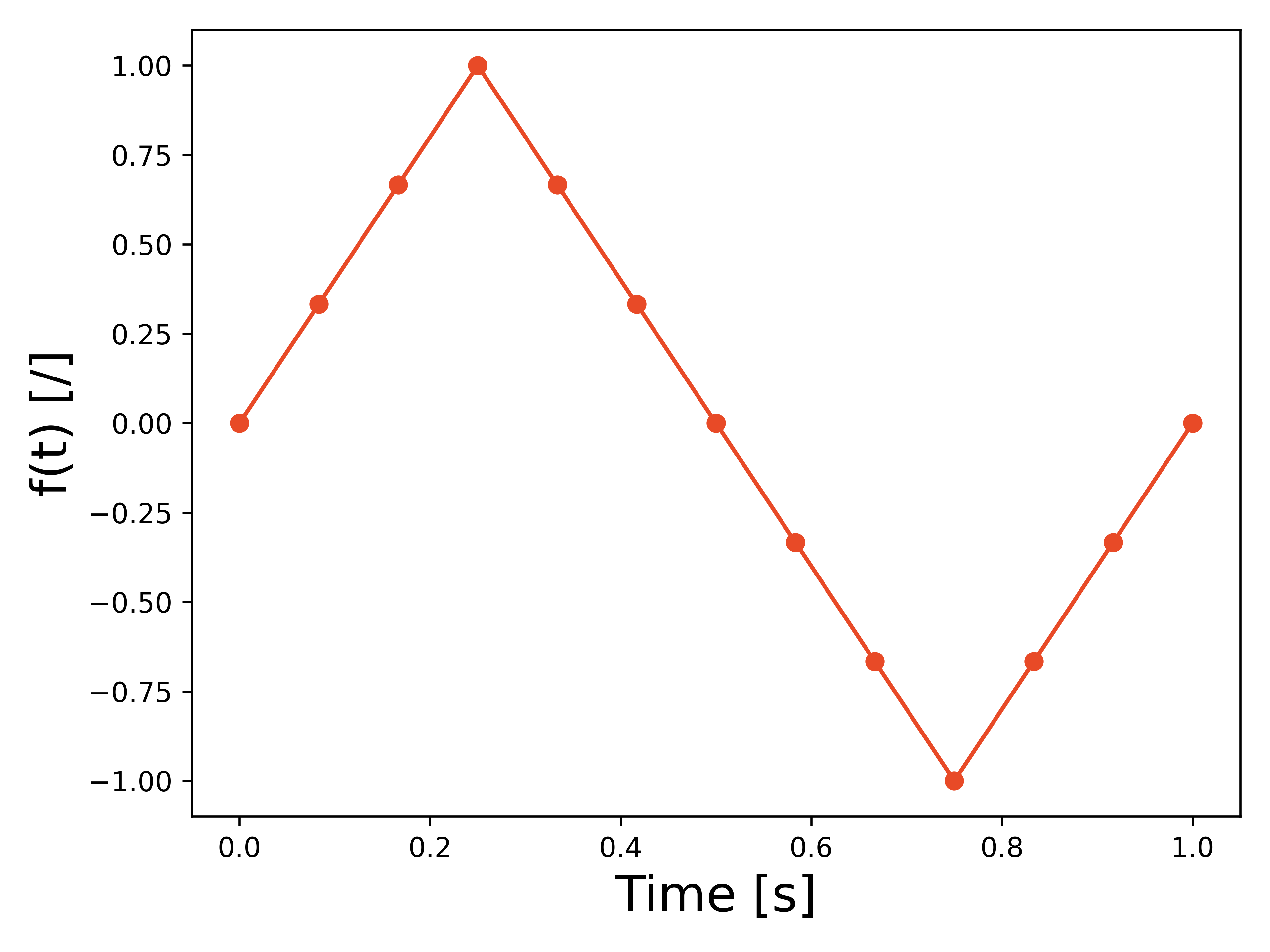}
         \label{ft}
     }
    \caption{Fully-reversible simple shear test: \psubref{uniform_shear} Geometry and applied displacements. \psubref{ft} Load magnitude as a function of time, simulation load steps marked in circles.}
    \label{load1}
\end{figure}

Consider a rectangular plate with dimensions 4$\times$4$\times$1 mm$^3$, discretized by a structured hexahedral mesh of 100$\times$100$\times$1 elements. A state of plane strain in the thickness (Z) direction was enforced by setting $u_z = 0$ on the front and back faces. The other four external faces were subjected to a time-dependent, fully reversible displacement field given by: $u_x (y,t) = \frac{y}{4} f(t)$ and $u_y = 0$. The applied boundary conditions induced a constant state of shear stress in the plate. A schematic of the loading condition and the time variation function $f(t)$ are depicted in \fref{load1}. A total of 12 load steps were used in the solution process for DEM and FEM. Two different types of hardening models were considered separately, one is a linear isotropic hardening model with $H=500$ MPa, and the other is a linear Ziegler kinematic hardening model with $C=500$ MPa. A convergence tolerance of $1\times 10^{-6}$ was used.

\begin{figure}[h!] 
    \centering
     \subfloat[]{
         \includegraphics[trim={0cm 0cm 0cm 1.4cm},clip,width=0.4\textwidth]{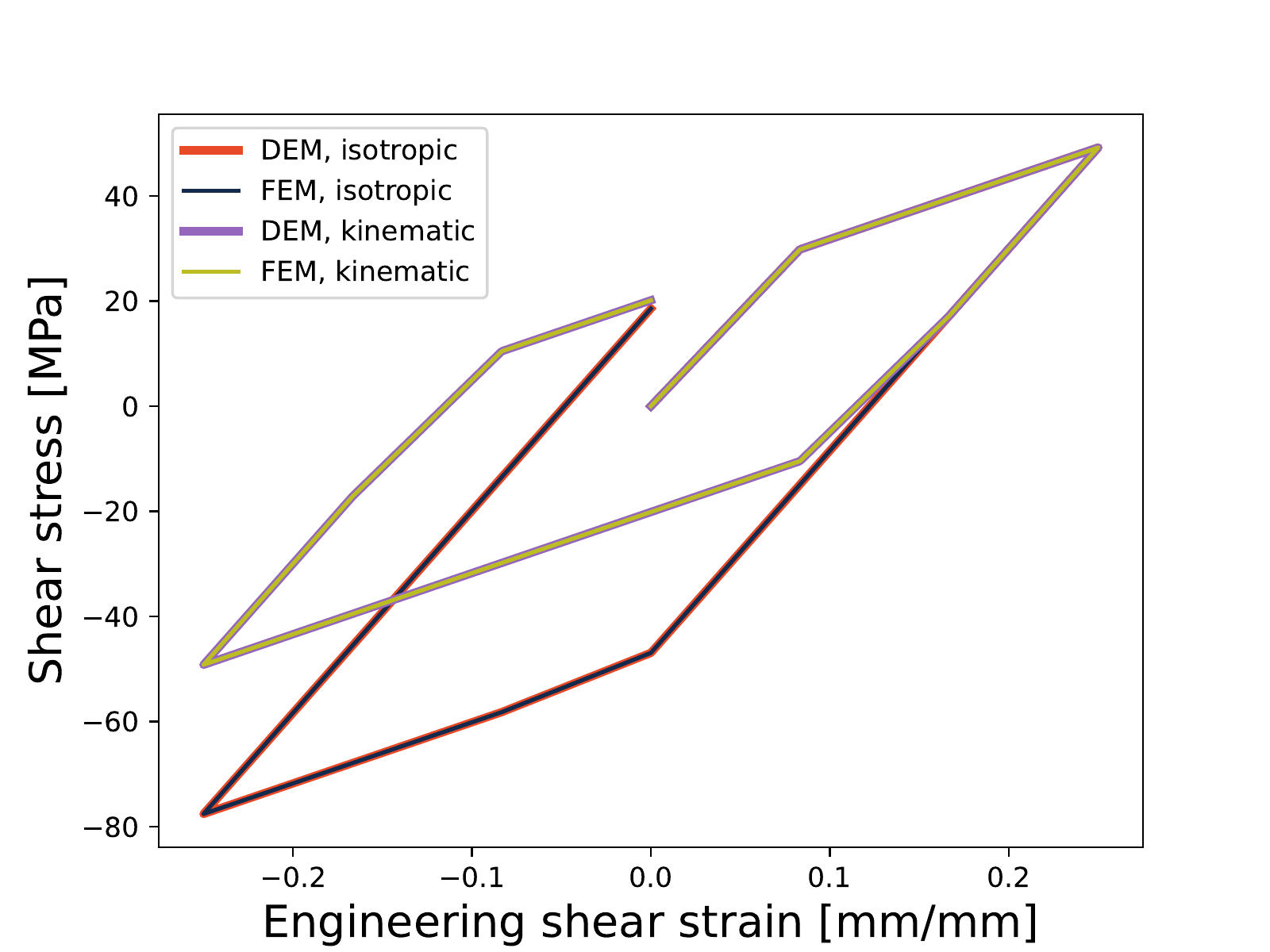}
         \label{ss}
     }
     \subfloat[]{
         \includegraphics[trim={0cm 0cm 0cm 1.4cm},clip,width=0.4\textwidth]{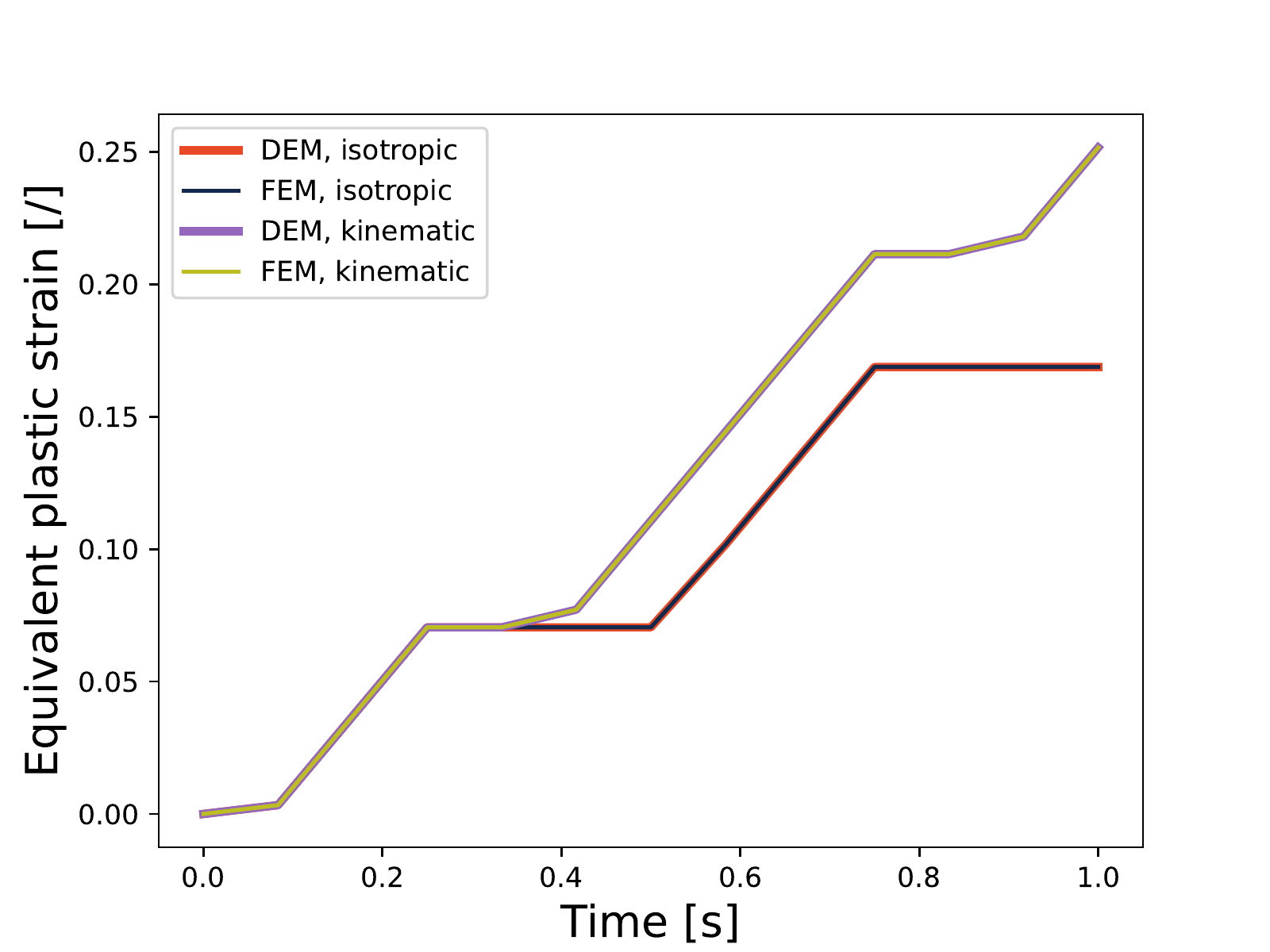}
         \label{peeq}
     }
    \caption{Comparing DEM with FEM: \psubref{ss} Simulated stress-strain curves. \psubref{peeq} Evolution of equivalent plastic strain.}
    \label{uniform}
\end{figure}

Since the analytical solution has a constant state of stress, the average (over all elements) shear stress and engineering shear strain are plotted for both DEM and FEM, the results are shown in \fref{ss}. The comparison for equivalent plastic strain as a function of time is shown in \fref{peeq}. The mean absolute differences over all 12 load steps and simulation times are shown in \tref{tab:accuracy1}.
\begin{table}[h!]
    \caption{Accuracy metrics and simulation times for stress-strain curves generation}
    \small
    \centering
    \begin{tabular}{cccccccc}
     Hardening model & \vline & $AD_{\sigma_{12}}$ [MPa] & $AD_{\Bar{\epsilon}^p}$ [/] & FEM time [s] & DEM time [s]\\
    \hline
    Isotropic & \vline  & 8.06$\times 10^{-5}$ & 9.48$\times 10^{-8}$ & 8 & 25\\
    Kinematic & \vline  & 4.11$\times 10^{-5}$ & 1.67$\times 10^{-7}$ & 9 & 33\\
    \end{tabular}
    \label{tab:accuracy1}
\end{table}
From the results, we see that the proposed DEM method accurately captures the stress-strain responses of the isotropic and kinematic hardening models in cyclic loading, generating results that match closely with FEM. However, we do note that in this case, the training time for DEM is about 2-3 times larger than the corresponding FEM solution time.

\subsection{Shear of a bi-material plate}
\label{bimat}
\begin{figure}[h!] 
    \centering
         \includegraphics[trim={0cm 0cm 0cm 0cm},clip,width=0.32\textwidth]{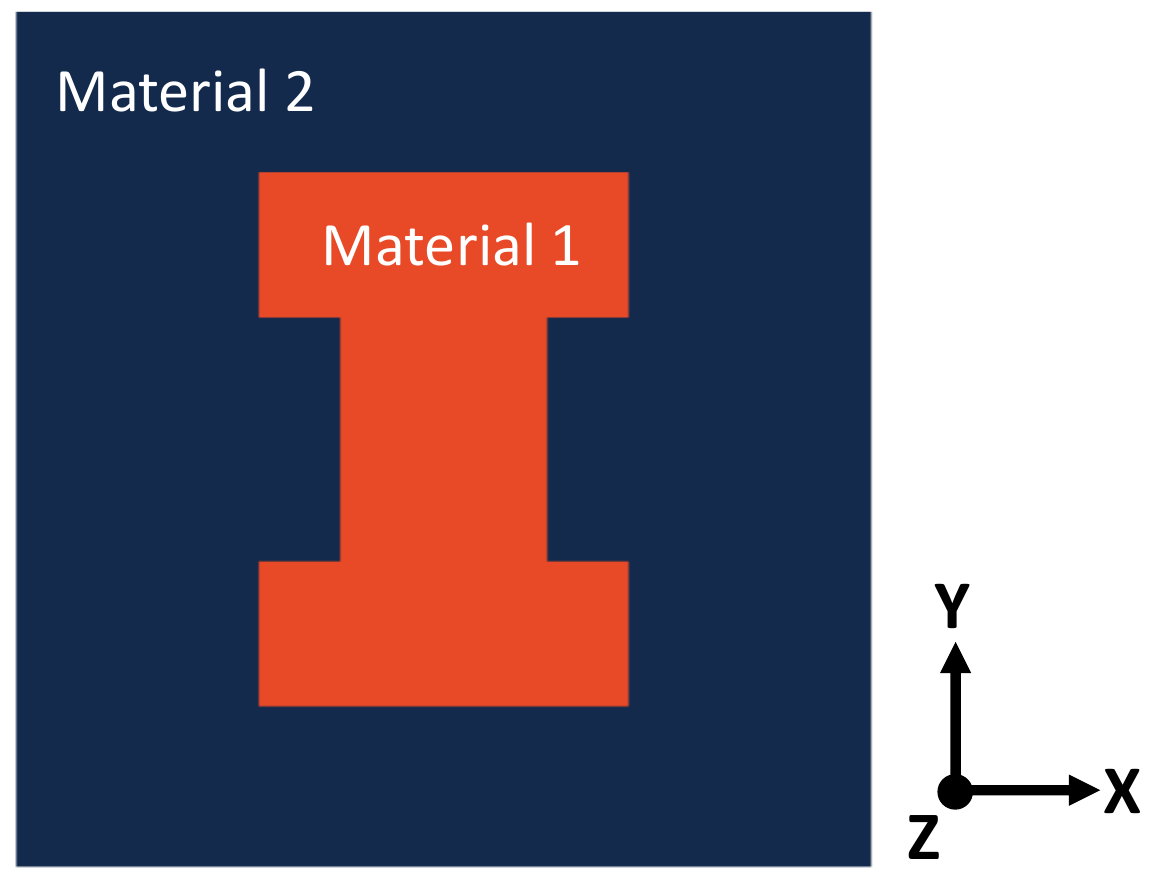}
    \caption{Material assignment in the bi-material plate problem.}
    \label{material_assignment}
\end{figure}

The previous example considered a single material, leading to a uniform stress distribution under the applied boundary conditions. In this example, we consider a bi-material configuration, shown in \fref{material_assignment}. Materials 1 and 2 share identical elastic properties and an isotropic hardening modulus of 500 MPa. The initial yield stress of material 1 is 50 MPa while material 2 is 60 MPa. All applied boundary conditions are identical to \sref{SS}, except that we set $f(t)=0.5/t$, a monotonic loading. The simple boundary conditions were adopted so that no stress concentration arises due to the boundary conditions, and all stress variations that appear are due to the heterogeneous material assignment. The DEM simulation was conducted using a single load step with a convergence tolerance of $1\times 10^{-6}$. 

\begin{figure}[h!]
\newcommand\x{0.22}
\captionsetup[subfigure]{labelformat=empty}
    \centering
    \begin{tabular}{ c c c }
    \begin{minipage}[c]{\x\textwidth}
       \centering 
        \subfloat[$U_x$, FEM]{\includegraphics[trim={24cm 5.5cm 17cm 7cm},clip,width=\textwidth]{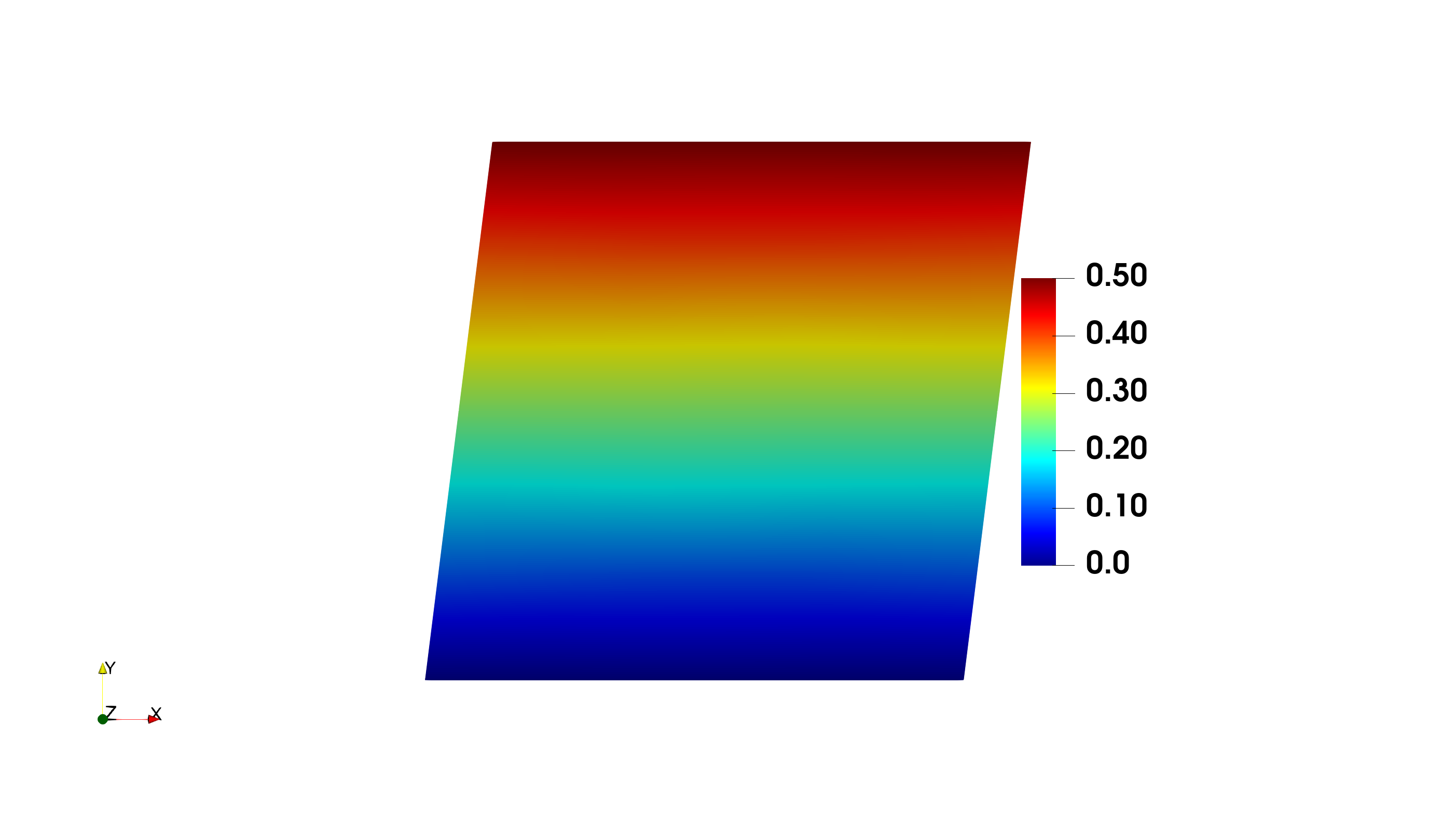}
        }
    \end{minipage}
    &
    \begin{minipage}[c]{\x\textwidth}
       \centering 
        \subfloat[$\Bar{\sigma}$, FEM]{\includegraphics[trim={24cm 5.5cm 17cm 7cm},clip,width=\textwidth]{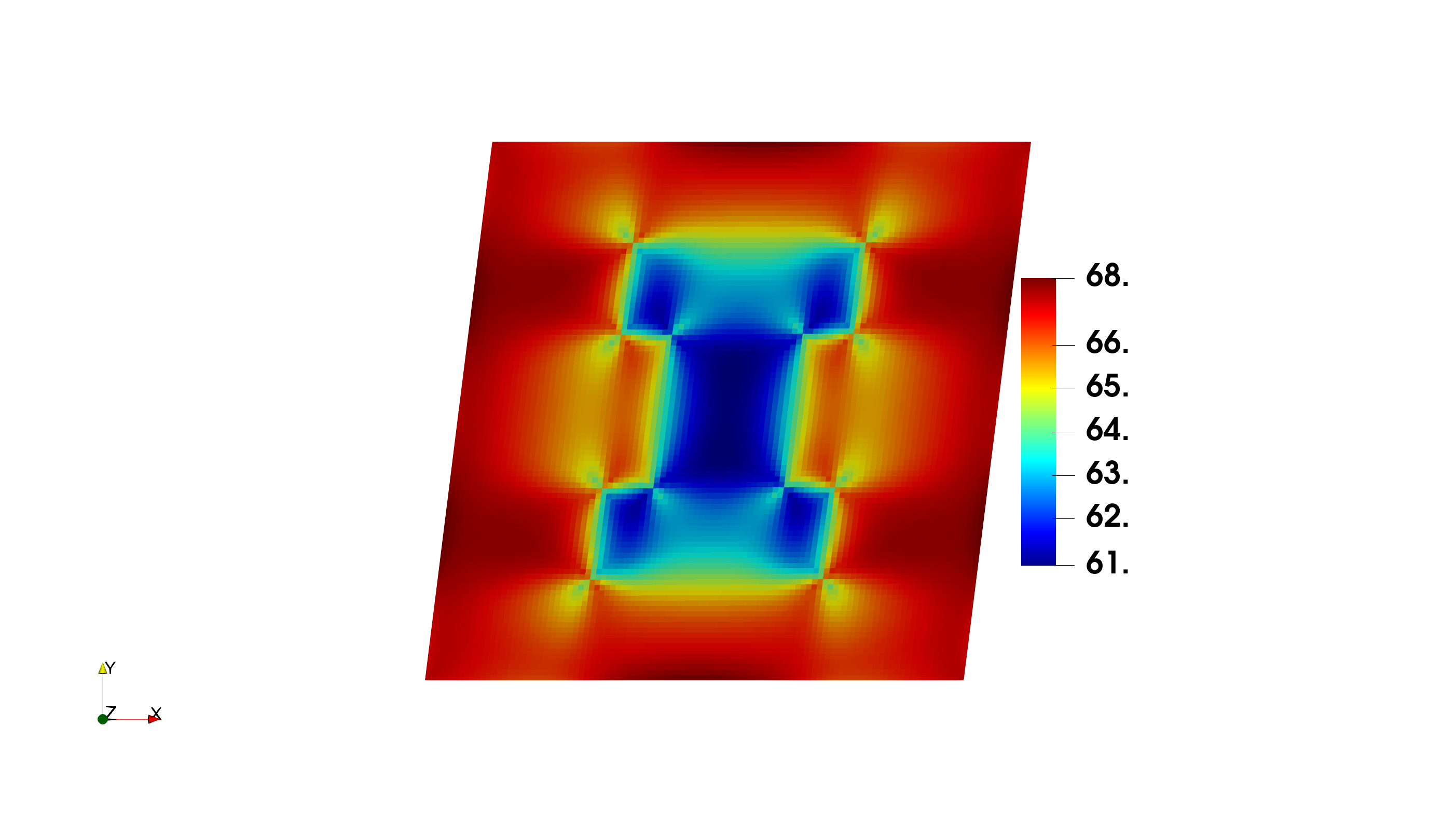}
        }
    \end{minipage}
    &
    \begin{minipage}[c]{\x\textwidth}
       \centering 
        \subfloat[$\Bar{\epsilon}^p$, FEM]{\includegraphics[trim={24cm 5.5cm 17cm 7cm},clip,width=\textwidth]{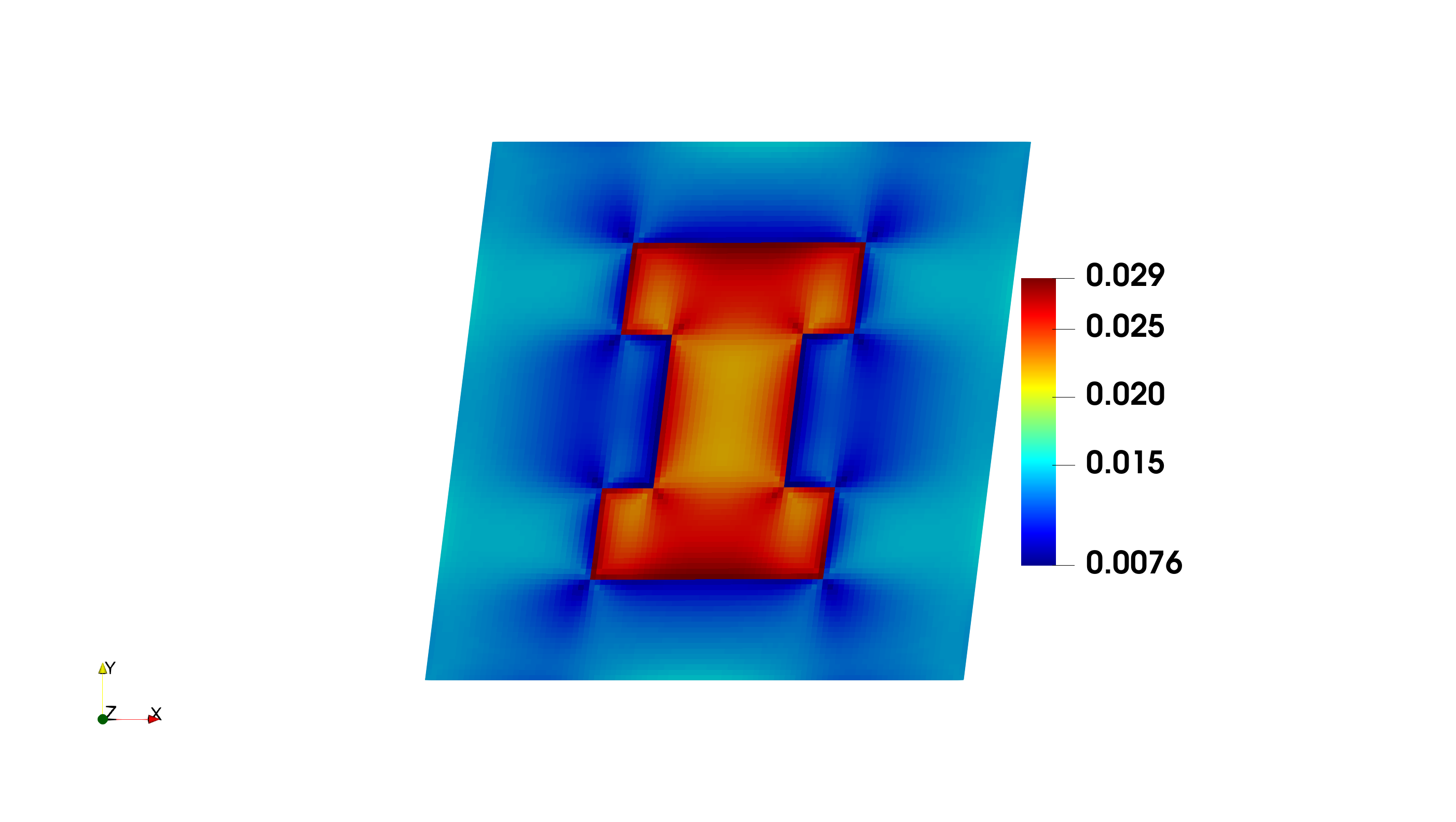}
        }
    \end{minipage}
    
    \\
    
    \begin{minipage}[c]{\x\textwidth}
       \centering 
        \subfloat[$AD_{U_x}$, DEM]{\includegraphics[trim={24cm 5.5cm 13cm 7cm},clip,width=\textwidth]{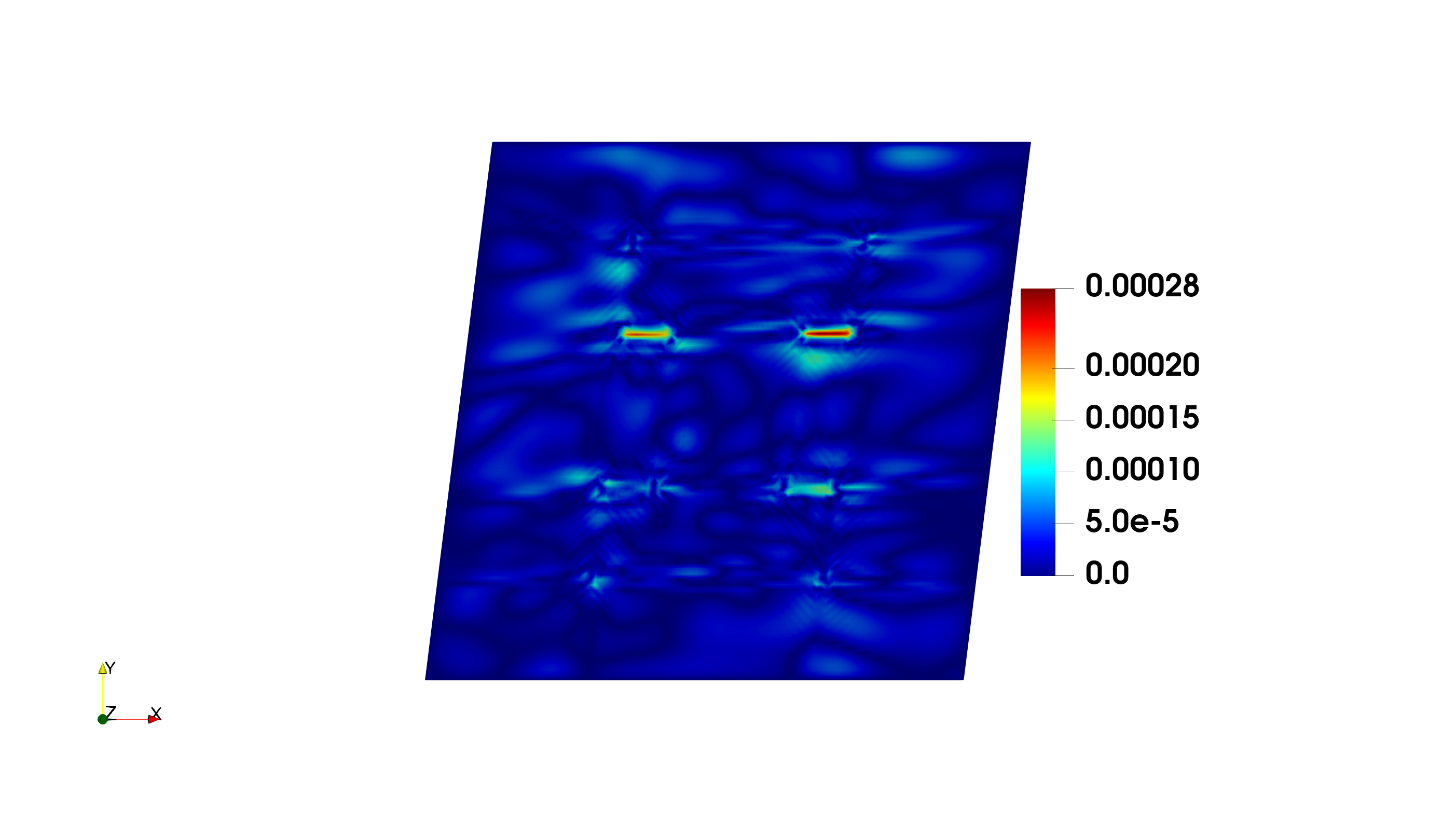}
        }
    \end{minipage}
    &
    \begin{minipage}[c]{\x\textwidth}
       \centering 
        \subfloat[$AD_{\Bar{\sigma}}$, DEM]{\includegraphics[trim={24cm 5.5cm 13cm 7cm},clip,width=\textwidth]{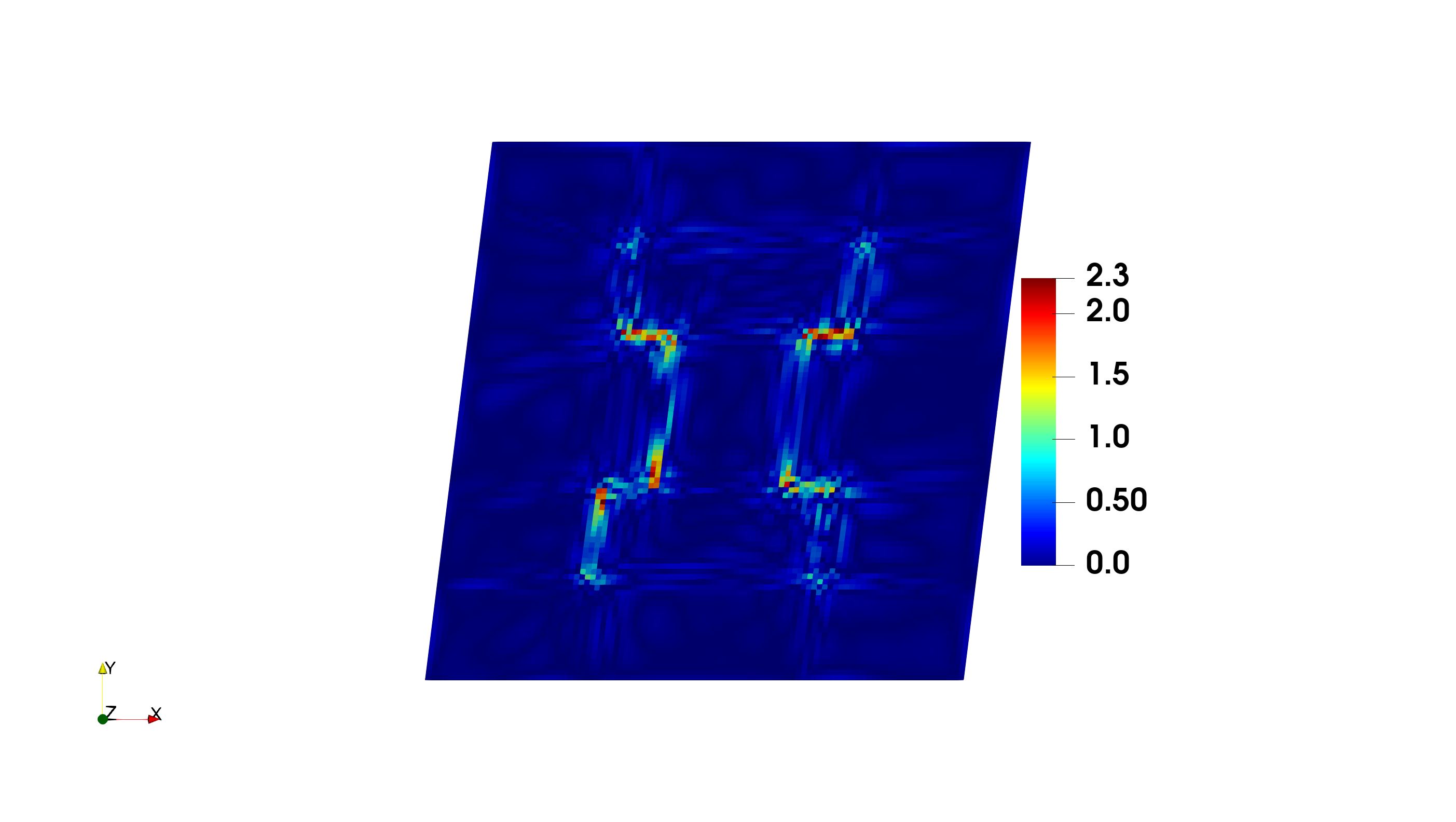}
        }
    \end{minipage}
    &
    \begin{minipage}[c]{\x\textwidth}
       \centering 
        \subfloat[$AD_{\Bar{\epsilon}^p}$, DEM]{\includegraphics[trim={24cm 5.5cm 13cm 7cm},clip,width=\textwidth]{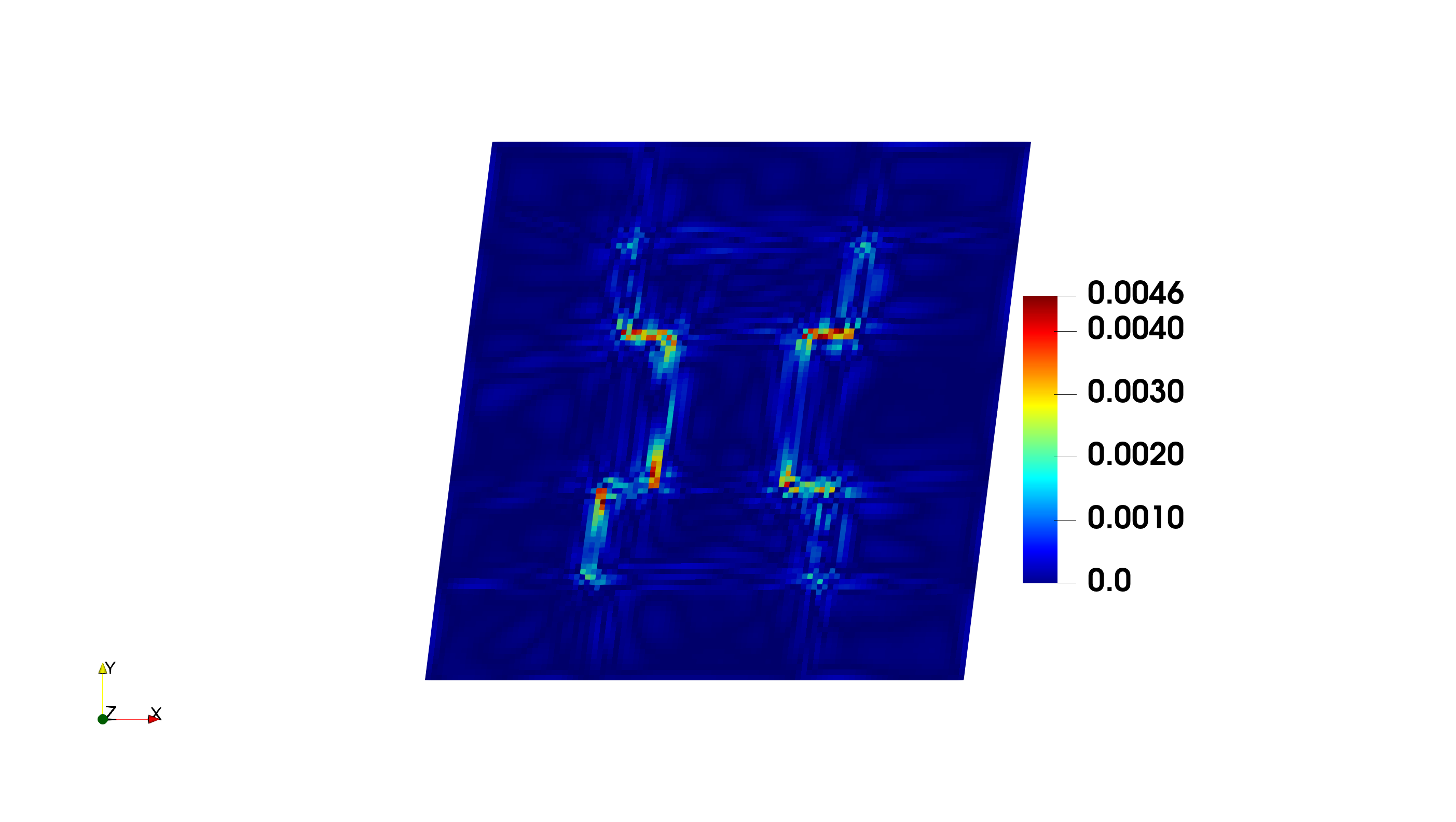}
        }
    \end{minipage}

    \end{tabular}
    \caption{Comparison between FEM and DEM for the bi-material example. Row 1: contour plots of $U_x$, $\Bar{\sigma}$ and $\Bar{\epsilon}^p$ predicted by the reference FEM solution. Row 2: absolute differences compared to FEM, DEM solution. The deformed shapes are computed from the DEM displacements with a scaling factor of 1.}
    \label{bimat_contours}
\end{figure}

The contour plots comparing the DEM predictions and FEM reference solutions are shown in \fref{bimat_contours}. Quantitative performance metrics are listed in \tref{bimat_metrics}. From the contour plots, we observe that the DEM prediction errors are localized at the material interface, where a stress jump occurs. However, the stress and equivalent plastic strain errors are relatively small compared to the magnitude of the reference FEM solution. The displacement solution is also highly accurate as can be seen in the low $L^2$ error norm for the displacement in \tref{bimat_metrics}. Even if the domain and boundary conditions as similar to those in \sref{SS}, the heterogeneous material assignment and the localized stress concentration it induced significantly increased the DEM training time. In this case, DEM training time is more than 15 times longer than the FEM solution time. 
\begin{table}[h!]
    \caption{Accuracy metrics and simulation times for the bi-material example}
    \small
    \centering
    \begin{tabular}{ccccccccc}
     Case & \vline & FEM [s] & DEM [s] &  $AD_{U_{x}}$ [mm] & $AD_{U_{y}}$ [mm] & $AD_{\Bar{\sigma}}$ [MPa] & $AD_{\Bar{\epsilon}^p}$ [/] & $||\bm{e}||_{L^2}$ [\%] \\
    \hline
    Bi-mat & \vline  & 10 & 182 & 2.03$\times 10^{-5}$ & 2.27$\times 10^{-8}$ & 9.82$\times 10^{-2}$ & 1.96$\times 10^{-4}$ & 1.48$\times 10^{-2}$ \\
    \end{tabular}
    \label{bimat_metrics}
\end{table}

\subsection{Comparison between DEM and DCM on an elastoplasticity problem}
\label{DCM}
DCM is another PINN that has been widely applied to solve PDEs in solid mechanics. Its loss function is based on \eref{strong}:
\begin{equation}
    \mathcal{L} = || \nabla \cdot \bm{\sigma} ||_2 + || \Bar{\bm{t}} - \bm{\sigma} \cdot \bm{n} ||_2,
    \label{DCM_loss}
\end{equation}
where the first term is evaluated using all nodes in the domain interior, and the second term is evaluated using all boundary nodes. Unlike DEM, where traction-free boundary conditions are automatically fulfilled (\eref{pext} vanishes on traction-free boundaries), they need to be explicitly included in the DCM loss function in the form of $|| \bm{\sigma} \cdot \bm{n} ||_2$ (setting $\Bar{\bm{t}}=\bm{0}$). The DCM loss function involves the second-order spatial gradients of the displacement field, which are computed using automatic differentiation. DCM has been used to solve elastoplasticity problems in the work of \cite{abueidda2021meshless}. The radial return algorithm was used to perform the plastic state update, but an extremely refined grid with 1.2 $\times 10^{11}$ nodes was used to discretize a rectangular domain for an accurate solution. 

To obtain a fair comparison of performance and accuracy between DCM and DEM on elastoplasticity problems, we simply modify the loss function implementation from \eref{freeE_iso} to \eref{DCM_loss} to generate a DCM model that uses the same MLP network with 3009 trainable parameters. The geometry is identical to that used in \sref{SS}, therefore a grid of 101$\times$101$\times$2 nodes was used in automatic differentiation when evaluating the DCM loss function. A nonlinear displacement field was applied:
\begin{equation}
\begin{aligned}
    u_x = \frac{y}{8} + 0.1 \, {\rm{sin}( \frac{ \pi }{ 2 } y )},\\
    u_y = 0.
\end{aligned}
\end{equation}
Dirichlet boundary conditions were applied to all external boundaries, so only the first term in \eref{DCM_loss} remains, greatly simplifying the loss function. A single load step and a convergence tolerance of $1\times 10^{-6}$ were used. Both the isotropic and kinematic hardening models were tested using the two methods.

\begin{figure}[h!]
\newcommand\x{0.22}
\captionsetup[subfigure]{labelformat=empty}
    \centering
    \begin{tabular}{ c c c }
    \begin{minipage}[c]{\x\textwidth}
       \centering 
        \subfloat[$U_x$, FEM]{\includegraphics[trim={24cm 5.5cm 17cm 7cm},clip,width=\textwidth]{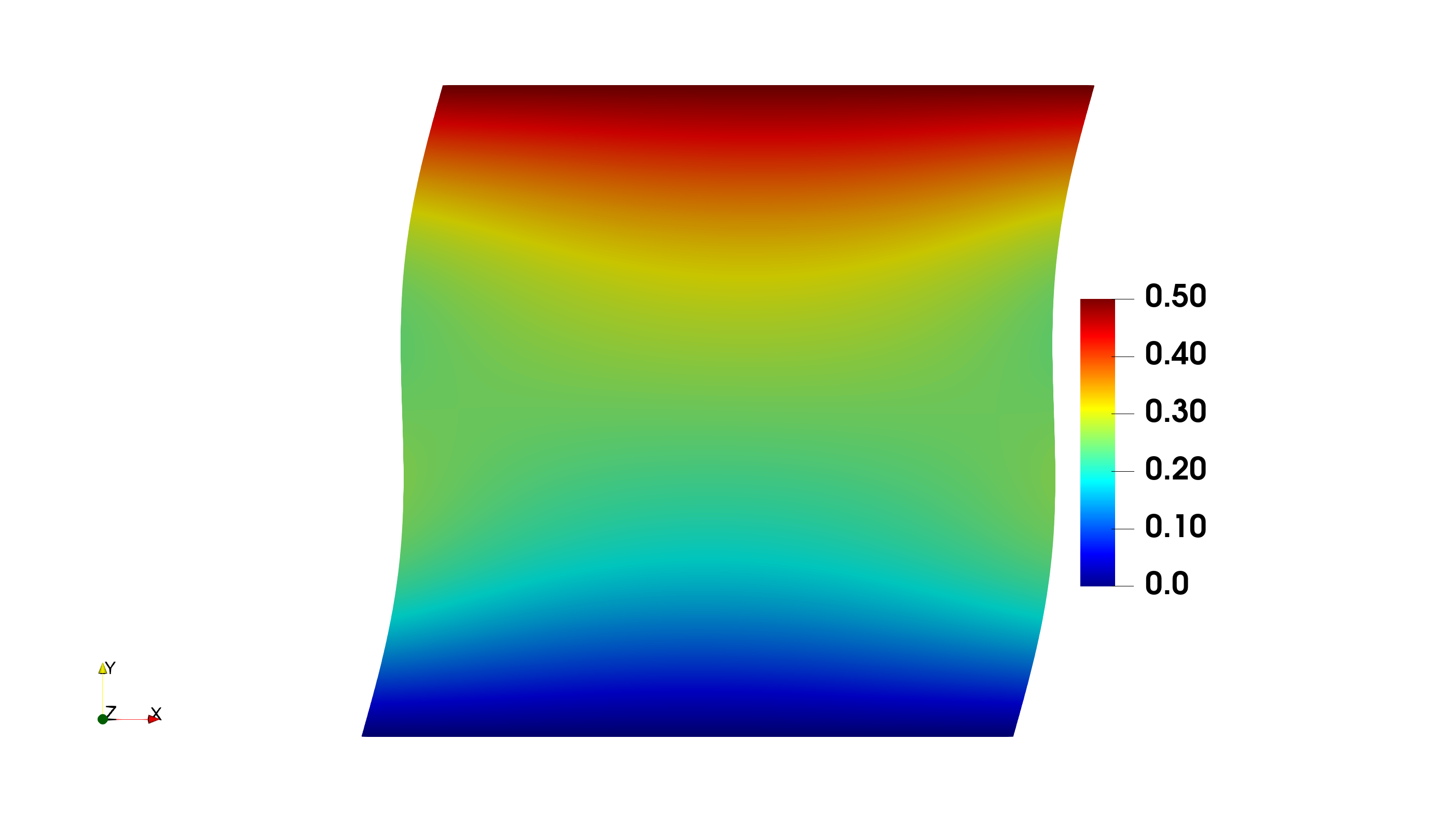}
        }
    \end{minipage}
    &
    \begin{minipage}[c]{\x\textwidth}
       \centering 
        \subfloat[$\Bar{\sigma}$, FEM]{\includegraphics[trim={24cm 5.5cm 17cm 7cm},clip,width=\textwidth]{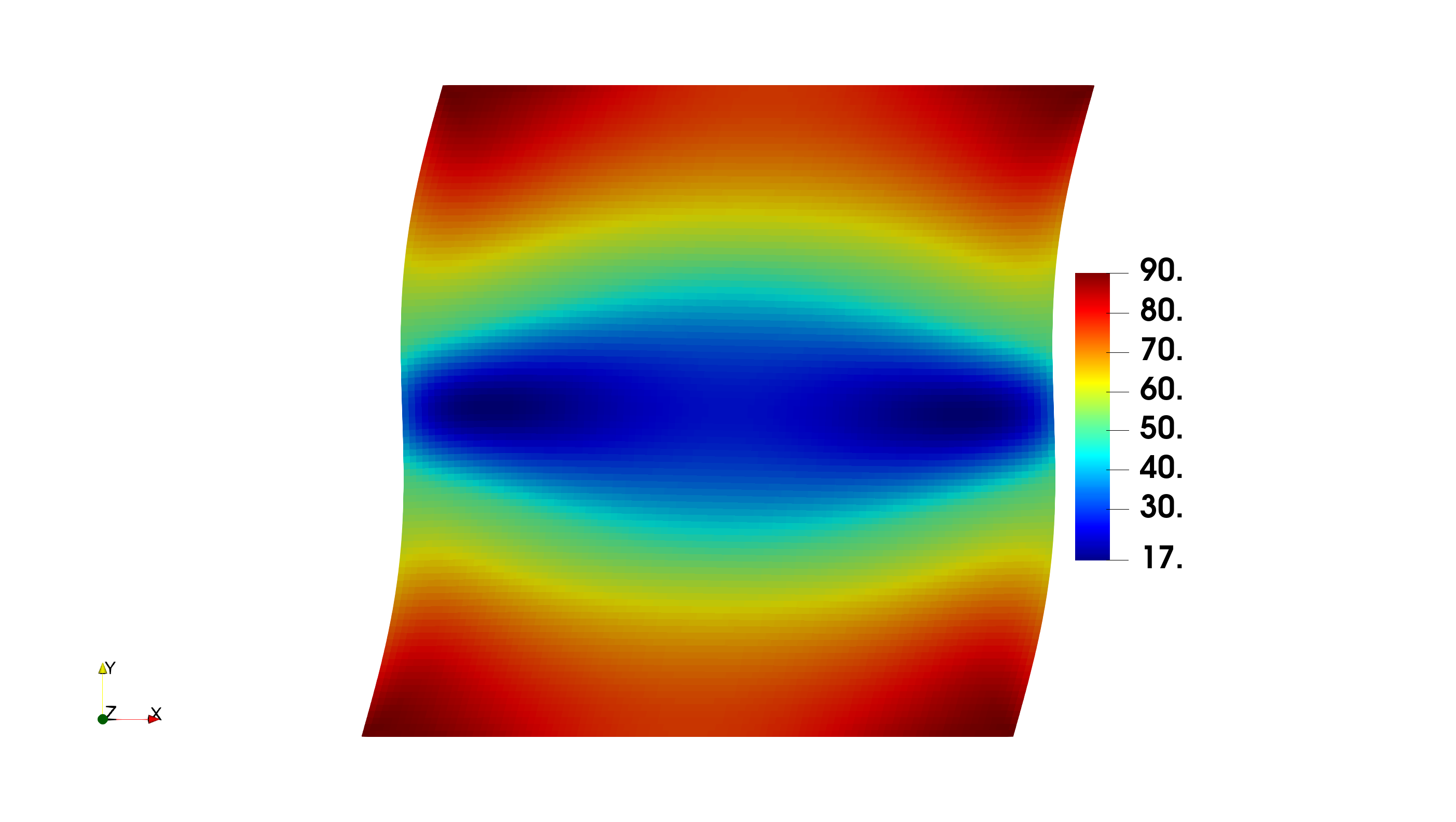}
        }
    \end{minipage}
    &
    \begin{minipage}[c]{\x\textwidth}
       \centering 
        \subfloat[$\Bar{\epsilon}^p$, FEM]{\includegraphics[trim={24cm 5.5cm 17cm 7cm},clip,width=\textwidth]{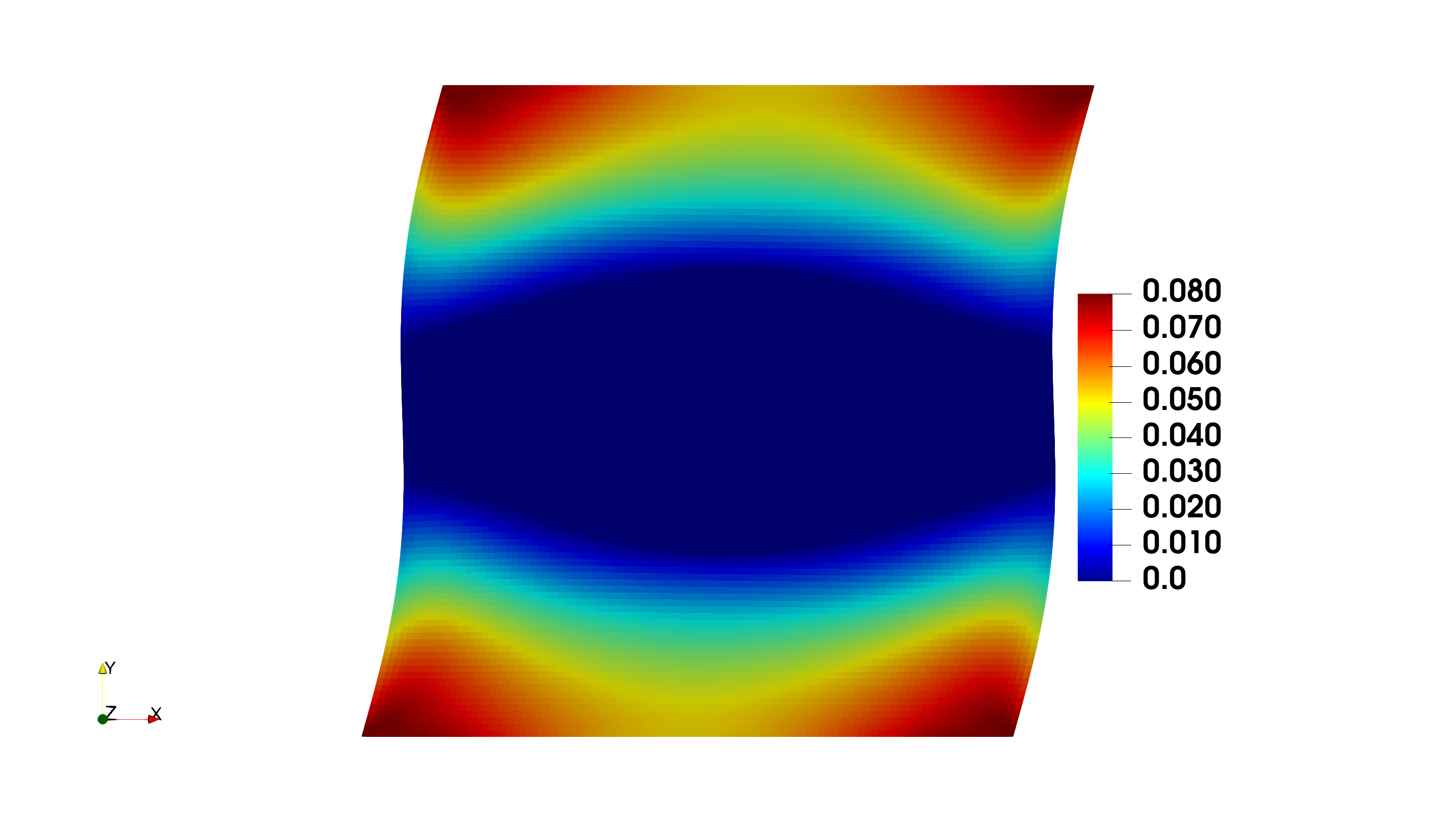}
        }
    \end{minipage}
    
    \\
    
    \begin{minipage}[c]{\x\textwidth}
       \centering 
        \subfloat[$AD_{U_x}$, DEM]{\includegraphics[trim={24cm 5.5cm 13cm 7cm},clip,width=\textwidth]{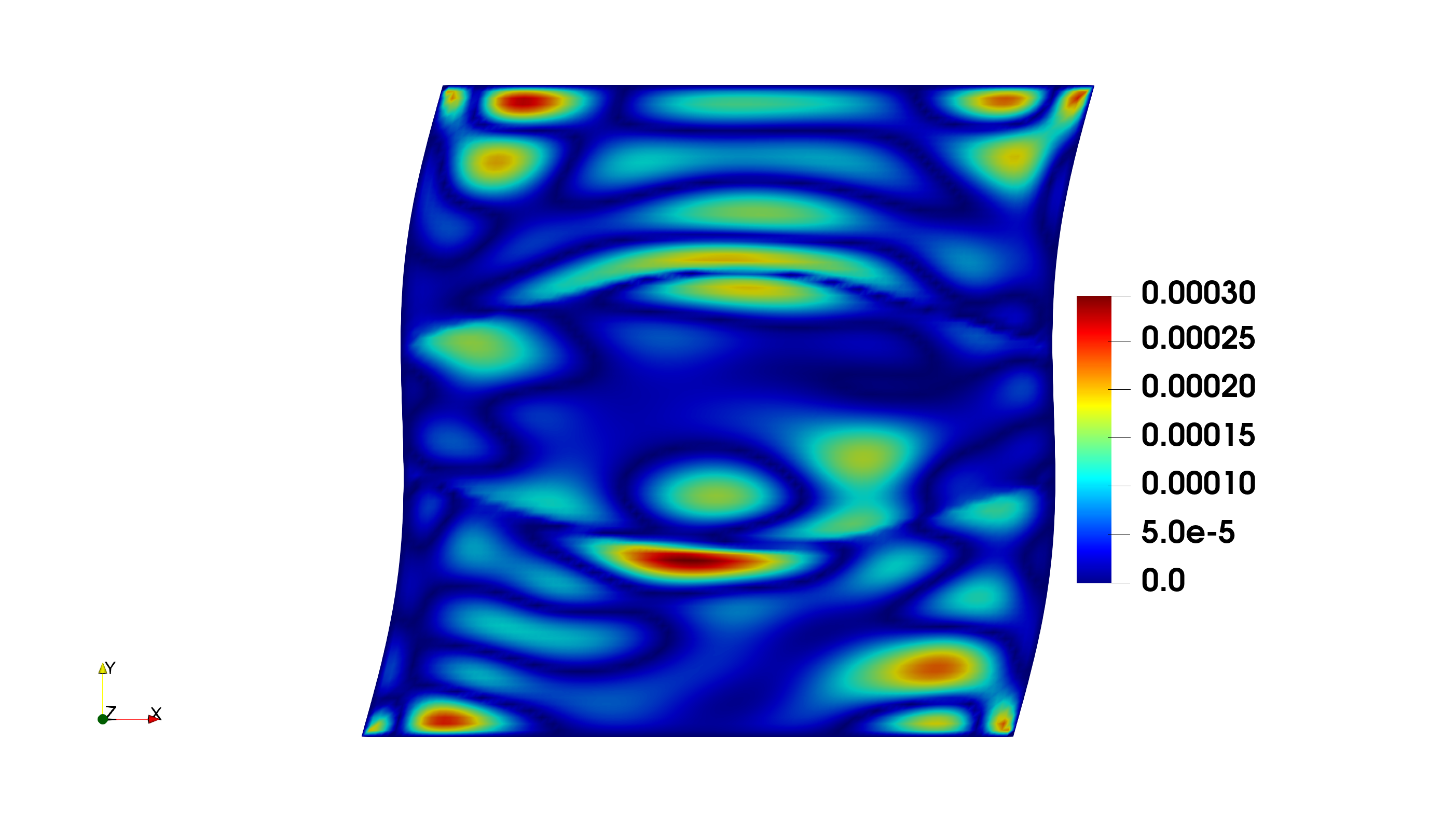}
        }
    \end{minipage}
    &
    \begin{minipage}[c]{\x\textwidth}
       \centering 
        \subfloat[$AD_{\Bar{\sigma}}$, DEM]{\includegraphics[trim={24cm 5.5cm 13cm 7cm},clip,width=\textwidth]{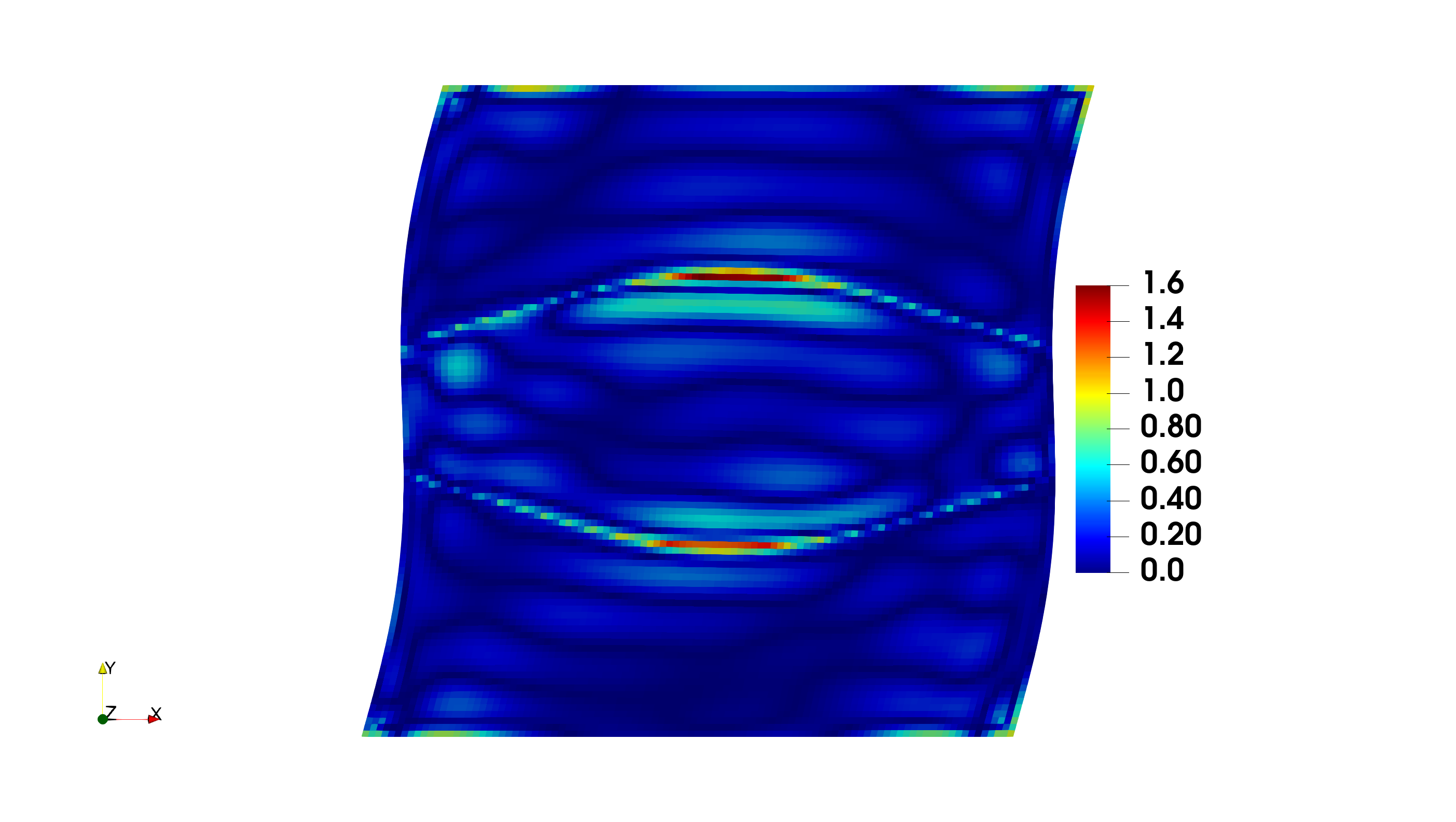}
        }
    \end{minipage}
    &
    \begin{minipage}[c]{\x\textwidth}
       \centering 
        \subfloat[$AD_{\Bar{\epsilon}^p}$, DEM]{\includegraphics[trim={24cm 5.5cm 13cm 7cm},clip,width=\textwidth]{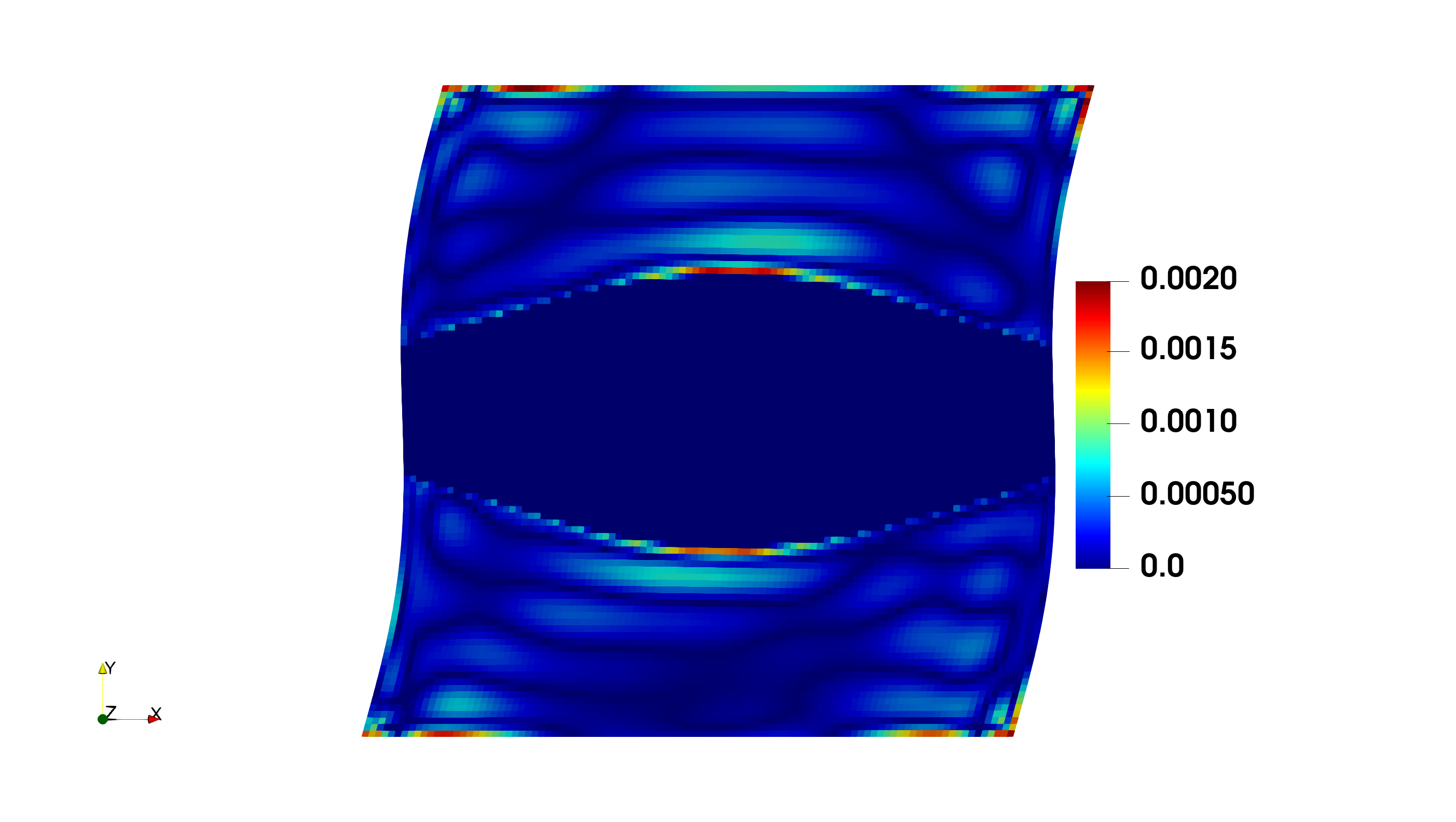}
        }
    \end{minipage}

    \\
    
    \begin{minipage}[c]{\x\textwidth}
       \centering 
        \subfloat[$AD_{U_x}$, DCM]{\includegraphics[trim={28cm 9cm 17cm 7cm},clip,width=\textwidth]{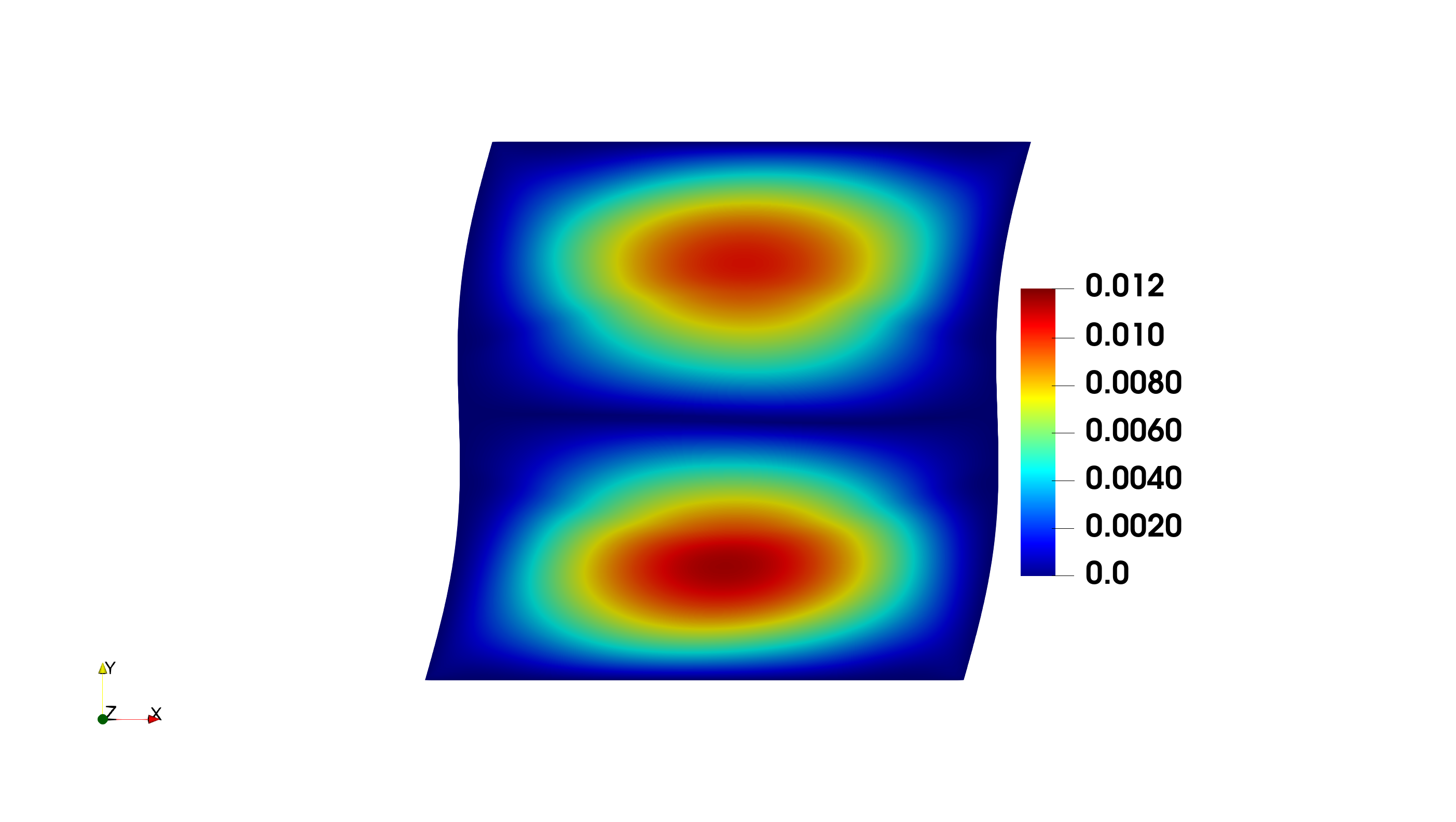}
        }
    \end{minipage}
    &
    \begin{minipage}[c]{\x\textwidth}
       \centering 
        \subfloat[$AD_{\Bar{\sigma}}$, DCM]{\includegraphics[trim={28cm 9cm 17cm 7cm},clip,width=\textwidth]{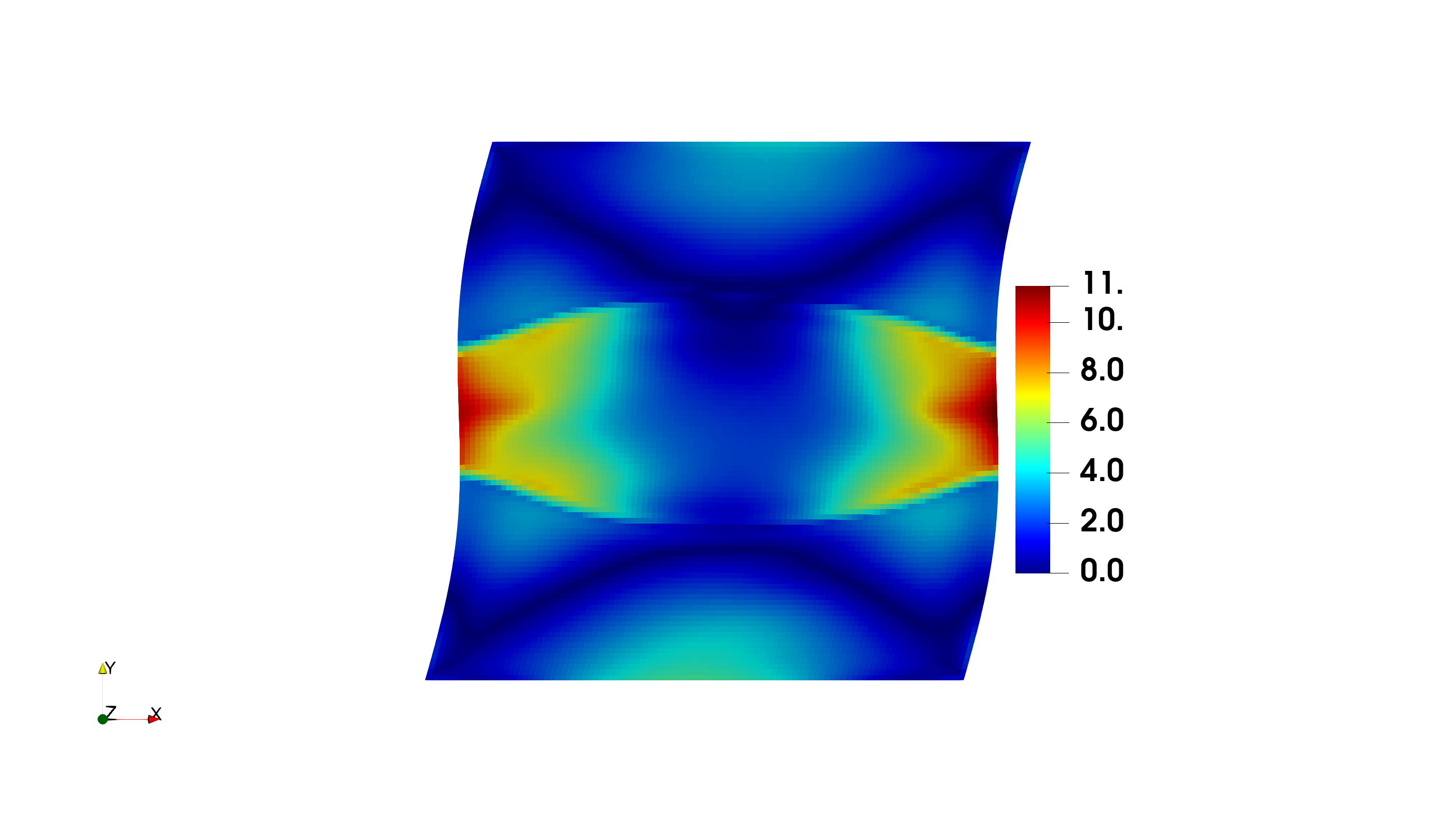}
        }
    \end{minipage}
    &
    \begin{minipage}[c]{\x\textwidth}
       \centering 
        \subfloat[$AD_{\Bar{\epsilon}^p}$, DCM]{\includegraphics[trim={28cm 9cm 17cm 7cm},clip,width=\textwidth]{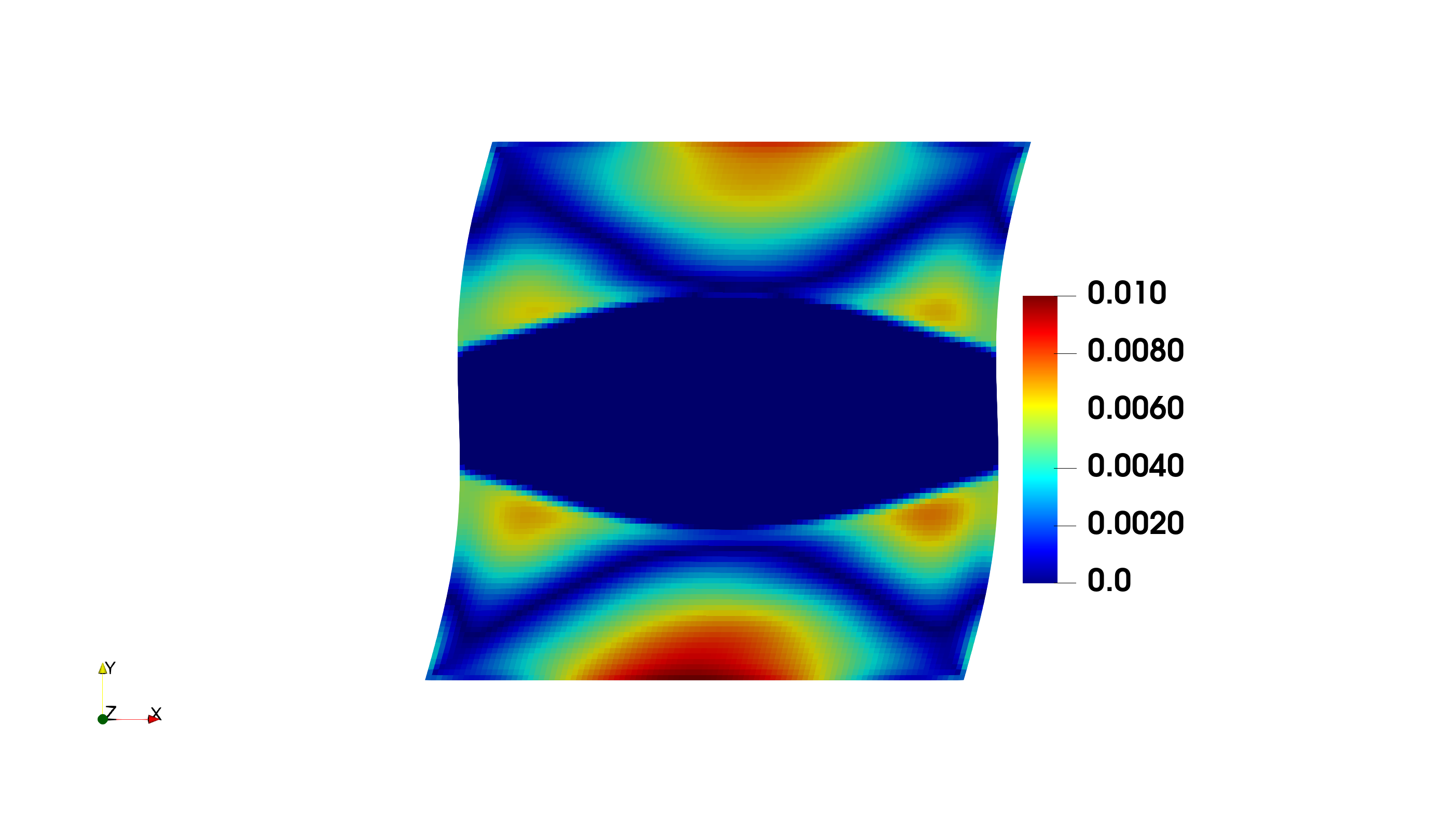}
        }
    \end{minipage}
 
    \end{tabular}
    \caption{Comparison between FEM, DEM, and DCM, isotropic hardening model. Row 1: FEM reference solution. Row 2: absolute differences in the DEM solution. Row 3: absolute differences in the DCM solution. The color scales for each row are different (indicated in the corresponding color bars), and deformed shapes are computed from the DEM displacements with a scaling factor of 1.}
    \label{Contour1}
\end{figure}

\begin{figure}[h!]
\newcommand\x{0.22}
\captionsetup[subfigure]{labelformat=empty}
    \centering
    \begin{tabular}{ c c c }
    \begin{minipage}[c]{\x\textwidth}
       \centering 
        \subfloat[$U_x$, FEM]{\includegraphics[trim={24cm 5.5cm 17cm 7cm},clip,width=\textwidth]{iso_fem_ux.png}
        }
    \end{minipage}
    &
    \begin{minipage}[c]{\x\textwidth}
       \centering 
        \subfloat[$\Bar{\sigma}$, FEM]{\includegraphics[trim={24cm 5.5cm 17cm 7cm},clip,width=\textwidth]{iso_fem_sigma.png}
        }
    \end{minipage}
    &
    \begin{minipage}[c]{\x\textwidth}
       \centering 
        \subfloat[$\Bar{\epsilon}^p$, FEM]{\includegraphics[trim={24cm 5.5cm 17cm 7cm},clip,width=\textwidth]{iso_fem_peeq.png}
        }
    \end{minipage}
    
    \\
    
    \begin{minipage}[c]{\x\textwidth}
       \centering 
        \subfloat[$AD_{U_x}$, DEM]{\includegraphics[trim={28cm 9cm 17cm 7cm},clip,width=\textwidth]{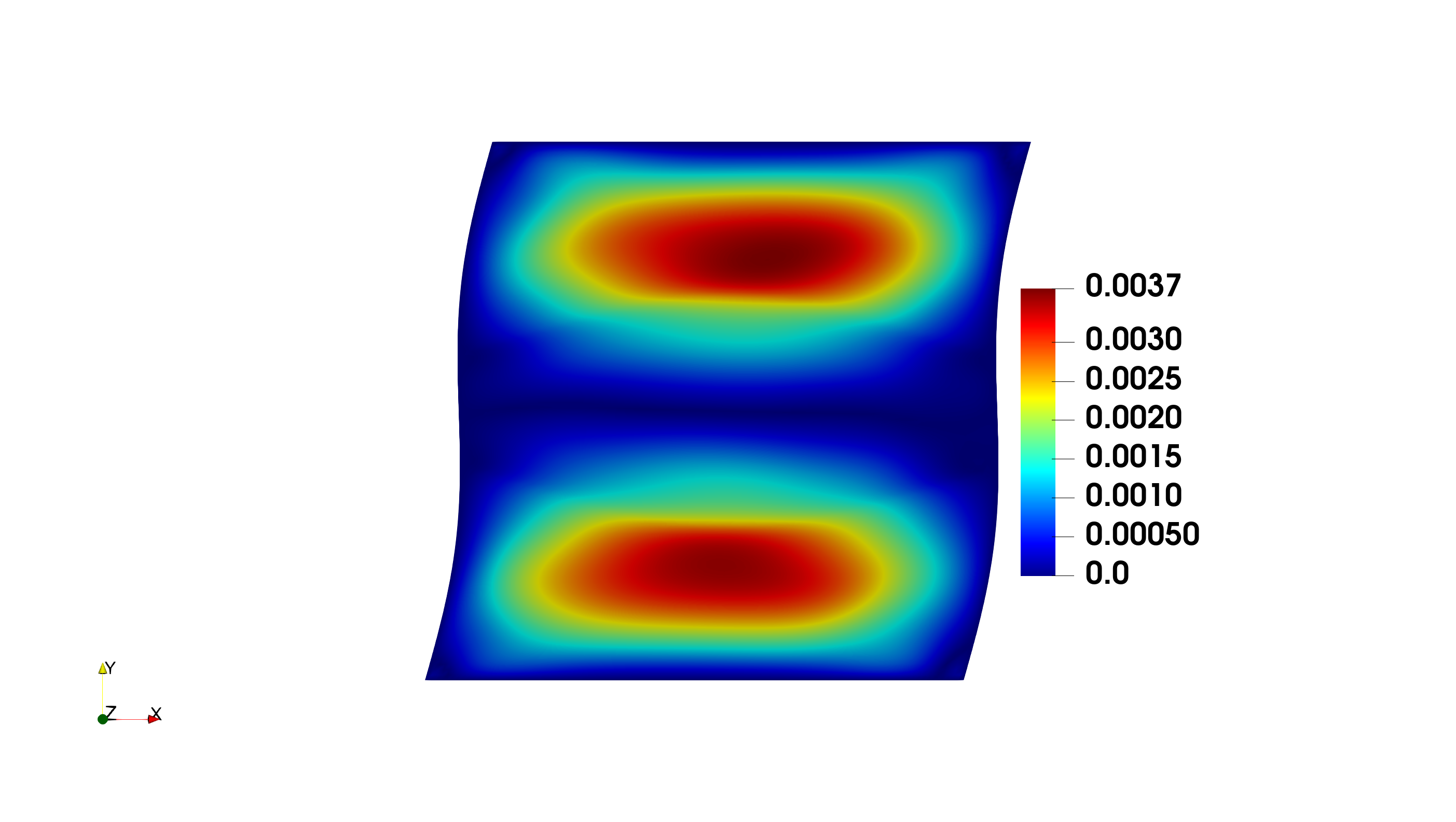}
        }
    \end{minipage}
    &
    \begin{minipage}[c]{\x\textwidth}
       \centering 
        \subfloat[$AD_{\Bar{\sigma}}$, DEM]{\includegraphics[trim={28cm 9cm 17cm 7cm},clip,width=\textwidth]{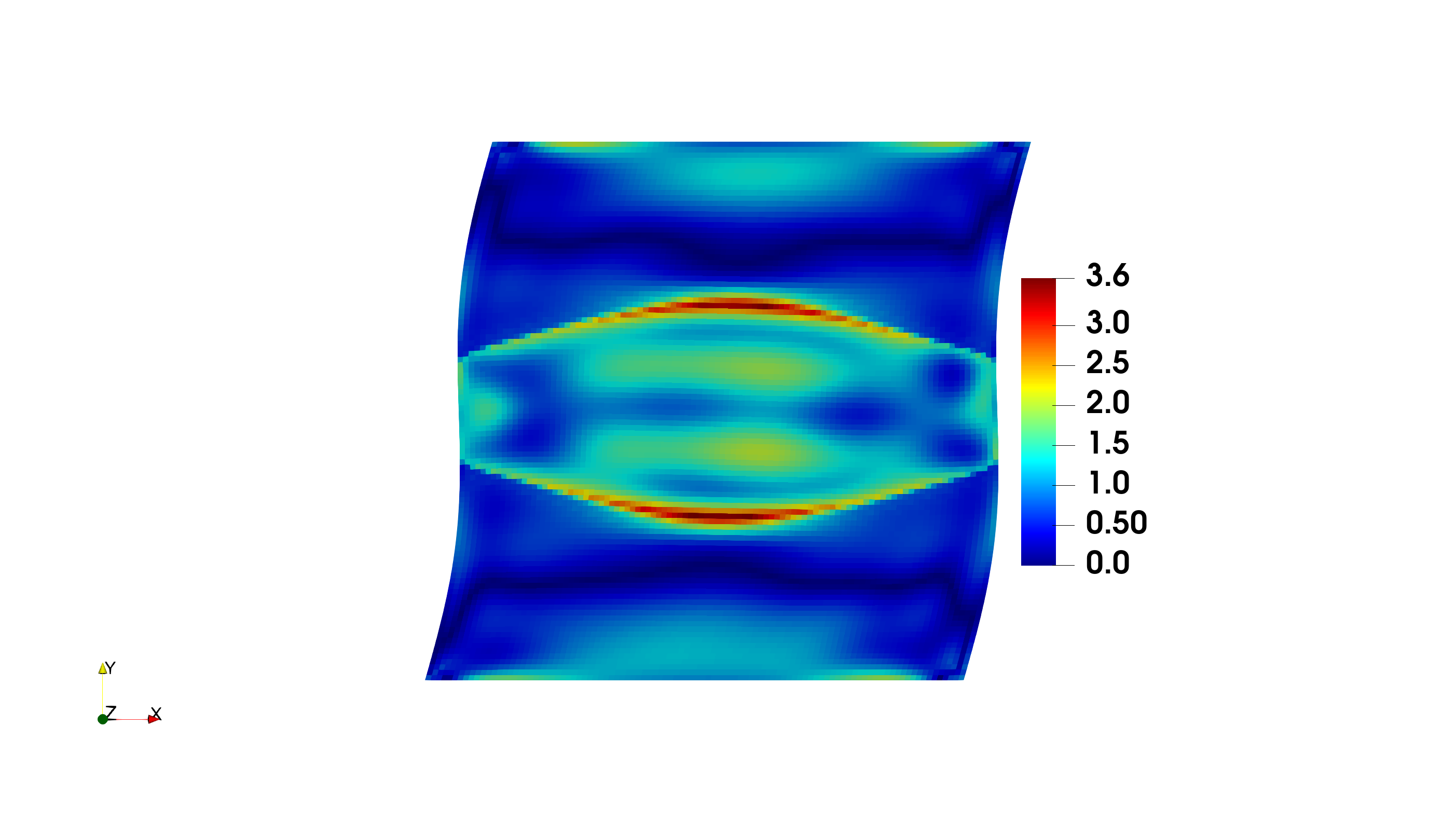}
        }
    \end{minipage}
    &
    \begin{minipage}[c]{\x\textwidth}
       \centering 
        \subfloat[$AD_{\Bar{\epsilon}^p}$, DEM]{\includegraphics[trim={28cm 9cm 17cm 7cm},clip,width=\textwidth]{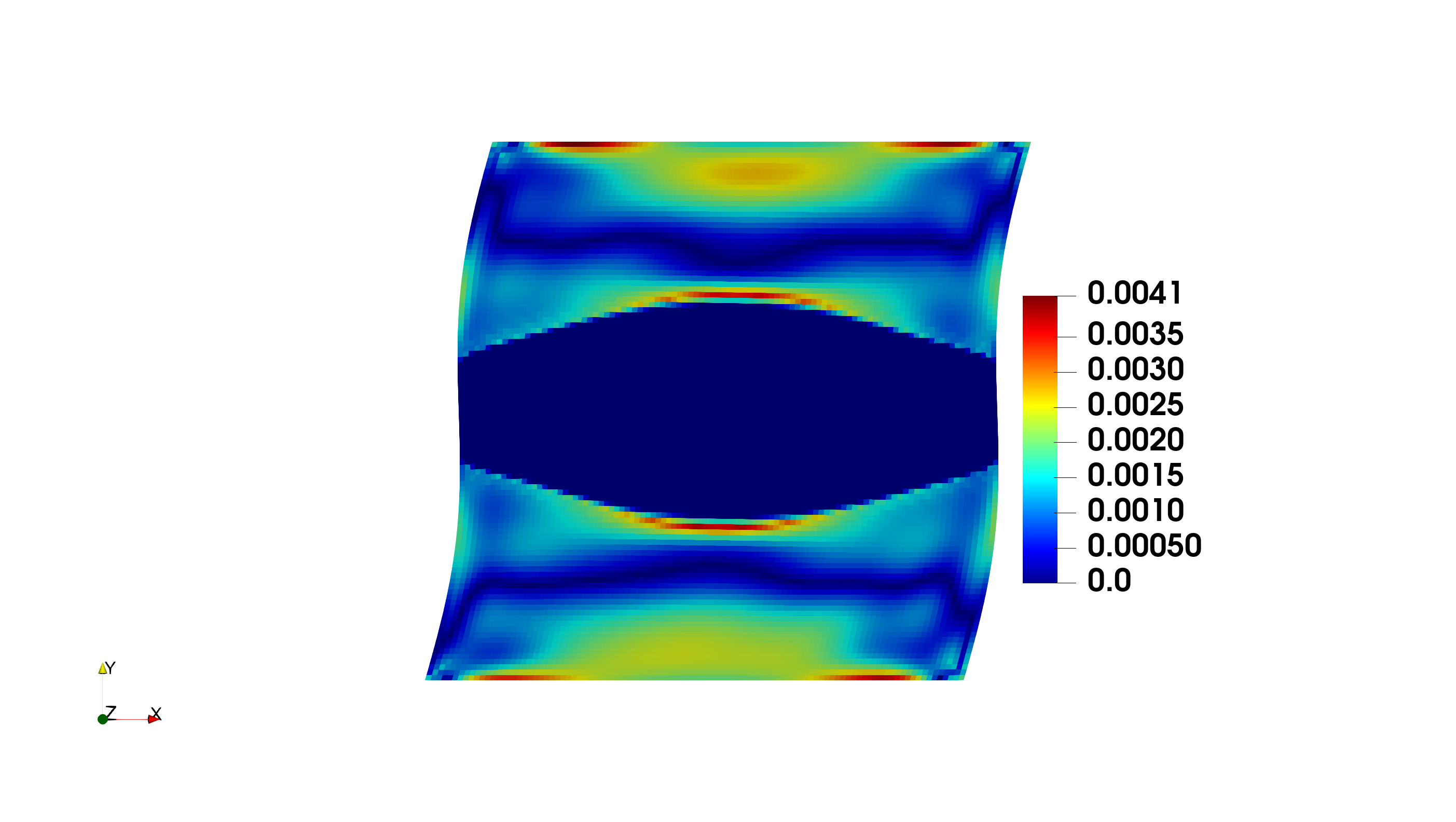}
        }
    \end{minipage}

    \\
    
    \begin{minipage}[c]{\x\textwidth}
       \centering 
        \subfloat[$AD_{U_x}$, DCM]{\includegraphics[trim={28cm 9cm 17cm 7cm},clip,width=\textwidth]{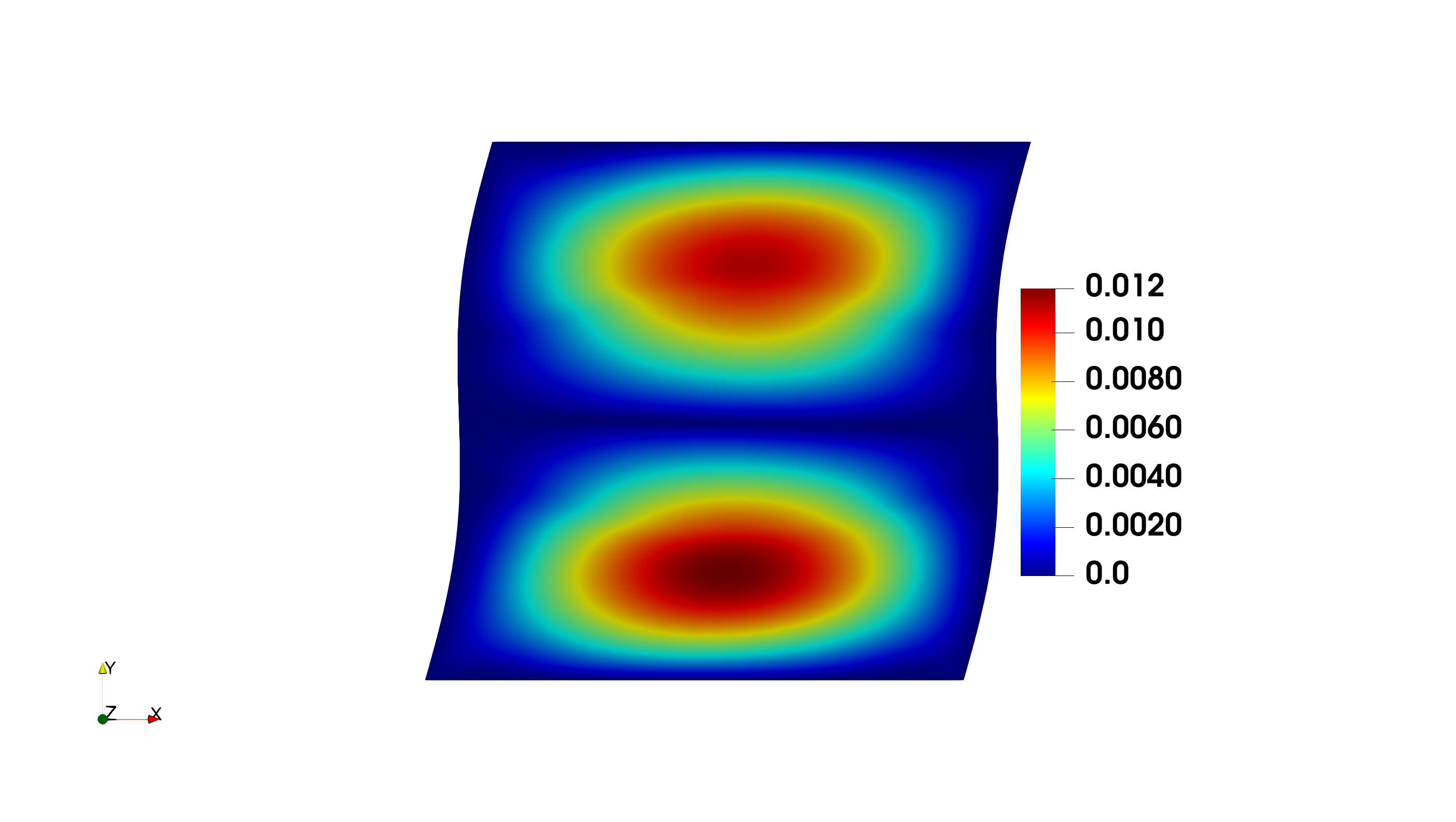}
        }
    \end{minipage}
    &
    \begin{minipage}[c]{\x\textwidth}
       \centering 
        \subfloat[$AD_{\Bar{\sigma}}$, DCM]{\includegraphics[trim={28cm 9cm 17cm 7cm},clip,width=\textwidth]{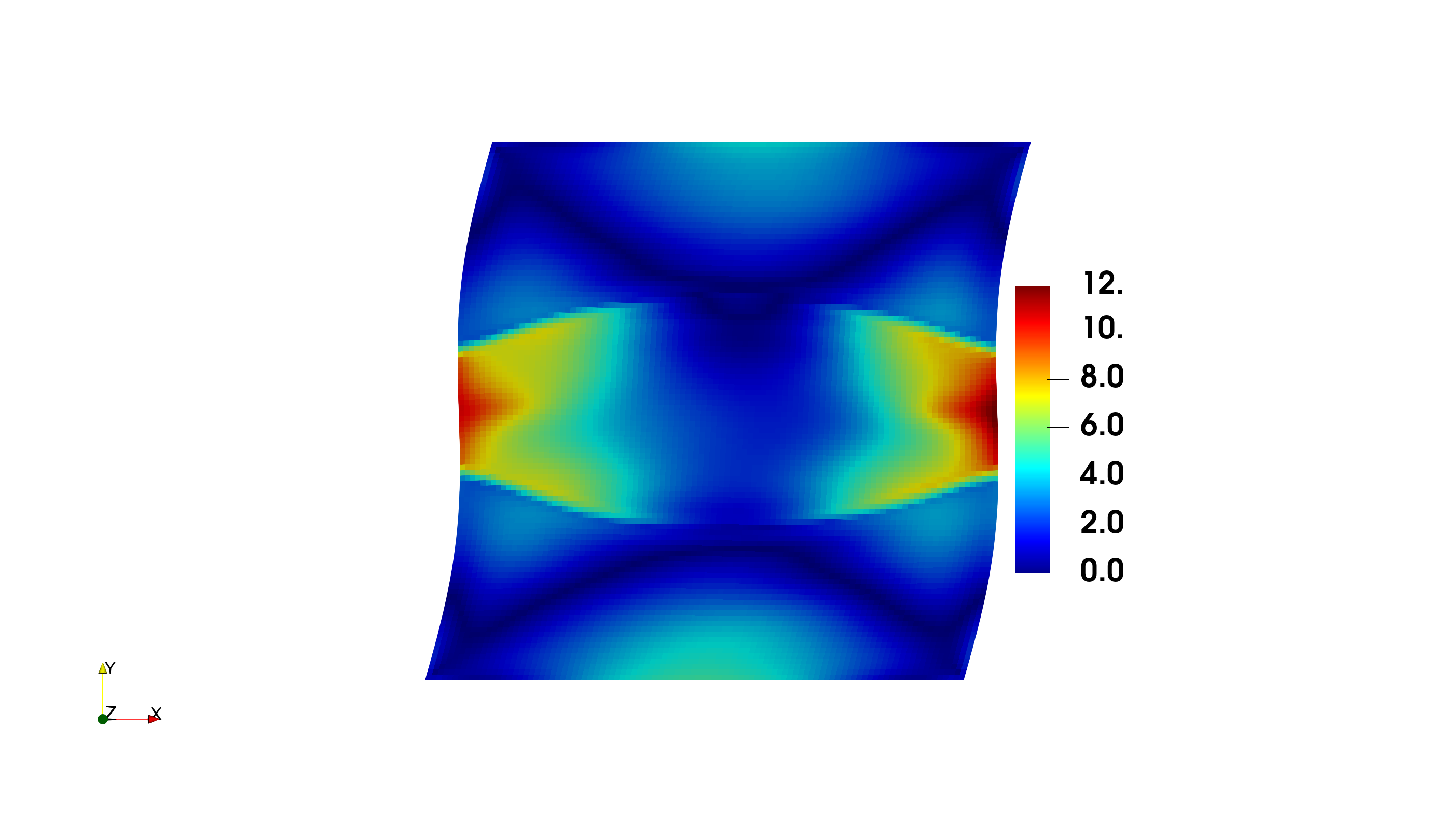}
        }
    \end{minipage}
    &
    \begin{minipage}[c]{\x\textwidth}
       \centering 
        \subfloat[$AD_{\Bar{\epsilon}^p}$, DCM]{\includegraphics[trim={28cm 9cm 17cm 7cm},clip,width=\textwidth]{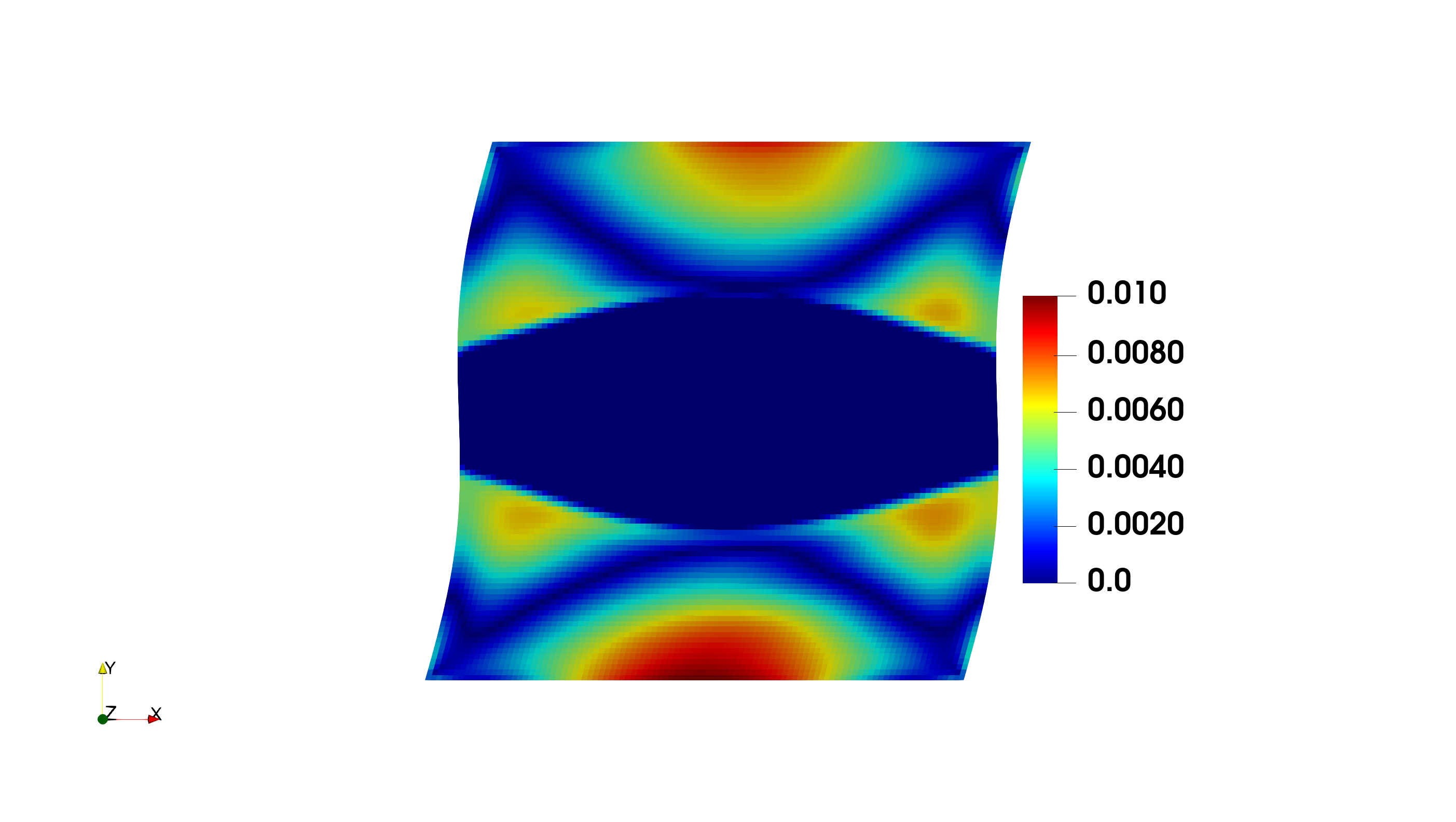}
        }
    \end{minipage}
 
    \end{tabular}
    \caption{Comparison between FEM, DEM, and DCM, kinematic hardening model. Row 1: FEM reference solution. Row 2: absolute differences in the DEM solution. Row 3: absolute differences in the DCM solution. The color scales for each row are different (indicated in the corresponding color bars), and deformed shapes are computed from the DEM displacements with a scaling factor of 1.}
    \label{Contour2}
\end{figure}

Contour plots of the FEM reference solutions and the absolute differences compared to the DEM and DCM predictions are shown in \fref{Contour1} and \fref{Contour2}. Since DCM is a point-wise method where stresses and strain are evaluated at the nodes, it cannot be directly compared to stress solutions from FEM, which are at element integration points. Therefore, when comparing integration point results (e.g., von Mises stress and equivalent plastic strain), we used the DCM-predicted displacements along with FE shape function gradients to generate the results at the integration point locations identical to FEM. Key accuracy metrics and solution time are presented in \tref{DCM-DEM}.

\begin{table}[h!]
    \caption{Comparison of DEM and DCM: mean absolute differences and simulation time}
    \small
    \centering
    \begin{tabular}{ccccccccc}
     Model & Method & \vline & Time [s] &  $AD_{U_{x}}$ [mm] & $AD_{U_{y}}$ [mm] & $AD_{\Bar{\sigma}}$ [MPa] & $AD_{\Bar{\epsilon}^p}$ [/] & $||\bm{e}||_{L^2}$ [\%] \\
    \hline
    \multirow{3}{*}{Iso.}
    & FEM & \vline  & 12 & / & /  & / & /  & / \\
    & DEM & \vline  & 48 & 5.53$\times 10^{-5}$ & 4.07$\times 10^{-5}$ & 0.14 & 1.51$\times 10^{-4}$  & 3.32$\times 10^{-2}$ \\
    & DCM & \vline  & 1082 & 3.94$\times 10^{-3}$ & 2.68$\times 10^{-3}$ & 2.69 & 2.30$\times 10^{-3}$  & 2.27 \\
    
    \hline
    \multirow{3}{*}{Kine.}
    & FEM & \vline  & 3 & / & /  & / & /  & / \\
    & DEM & \vline  & 43 & 1.33$\times 10^{-3}$ & 1.95$\times 10^{-4}$ & 0.86 & 8.34$\times 10^{-4}$  & 0.62 \\
    & DCM & \vline  & 1146 & 3.99$\times 10^{-3}$ & 2.72$\times 10^{-3}$ & 2.71 & 2.32$\times 10^{-3}$  & 2.29 \\

    \end{tabular}
    \label{DCM-DEM}
\end{table}

For DEM with isotropic hardening, the predicted solution is accurate, with an $L^2$ norm of the displacement error lower than 0.05\%. Contour plots in \fref{Contour1} show that the DEM prediction errors are localized near the elastic-plastic boundary. With a kinematic hardening model, DEM is less accurate, yielding displacement errors about an order of magnitude larger and stress and plastic strain errors about 5 times larger. However, the solution is still decently accurate with a $L^2$ displacement error norm less than 1\%. DEM prediction accuracy can be enhanced by using a lower convergence tolerance, at the cost of longer training time. Comparing \fref{Contour1} and \fref{Contour2}, error contour distributions are different for the two material models, with the kinematic hardening results showing errors concentrated in two elliptical regions. For both material models, DEM finishes in less than 50 seconds, while FEM finishes about 3-13 times faster.

The performance of DCM is similar for the two material models, both resulted in a solution that is much less accurate than the DEM ones. Displacement errors are localized at the interior of the domain in two elliptical regions, while stress error is highest on the left and right boundaries of the elastic portion of the domain. High plastic strain errors can be seen near the elastic-plastic boundary and on the top and bottom of the domain. DCM also takes much longer than DEM to train, both taking more than 1000 seconds and more than 20 times slower than DEM.

The results indicate that when the neural network architecture, grid density, and convergence tolerance are identical, DEM is superior to DCM, yielding a faster and more accurate solution. Although it might not be immediately obvious, DEM is also more memory efficient. For each element, DEM requires the storage of 9 (stored as full tensors) total strains, 9 plastic strains, and 1 equivalent plastic strain (and 9 back stresses, depending on the hardening model). For each node, DCM needs identical variable storage. However, when evaluating the loss function, DCM requires an additional 18 components to store the full (e.g., X, Y, and Z) spatial gradients of each of the 6 independent stress components, thus rendering it much more memory intensive.

\subsection{Cyclic loading of a plate with a circular hole and DEM inference}
\label{Cyclic}
\begin{figure}[h!] 
    \centering
     \subfloat[]{
         \includegraphics[trim={0cm 0cm 0cm 0cm},clip,width=0.3\textwidth]{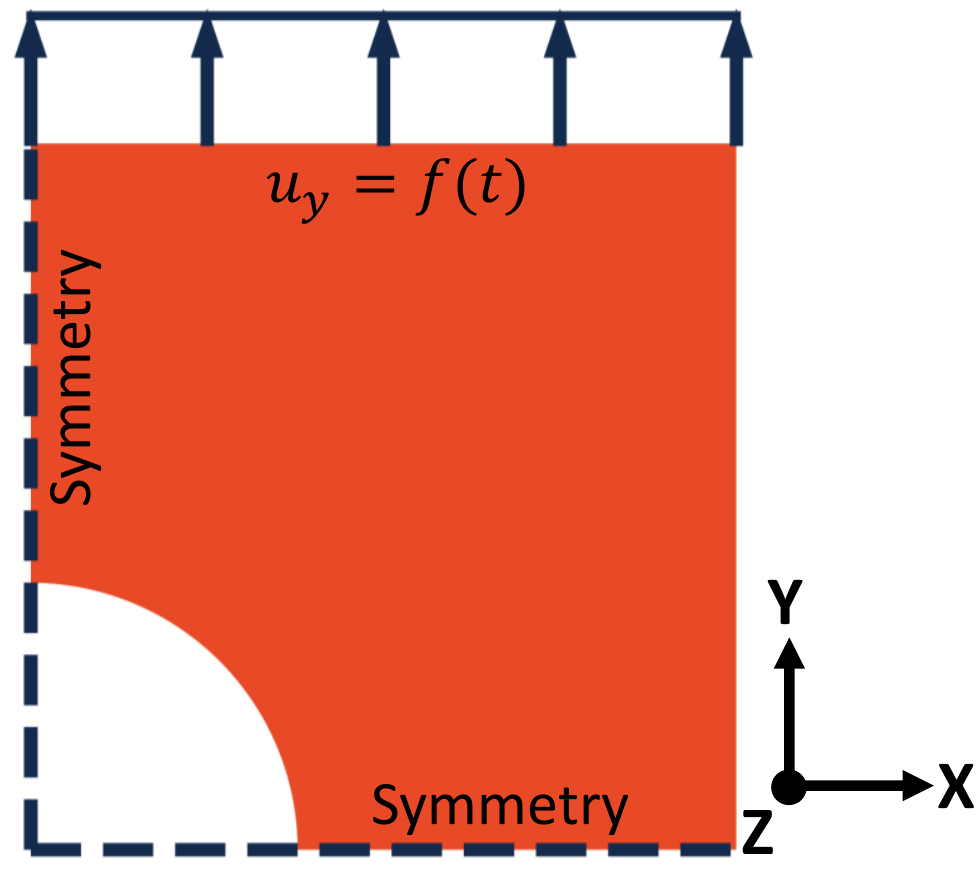}
         \label{plate_BCs}
     }
     \subfloat[]{
         \includegraphics[trim={0cm 0.45cm 0cm 0.2cm},clip,width=0.38\textwidth]{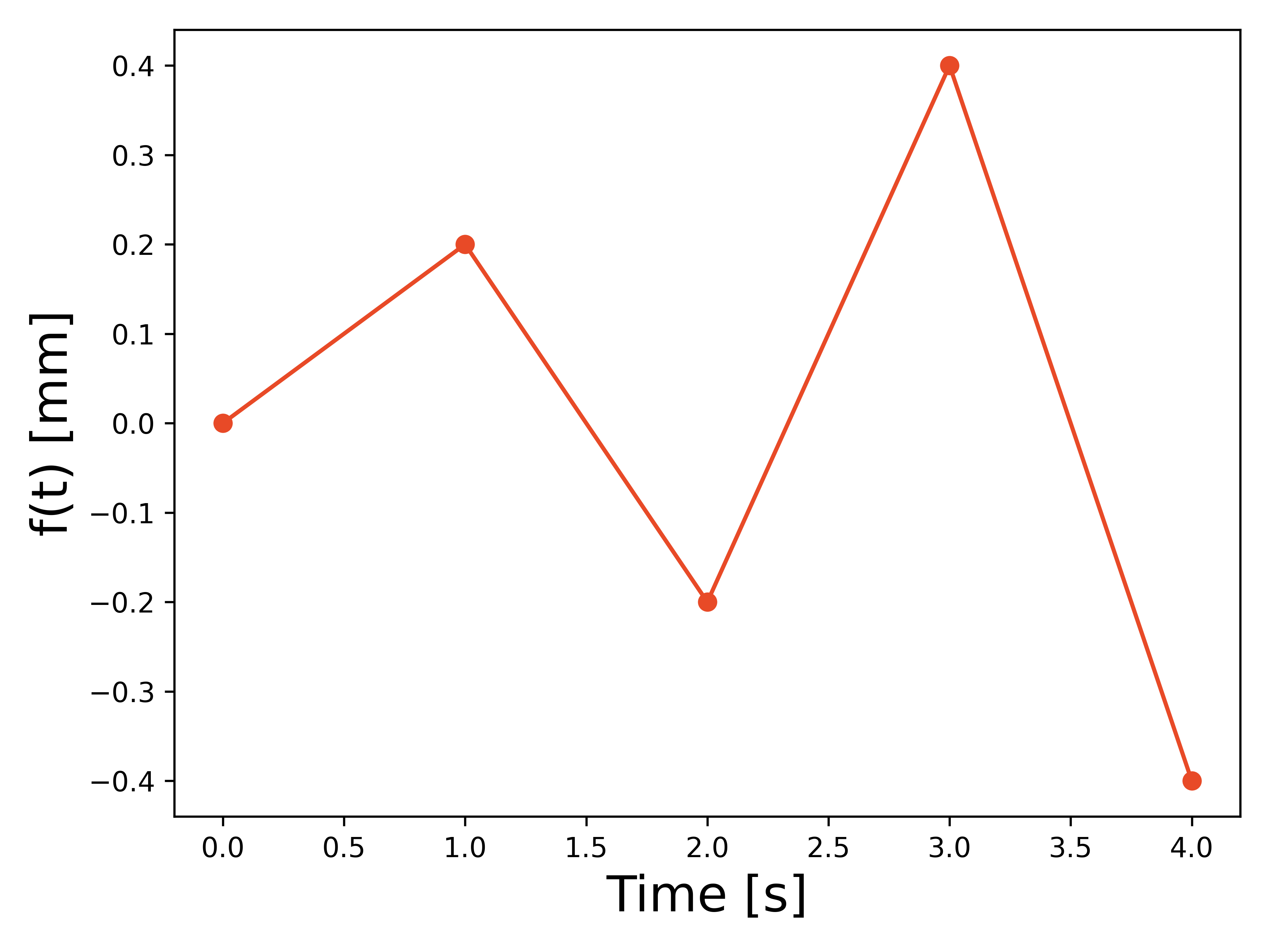}
         \label{ft3}
     }
    \caption{Plate with a hole: \psubref{plate_BCs} Geometry and applied displacements. The symmetry boundary condition is also applied to the $Z=0$ face. \psubref{ft3} Load magnitude as a function of time, simulation load steps marked in circles.}
    \label{plate_hole}
\end{figure}

\begin{figure}[h!] 
    \centering
     \subfloat[]{
         \includegraphics[trim={18cm 4cm 18cm 2cm},clip,width=0.25\textwidth]{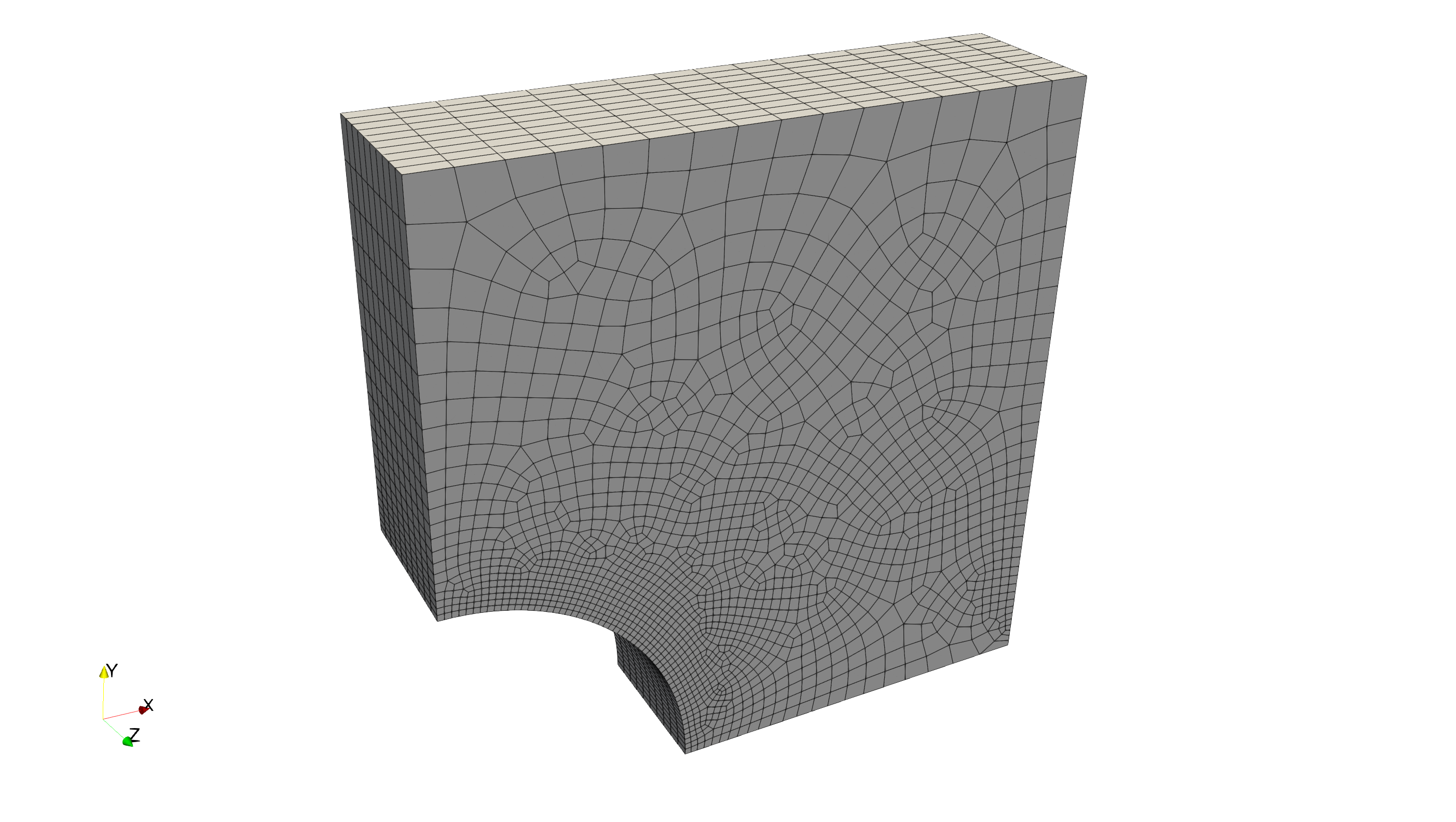}
         \label{m1}
     }
     \subfloat[]{
         \includegraphics[trim={18cm 4cm 18cm 2cm},clip,width=0.25\textwidth]{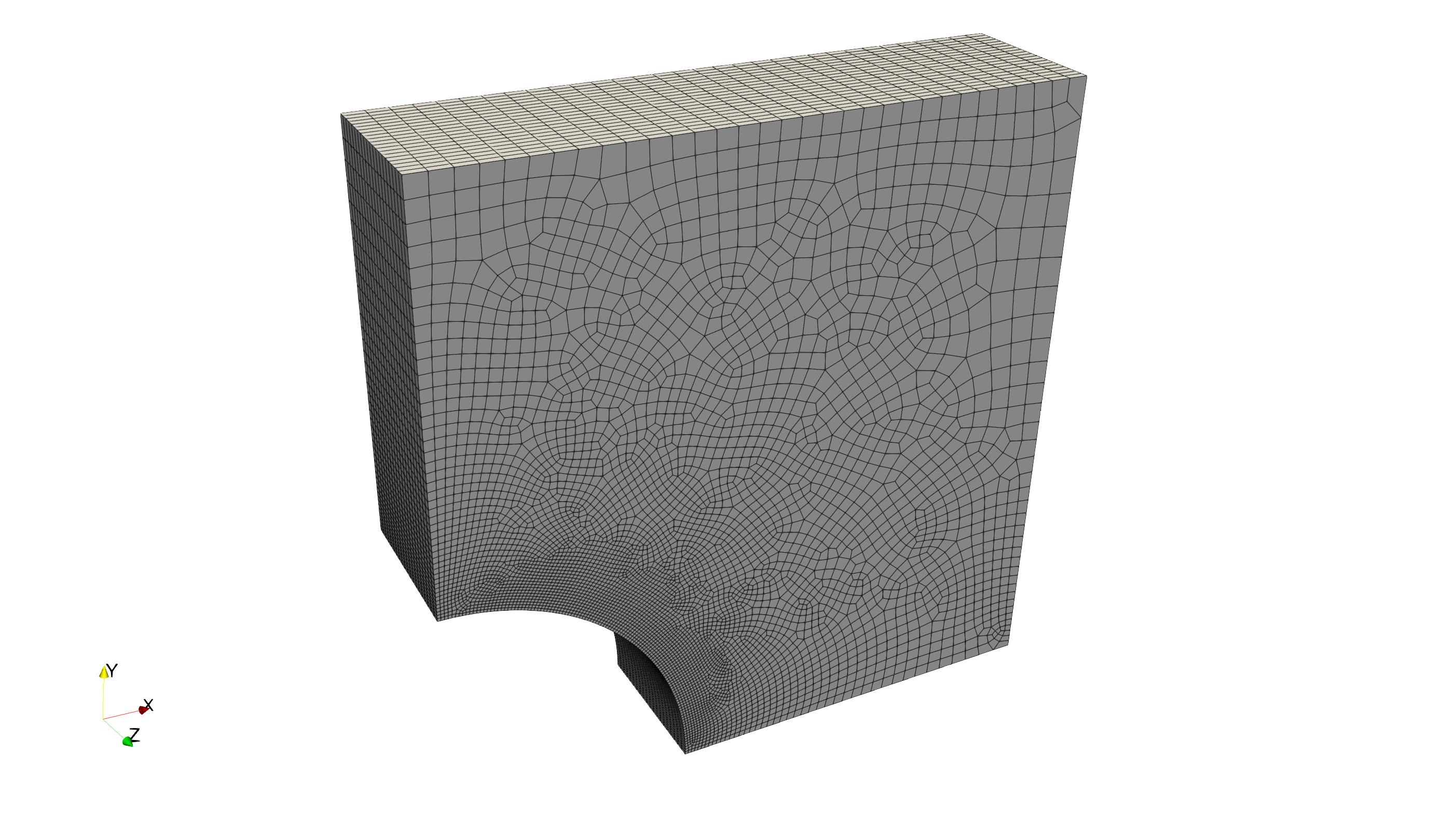}
         \label{m2}
     }
     \subfloat[]{
         \includegraphics[trim={18cm 4cm 18cm 2cm},clip,width=0.25\textwidth]{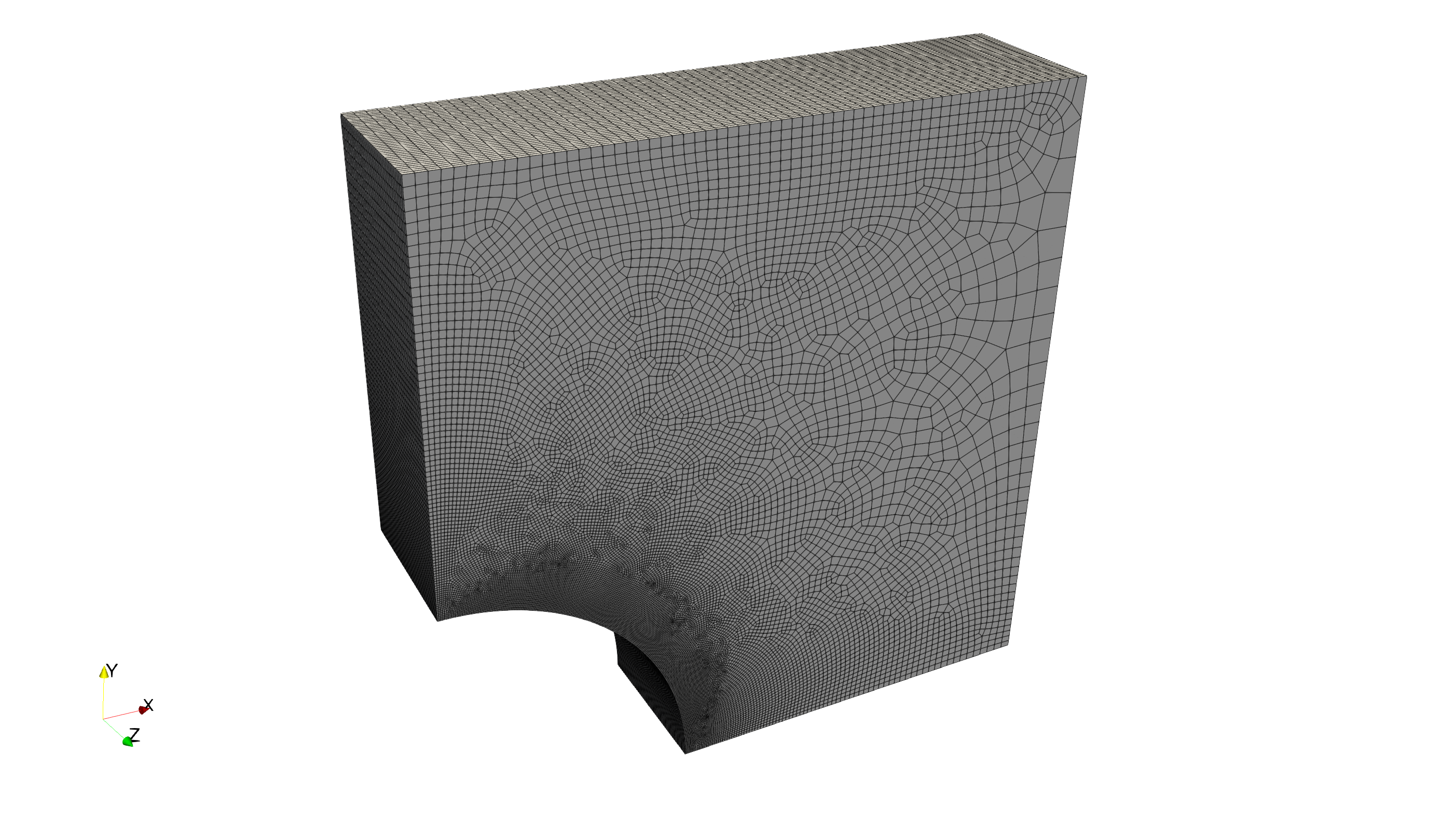}
         \label{m3}
     }
    \caption{Progressively refined meshes used in this example: \psubref{m1} Mesh1, 16670 elements. \psubref{m2} Mesh2, 99100 elements. \psubref{m3} Mesh3, 662120 elements. }
    \label{meshes}
\end{figure}

In the previous examples, we demonstrated our model on domains discretized by structured hexahedral elements. Our implementation is not limited to structured meshes and can be used with unstructured meshes (hexahedrons or tetrahedrons). Consider a rectangular plate with a circular hole. The plate has dimensions 8$\times$8$\times$2 mm$^3$, and the center hole has a radius of 1.5 mm. Symmetry was exploited to only model 1/8 of the domain, and a cyclic displacement was applied on the top surface of the domain. The applied boundary conditions and the time dependence of the applied displacement magnitude are shown in \fref{plate_hole}. The isotropic hardening model was used, and the DEM solution was done in 4 load steps using a convergence tolerance of $2\times 10^{-5}$. Three progressively refined meshes were used and are shown in \fref{meshes}; they are denoted by Mesh 1, Mesh 2, and Mesh 3 in subsequent discussions.

\begin{figure}[ht!]
\newcommand\x{0.13}
\captionsetup[subfigure]{labelformat=empty}
    \centering
    \begin{tabular}{ c c c c c }
    \begin{minipage}[c]{\x\textwidth}
       \centering 
        \subfloat[Step 1, $U_x$]{\includegraphics[trim={18cm 0cm 18cm 0cm},clip,width=\textwidth]{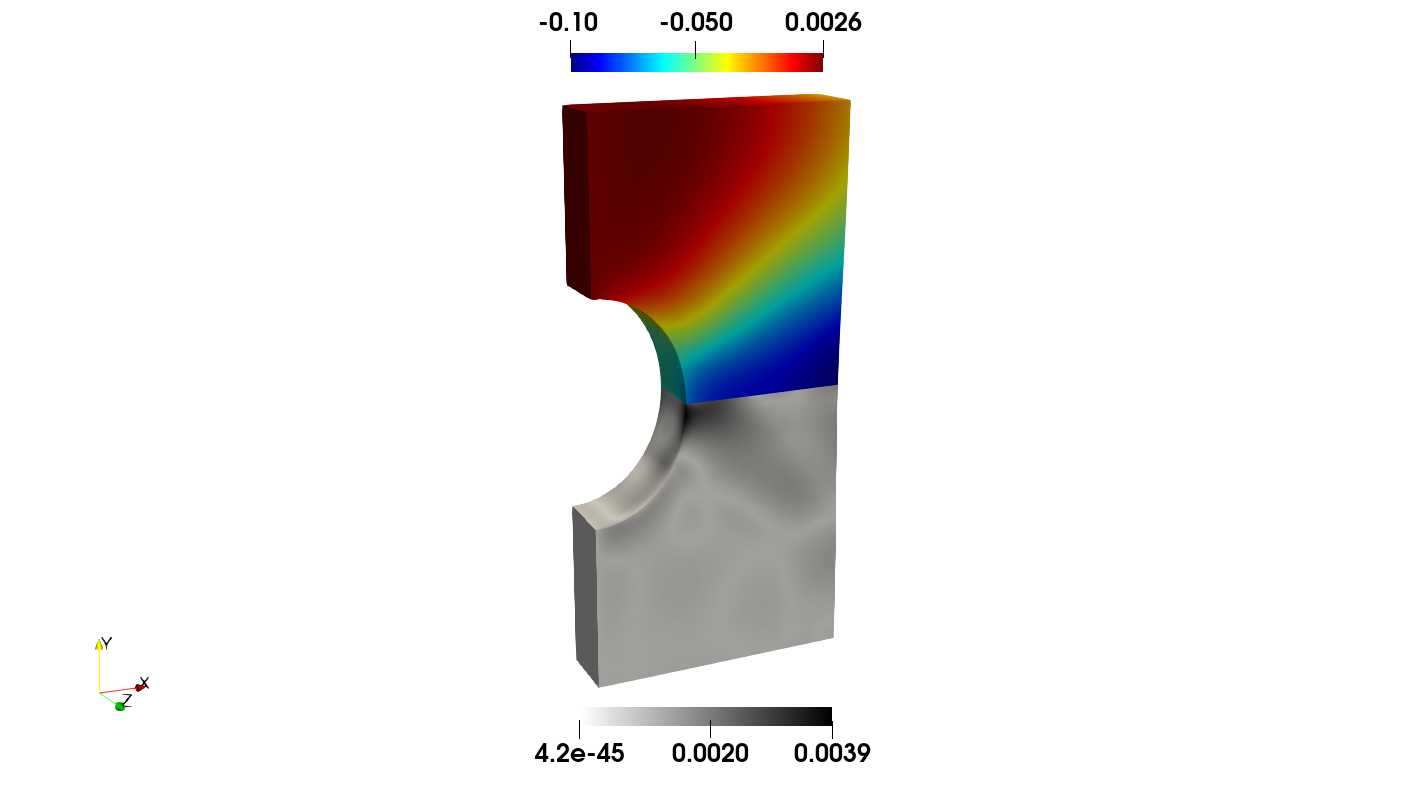}
        }
    \end{minipage}
    &
    \begin{minipage}[c]{\x\textwidth}
       \centering 
        \subfloat[Step 1, $U_y$]{\includegraphics[trim={18cm 0cm 18cm 0cm},clip,width=\textwidth]{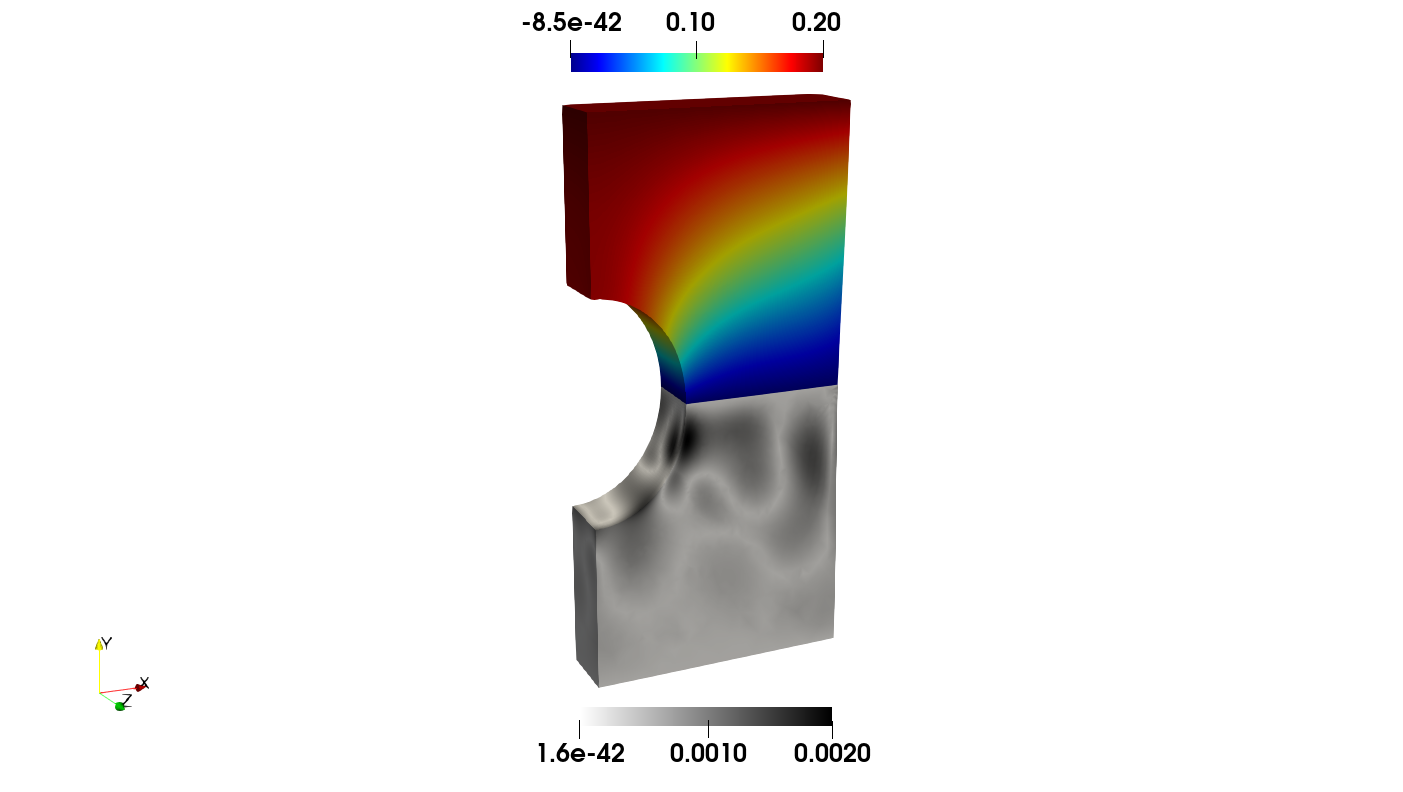}
        }
    \end{minipage}
    &
    \begin{minipage}[c]{\x\textwidth}
       \centering 
        \subfloat[Step 1, $U_z$]{\includegraphics[trim={18cm 0cm 18cm 0cm},clip,width=\textwidth]{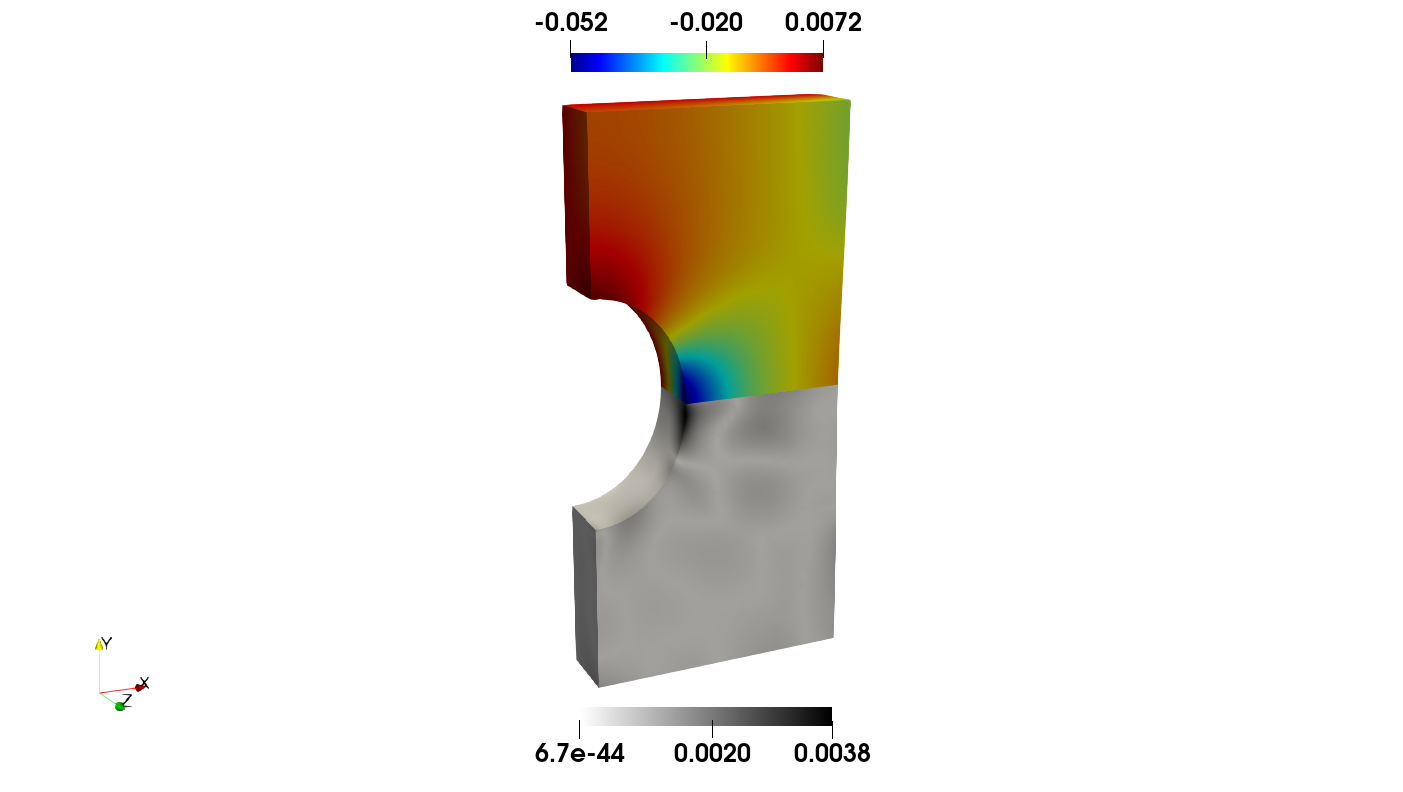}
        }
    \end{minipage}
    &
    \begin{minipage}[c]{\x\textwidth}
       \centering 
        \subfloat[Step 1, $\Bar{\sigma}$]{\includegraphics[trim={18cm 0cm 18cm 0cm},clip,width=\textwidth]{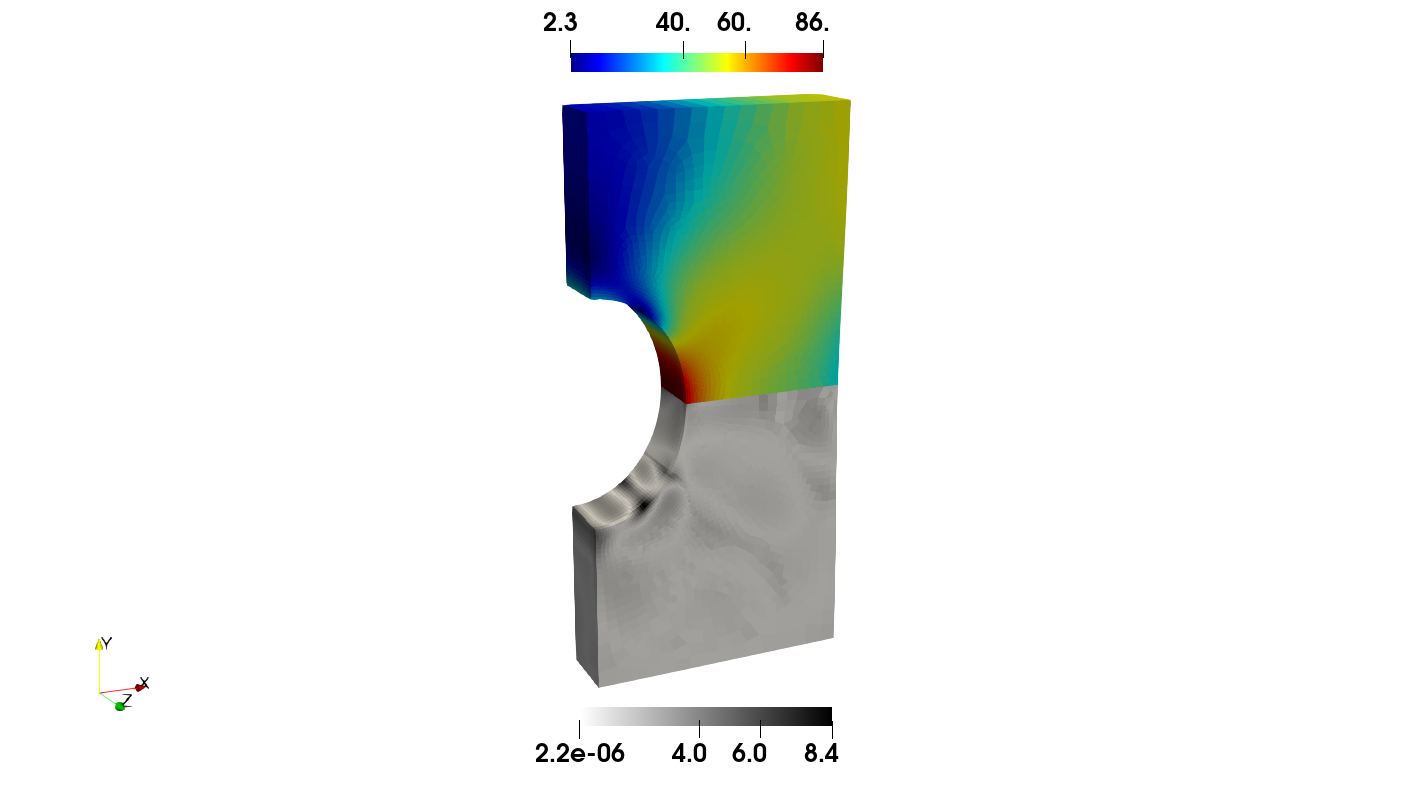}
        }
    \end{minipage}
    &
    \begin{minipage}[c]{\x\textwidth}
       \centering 
        \subfloat[Step 1, $\Bar{\epsilon}^p$]{\includegraphics[trim={18cm 0cm 18cm 0cm},clip,width=\textwidth]{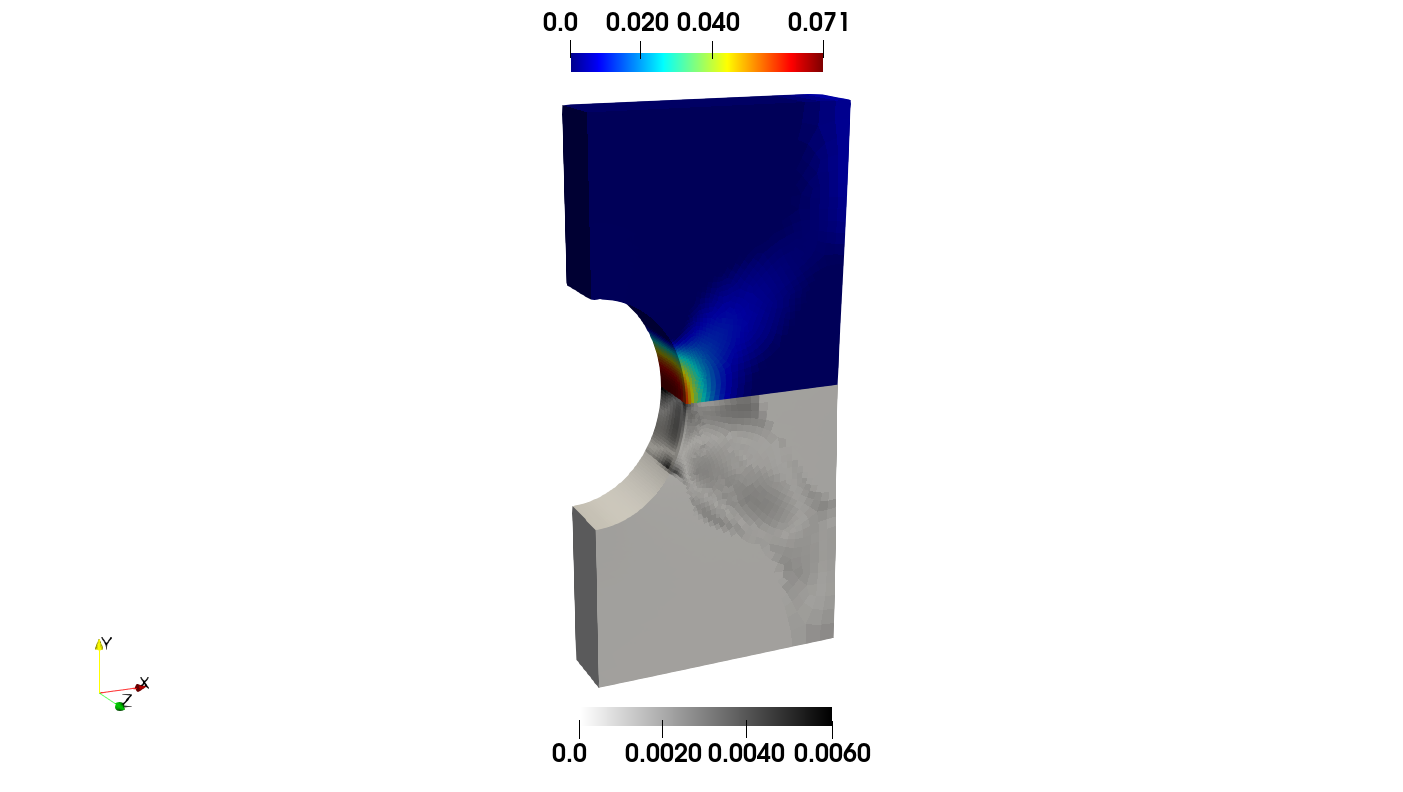}
        }
    \end{minipage}
    \\

     \begin{minipage}[c]{\x\textwidth}
       \centering 
        \subfloat[Step 2, $U_x$]{\includegraphics[trim={18cm 0cm 18cm 0cm},clip,width=\textwidth]{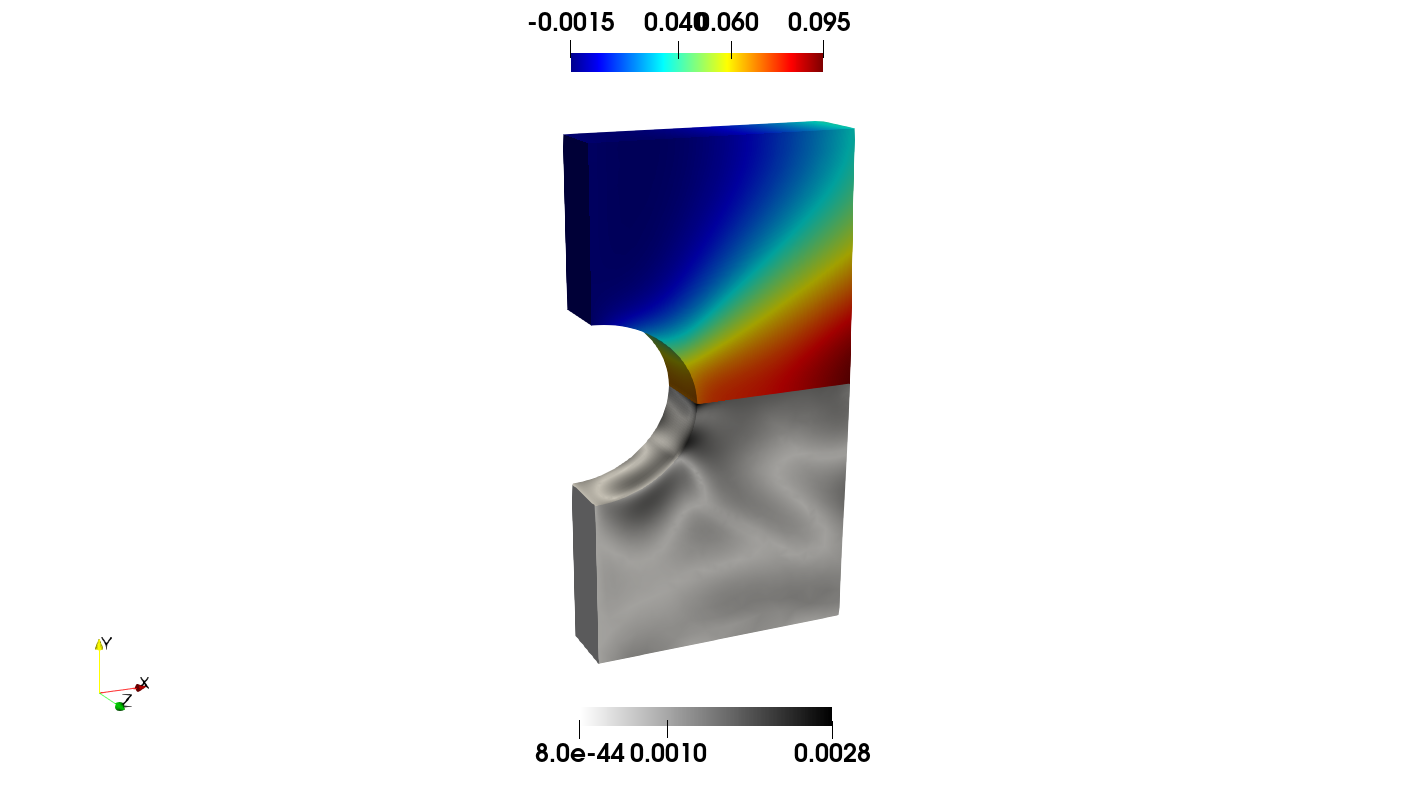}
        }
    \end{minipage}
    &
    \begin{minipage}[c]{\x\textwidth}
       \centering 
        \subfloat[Step 2, $U_y$]{\includegraphics[trim={18cm 0cm 18cm 0cm},clip,width=\textwidth]{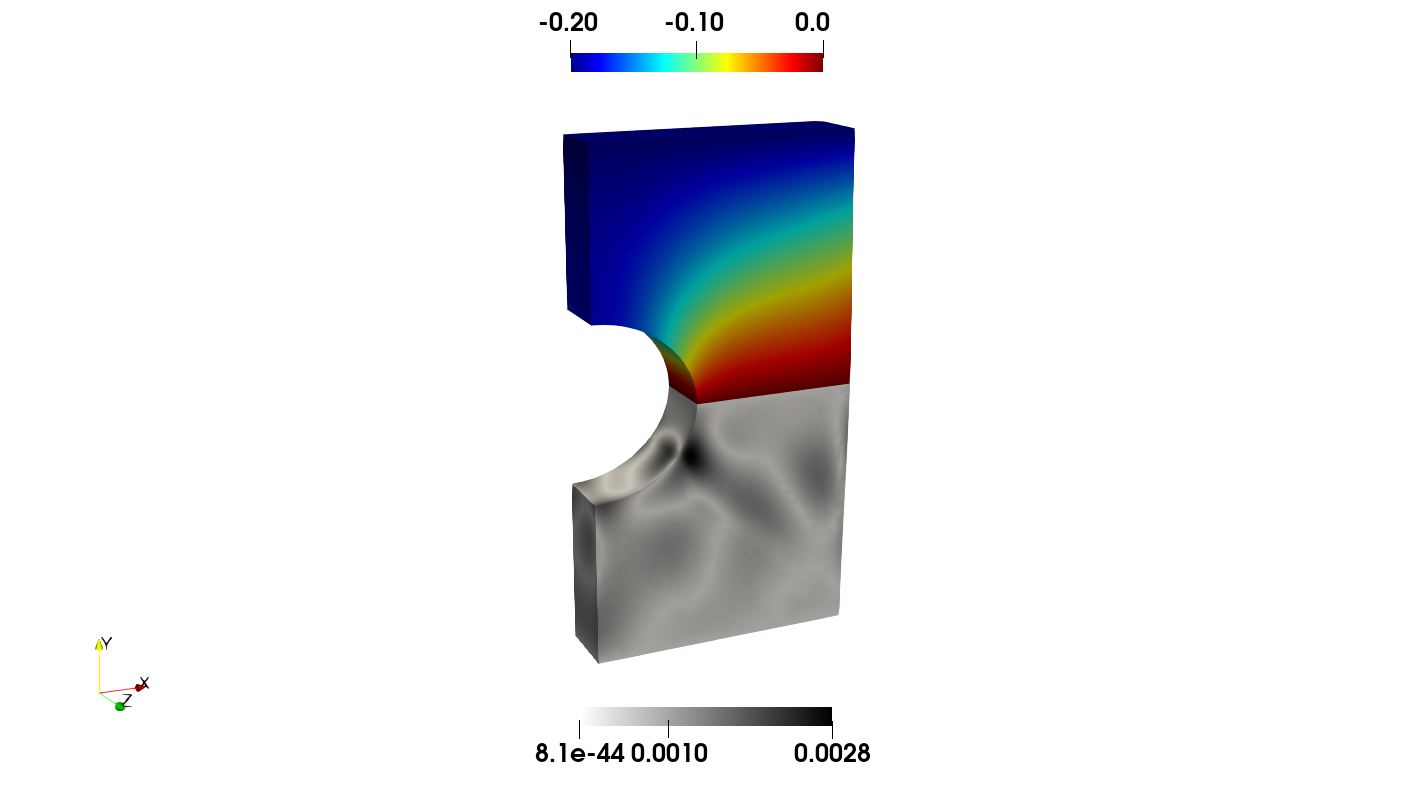}
        }
    \end{minipage}
    &
    \begin{minipage}[c]{\x\textwidth}
       \centering 
        \subfloat[Step 2, $U_z$]{\includegraphics[trim={18cm 0cm 18cm 0cm},clip,width=\textwidth]{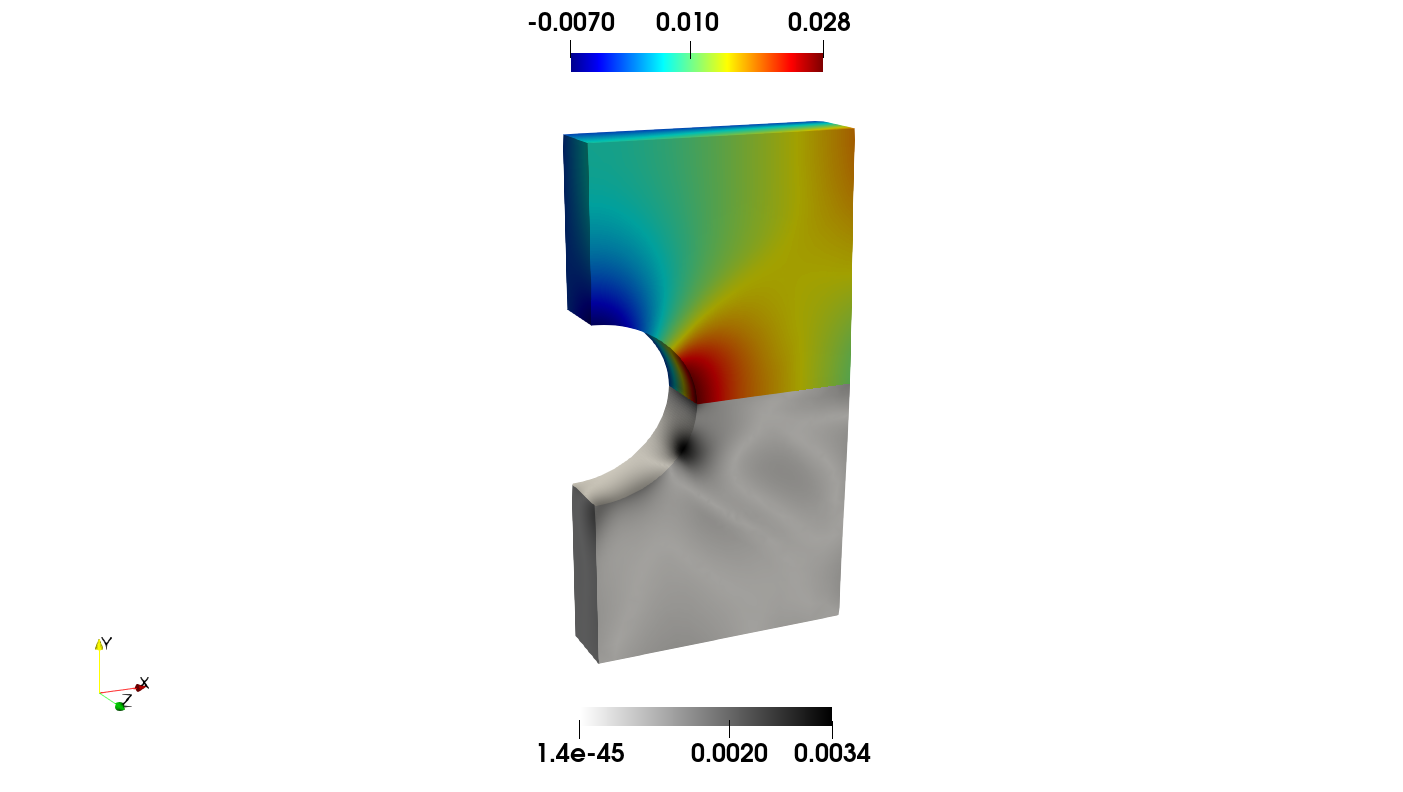}
        }
    \end{minipage}
    &
    \begin{minipage}[c]{\x\textwidth}
       \centering 
        \subfloat[Step 2, $\Bar{\sigma}$]{\includegraphics[trim={18cm 0cm 18cm 0cm},clip,width=\textwidth]{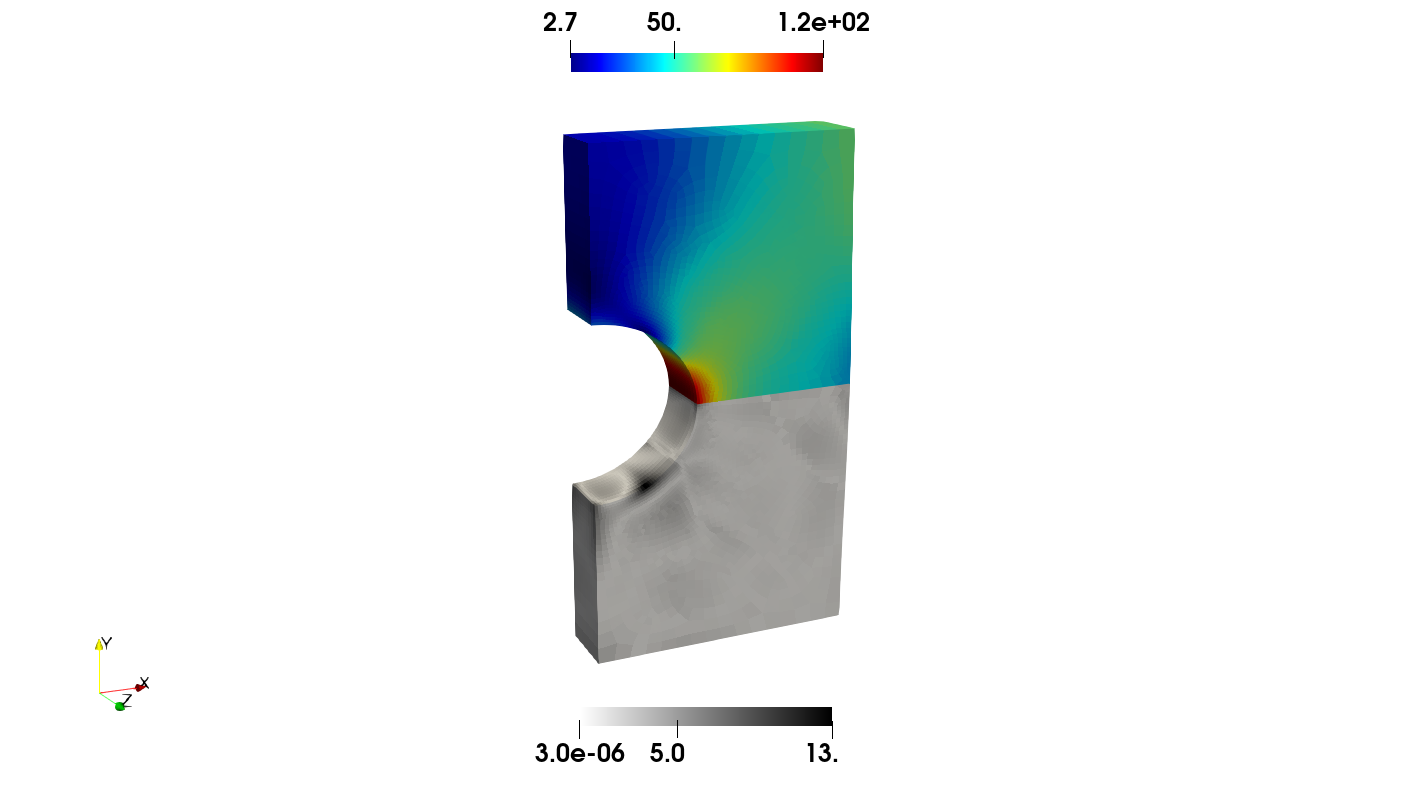}
        }
    \end{minipage}
    &
    \begin{minipage}[c]{\x\textwidth}
       \centering 
        \subfloat[Step 2, $\Bar{\epsilon}^p$]{\includegraphics[trim={18cm 0cm 18cm 0cm},clip,width=\textwidth]{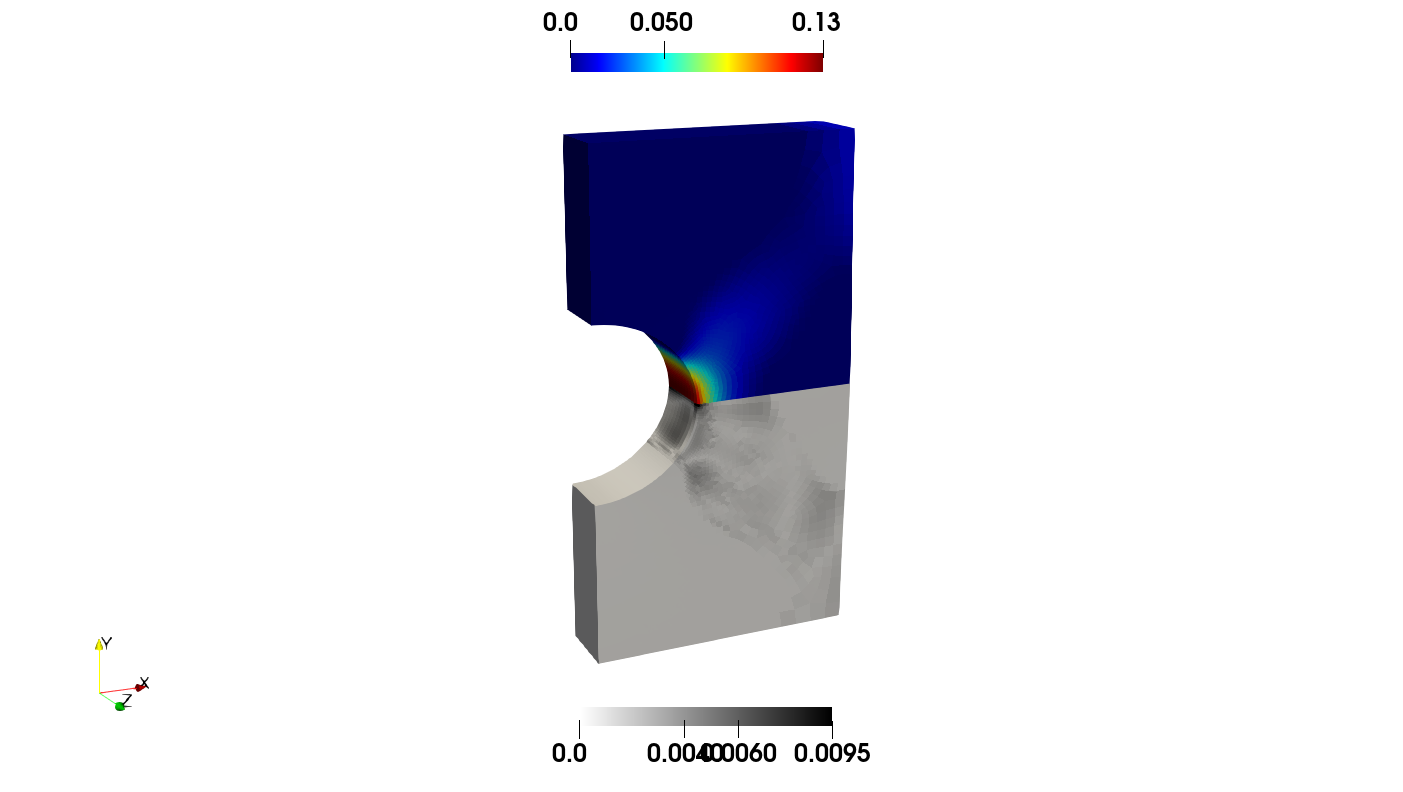}
        }
    \end{minipage}
    \\

    \begin{minipage}[c]{\x\textwidth}
       \centering 
        \subfloat[Step 3, $U_x$]{\includegraphics[trim={18cm 0cm 18cm 0cm},clip,width=\textwidth]{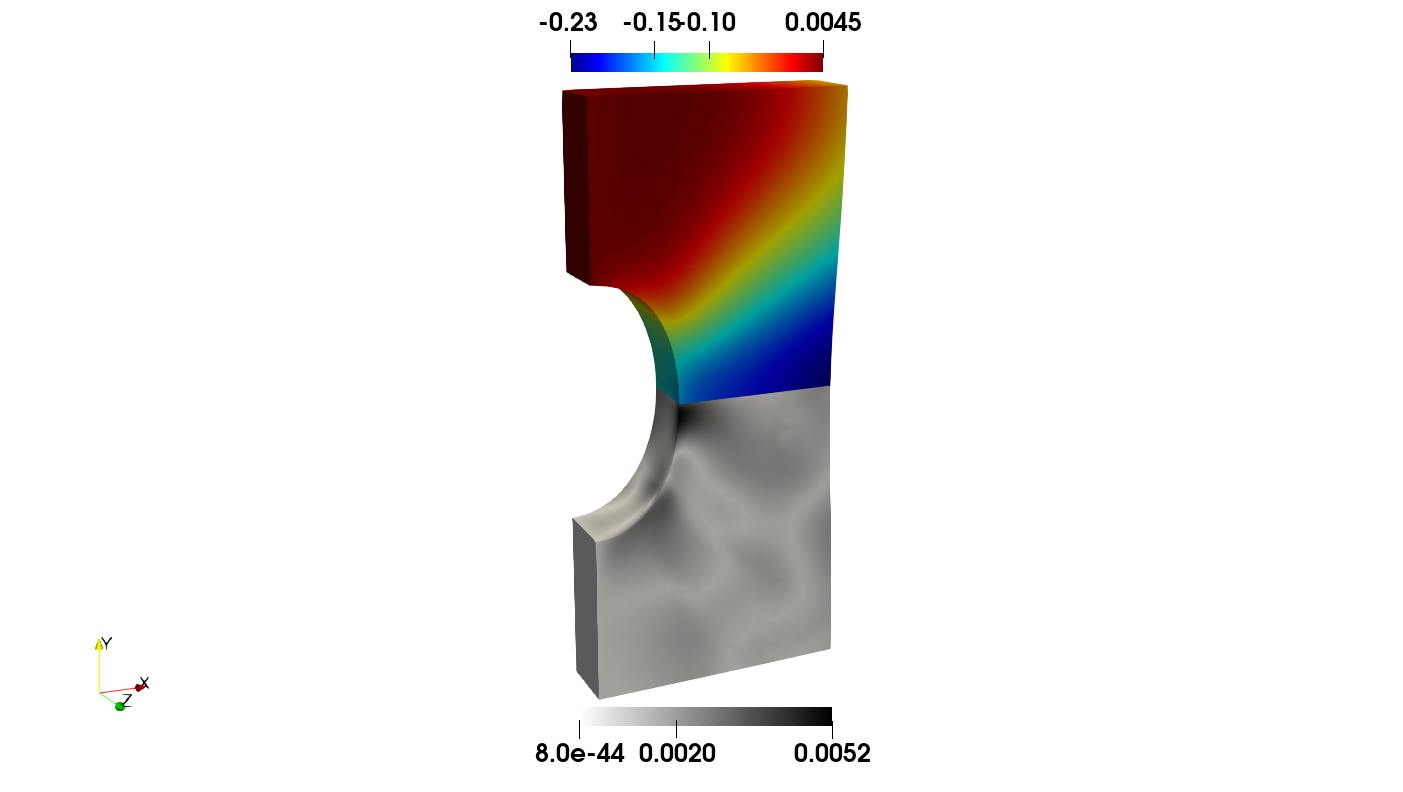}
        }
    \end{minipage}
    &
    \begin{minipage}[c]{\x\textwidth}
       \centering 
        \subfloat[Step 3, $U_y$]{\includegraphics[trim={18cm 0cm 18cm 0cm},clip,width=\textwidth]{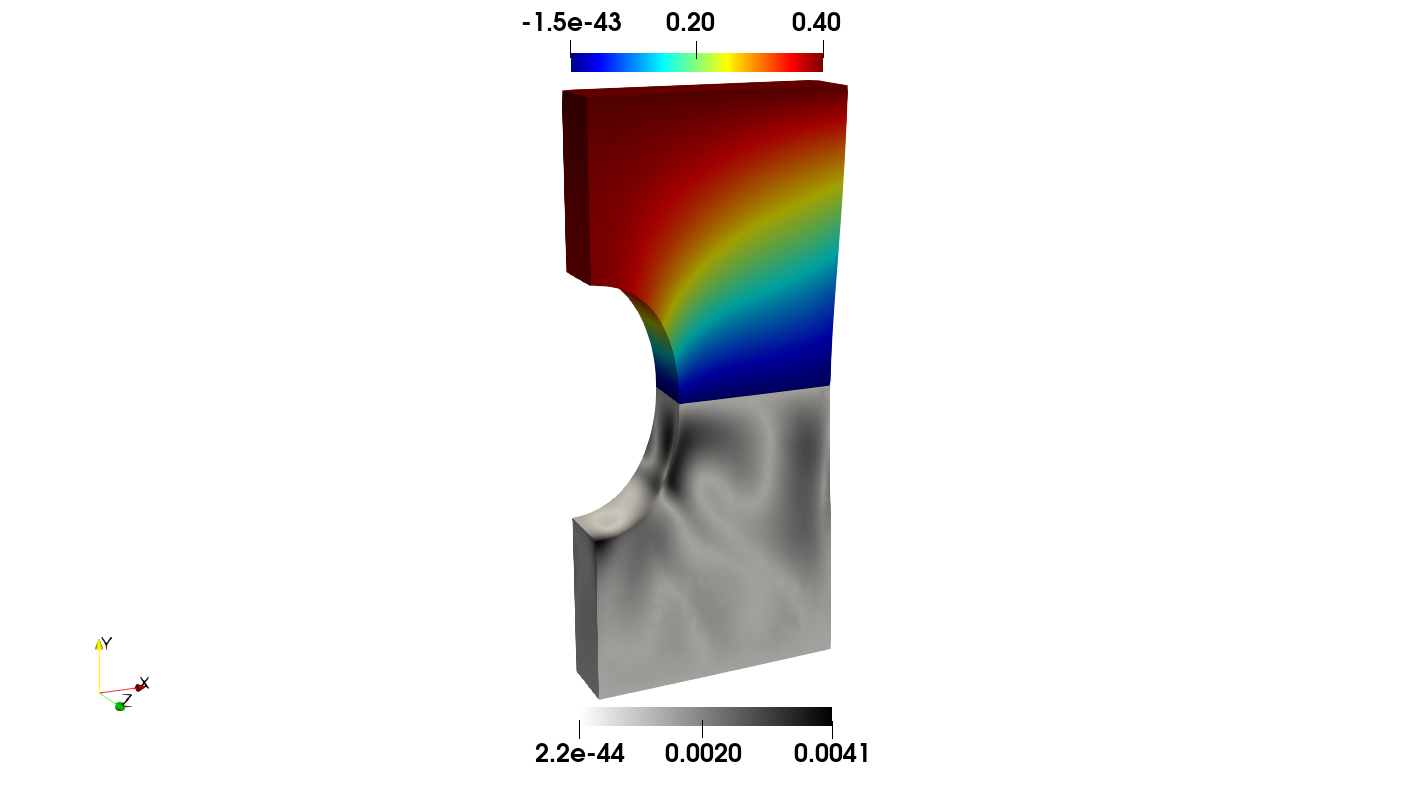}
        }
    \end{minipage}
    &
    \begin{minipage}[c]{\x\textwidth}
       \centering 
        \subfloat[Step 3, $U_z$]{\includegraphics[trim={18cm 0cm 18cm 0cm},clip,width=\textwidth]{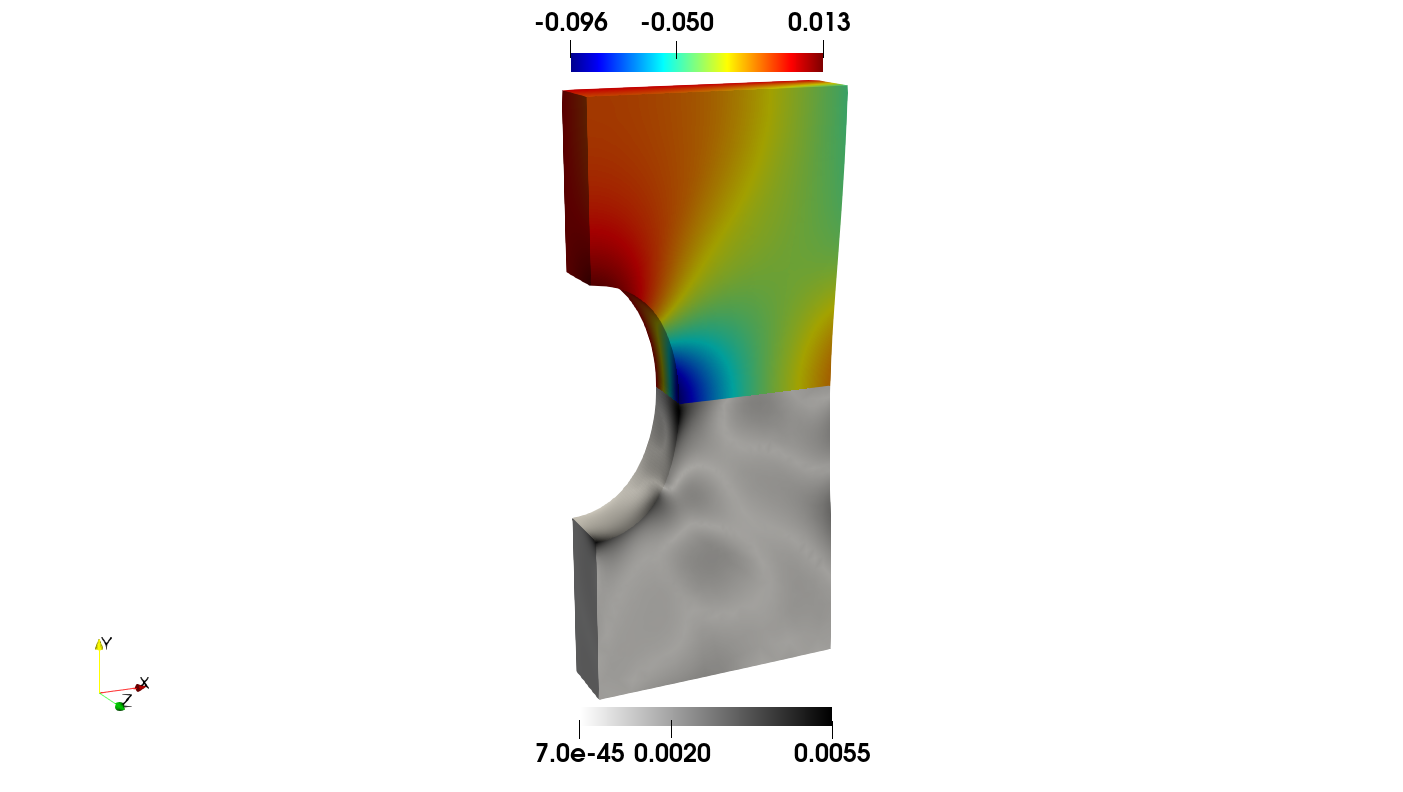}
        }
    \end{minipage}
    &
    \begin{minipage}[c]{\x\textwidth}
       \centering 
        \subfloat[Step 3, $\Bar{\sigma}$]{\includegraphics[trim={18cm 0cm 18cm 0cm},clip,width=\textwidth]{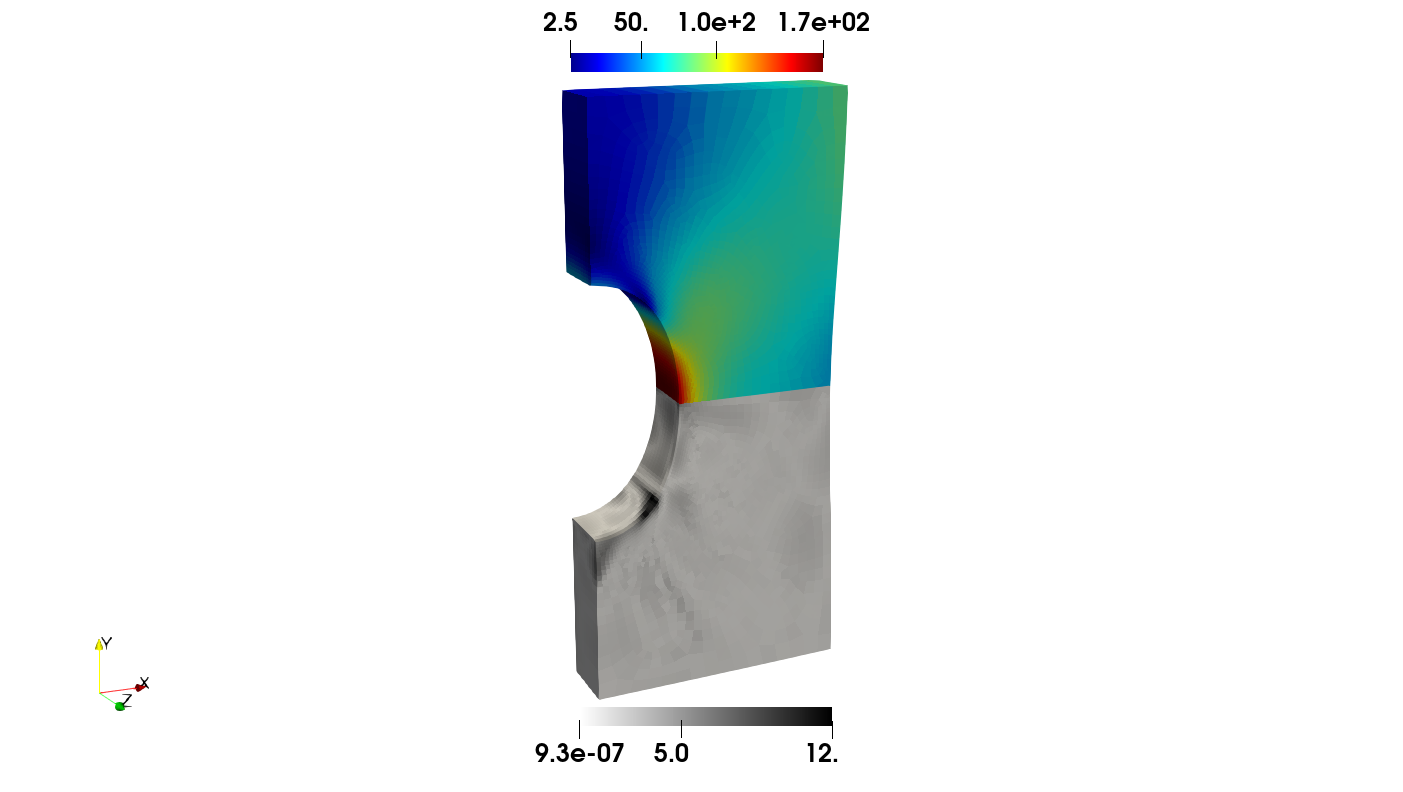}
        }
    \end{minipage}
    &
    \begin{minipage}[c]{\x\textwidth}
       \centering 
        \subfloat[Step 3, $\Bar{\epsilon}^p$]{\includegraphics[trim={18cm 0cm 18cm 0cm},clip,width=\textwidth]{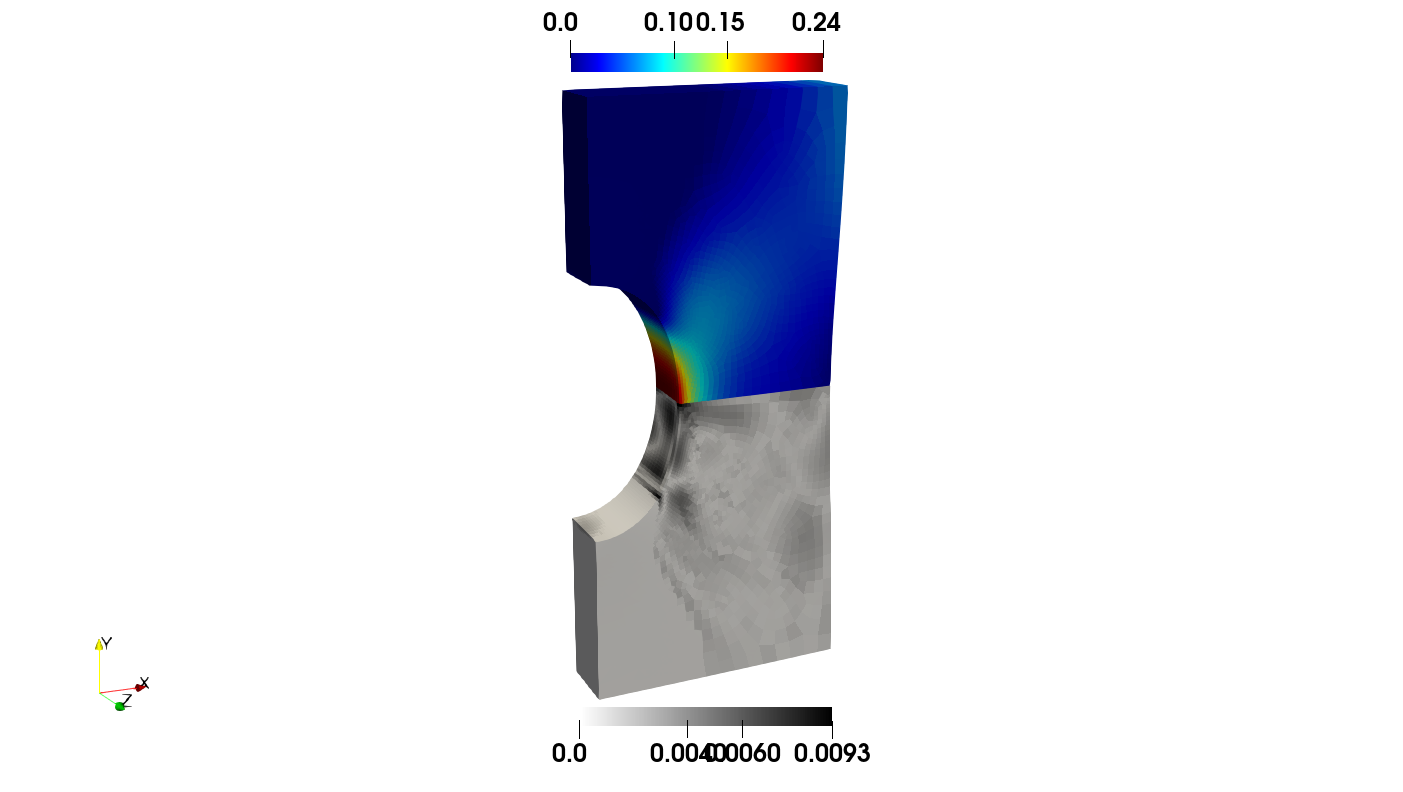}
        }
    \end{minipage}
    \\

    \begin{minipage}[c]{\x\textwidth}
       \centering 
        \subfloat[Step 4, $U_x$]{\includegraphics[trim={18cm 0cm 18cm 0cm},clip,width=\textwidth]{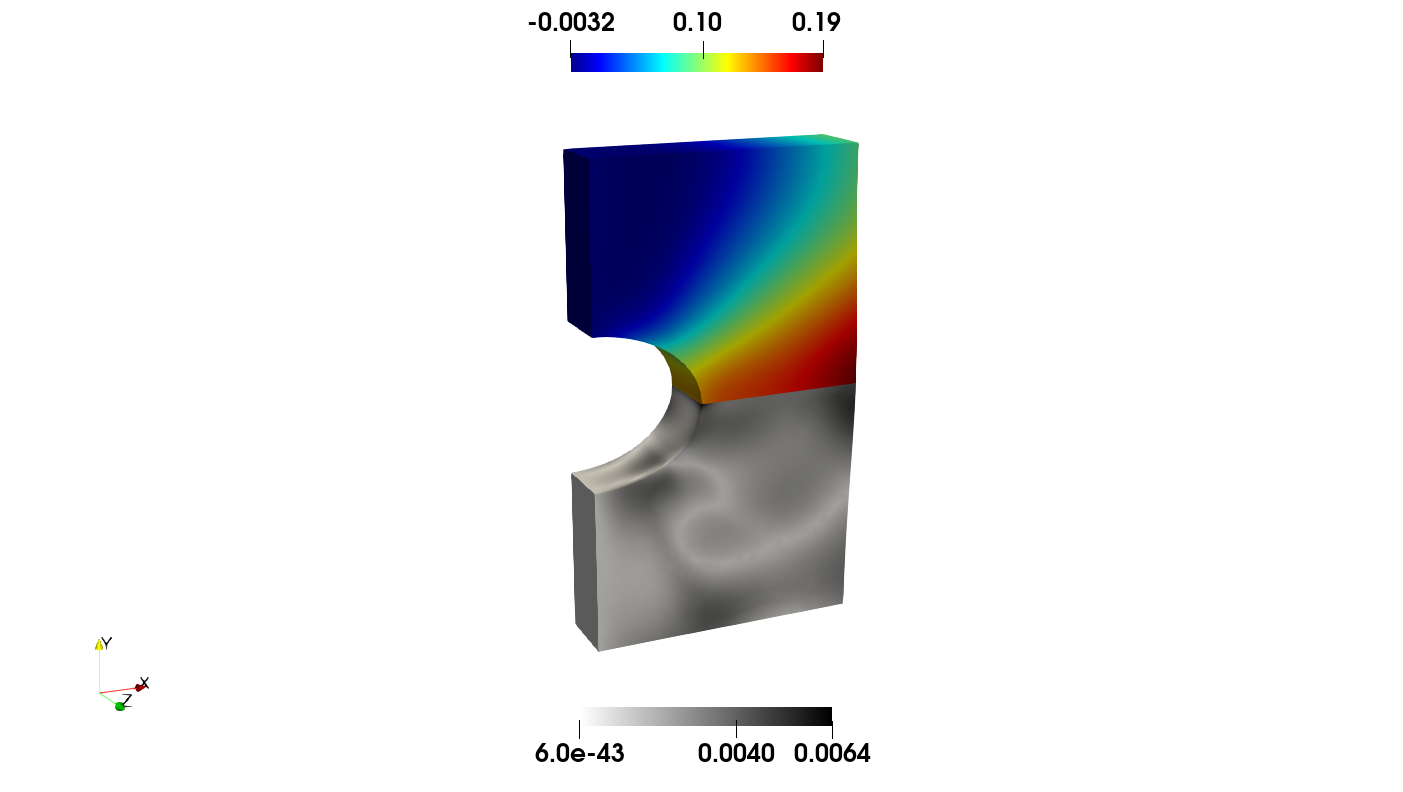}
        }
    \end{minipage}
    &
    \begin{minipage}[c]{\x\textwidth}
       \centering 
        \subfloat[Step 4, $U_y$]{\includegraphics[trim={18cm 0cm 18cm 0cm},clip,width=\textwidth]{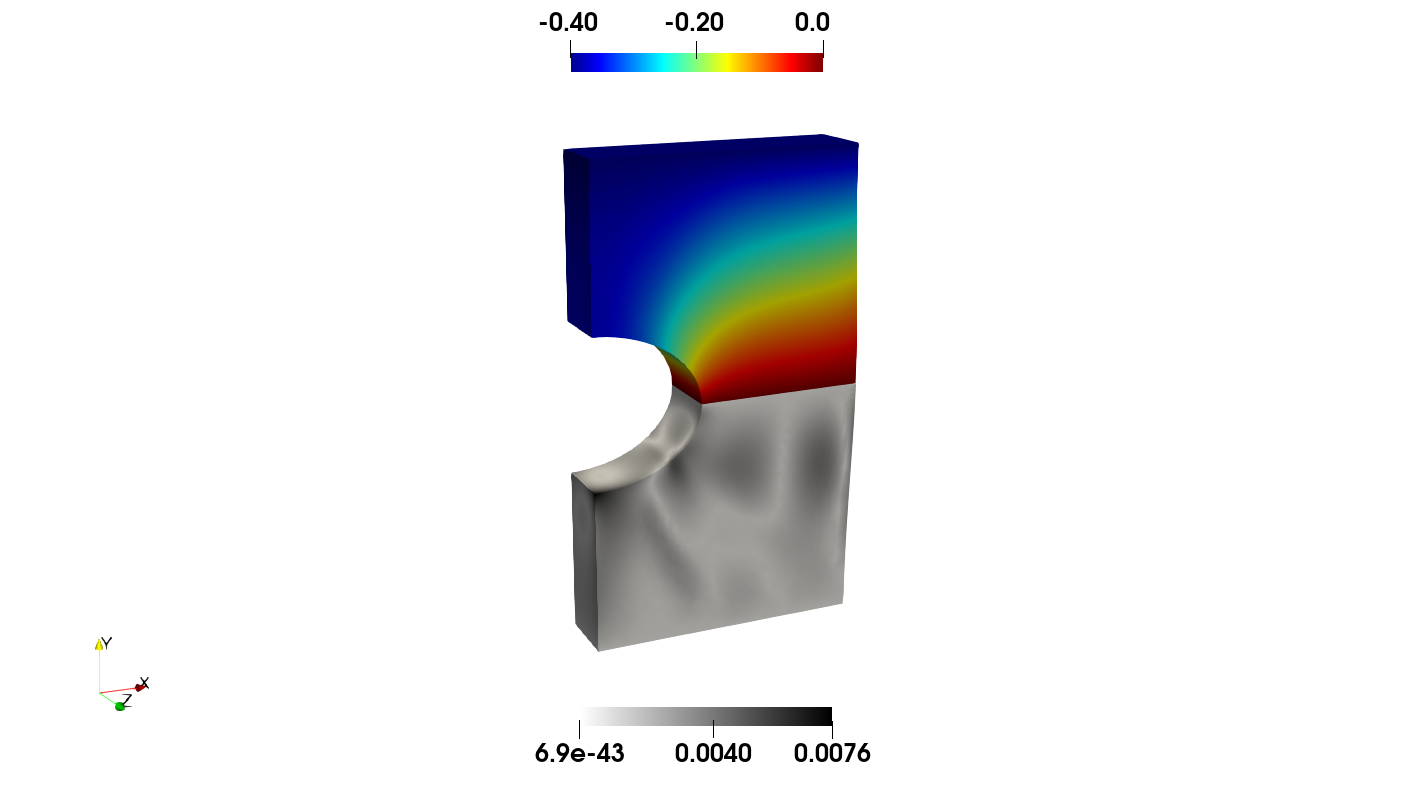}
        }
    \end{minipage}
    &
    \begin{minipage}[c]{\x\textwidth}
       \centering 
        \subfloat[Step 4, $U_z$]{\includegraphics[trim={18cm 0cm 18cm 0cm},clip,width=\textwidth]{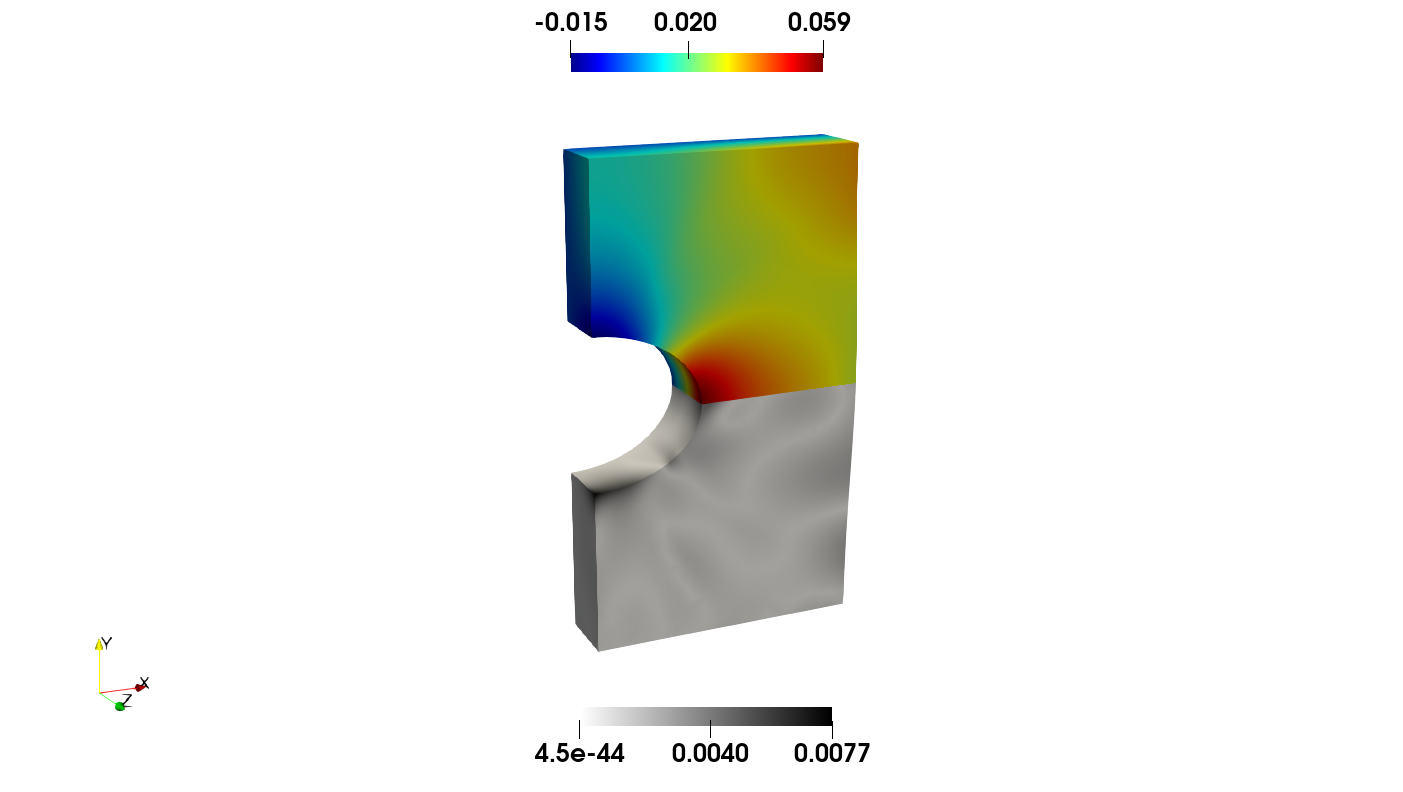}
        }
    \end{minipage}
    &
    \begin{minipage}[c]{\x\textwidth}
       \centering 
        \subfloat[Step 4, $\Bar{\sigma}$]{\includegraphics[trim={18cm 0cm 18cm 0cm},clip,width=\textwidth]{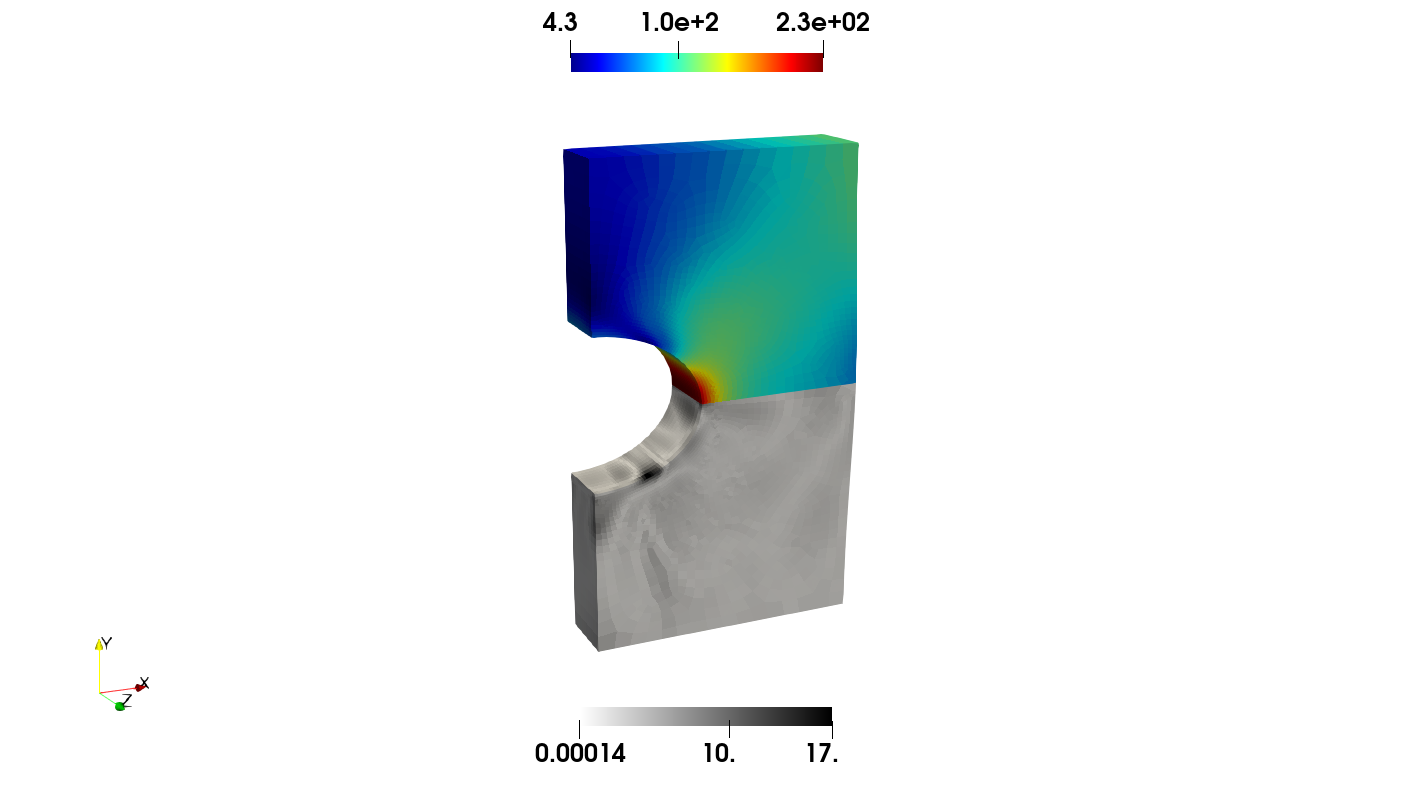}
        }
    \end{minipage}
    &
    \begin{minipage}[c]{\x\textwidth}
       \centering 
        \subfloat[Step 4, $\Bar{\epsilon}^p$]{\includegraphics[trim={18cm 0cm 18cm 0cm},clip,width=\textwidth]{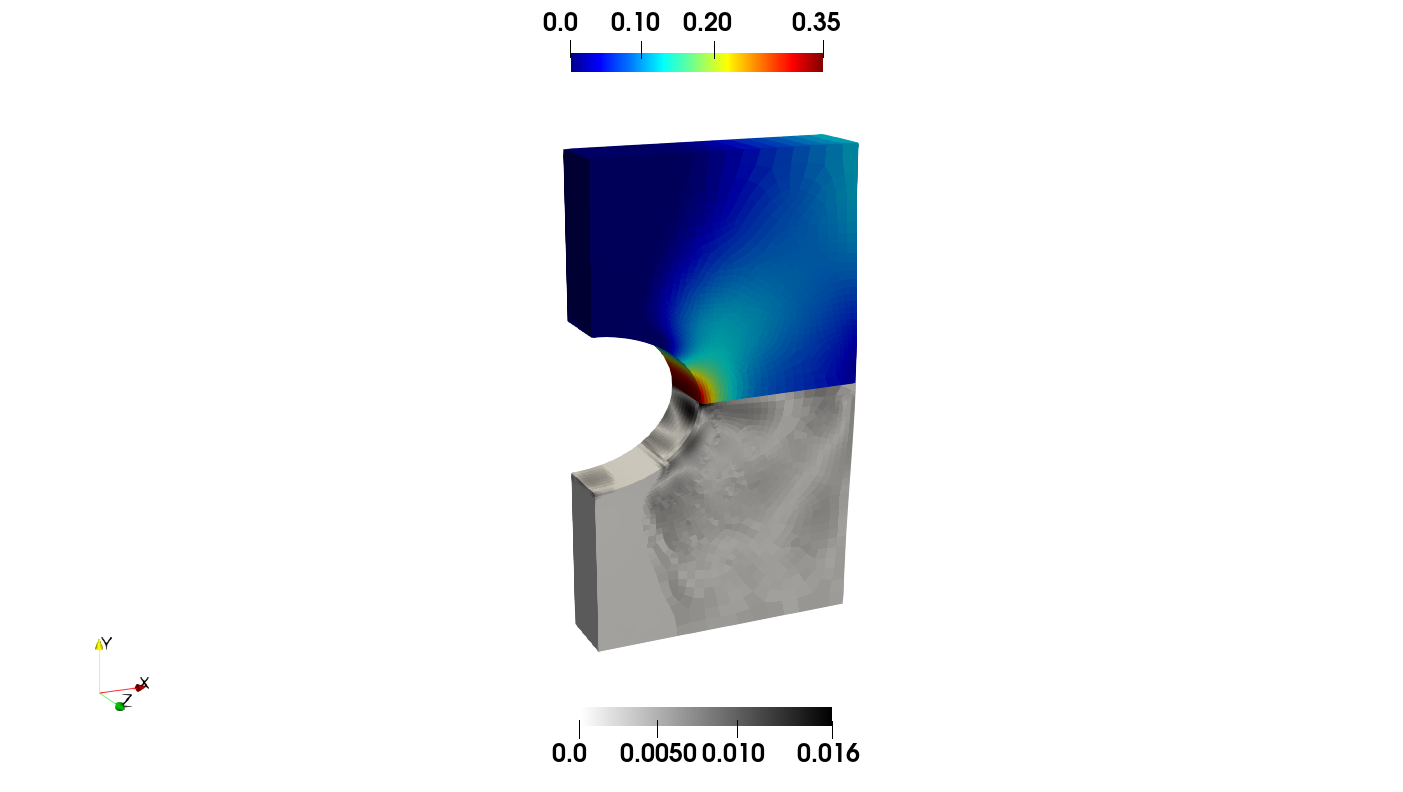}
        }
    \end{minipage}
    \\
    
    \end{tabular}
    \caption{Comparison between FEM and DEM in 4 load steps. For each sub-figure, the top half, in the rainbow color map, shows the FEM reference solution. While the bottom half, in a white-to-black color map, shows the absolute difference in the DEM solution. The deformed shapes are computed from the DEM displacements with a scaling factor of 1.}
    \label{Contour3}
\end{figure}

First, DEM training was done using Mesh 1, and the trained DEM model was saved. Contour plots comparing the DEM predictions to the reference FEM solutions are shown in \fref{Contour3}. Quantitative metrics for solution accuracy are provided in \tref{train_accuracy}.
\begin{table}[h!]
    \caption{DEM performance metrics in 4 load steps}
    \small
    \centering
    \begin{tabular}{cccccccc}
     Step & \vline &  $AD_{U_{x}}$ [mm] & $AD_{U_{y}}$ [mm] & $AD_{U_{y}}$ [mm] & $AD_{\Bar{\sigma}}$ [MPa] & $AD_{\Bar{\epsilon}^p}$ [/] & $||\bm{e}||_{L^2}$ [\%] \\
    \hline
    1 & \vline & 3.61$\times 10^{-4}$ & 3.42$\times 10^{-4}$ & 1.51$\times 10^{-4}$ & 0.56 & 2.84$\times 10^{-4}$  & 0.56 \\
    2 & \vline & 4.62$\times 10^{-4}$ & 4.29$\times 10^{-4}$ & 2.64$\times 10^{-4}$ & 0.72 & 4.21$\times 10^{-4}$  & 0.72 \\
    3 & \vline & 6.17$\times 10^{-4}$ & 6.25$\times 10^{-4}$ & 4.09$\times 10^{-4}$ & 0.69 & 8.42$\times 10^{-4}$  & 0.51 \\
    4 & \vline & 1.31$\times 10^{-3}$ & 1.00$\times 10^{-3}$ & 5.20$\times 10^{-4}$ & 1.03 & 1.46$\times 10^{-3}$  & 0.89 \\
    \end{tabular}
    \label{train_accuracy}
\end{table}

\begin{figure}[ht!]
\newcommand\x{0.133}
\captionsetup[subfigure]{labelformat=empty}
    \centering
    \begin{tabular}{ c c c c c }
    \begin{minipage}[c]{\x\textwidth}
       \centering 
        \subfloat[Mesh 1, $U_x$]{\includegraphics[trim={17cm 0cm 17cm 0cm},clip,width=\textwidth]{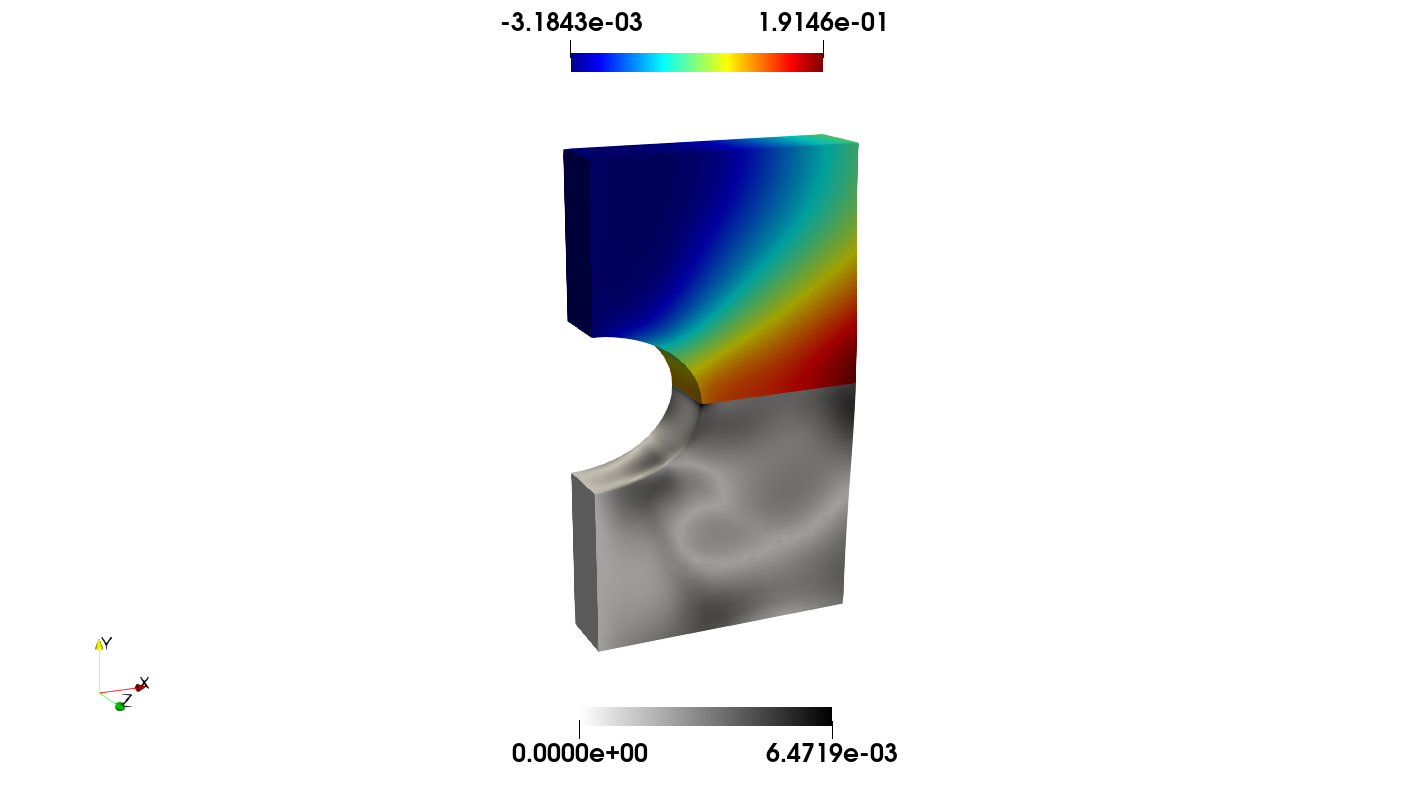}
        }
    \end{minipage}
    &
    \begin{minipage}[c]{\x\textwidth}
       \centering 
        \subfloat[Mesh 1, $U_y$]{\includegraphics[trim={17cm 0cm 17cm 0cm},clip,width=\textwidth]{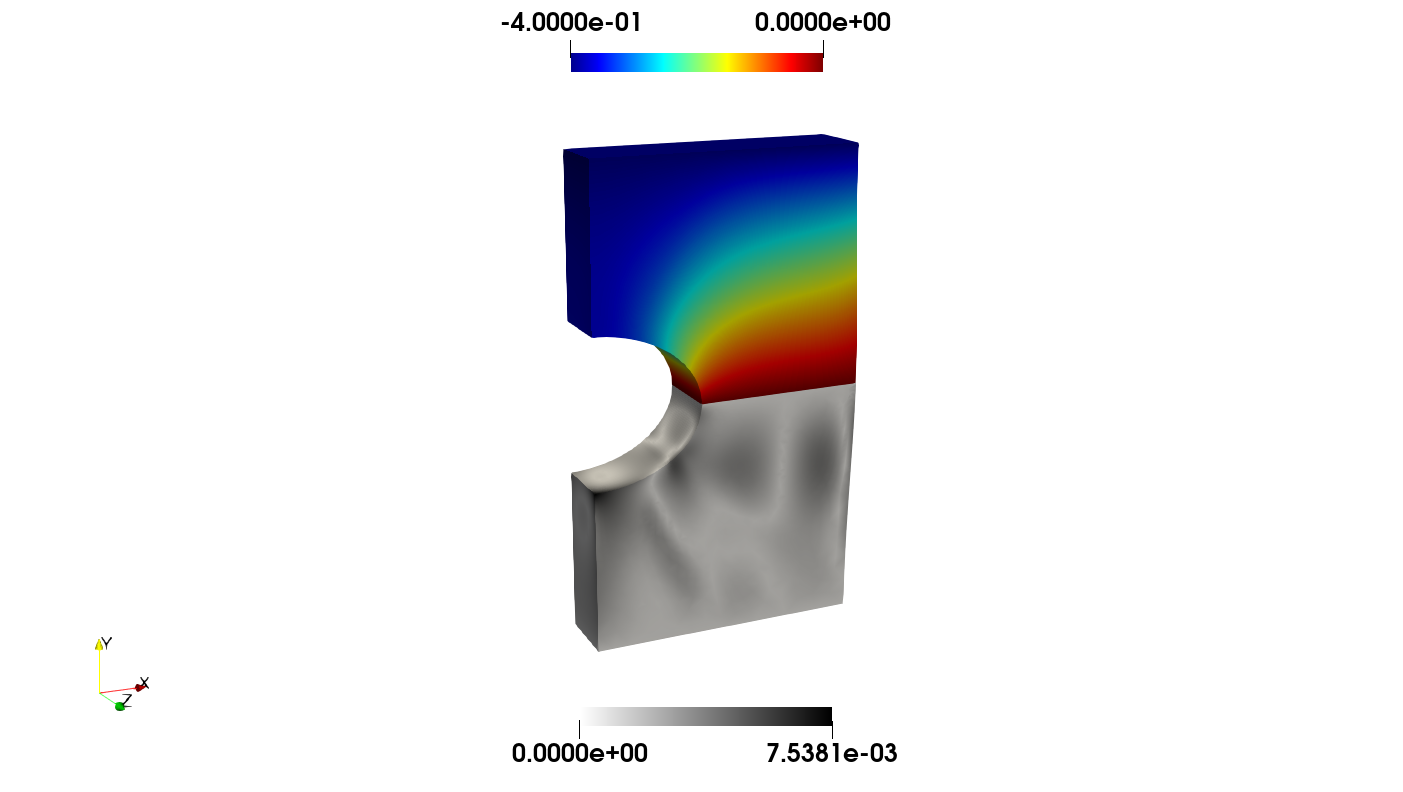}
        }
    \end{minipage}
    &
    \begin{minipage}[c]{\x\textwidth}
       \centering 
        \subfloat[Mesh 1, $U_z$]{\includegraphics[trim={17cm 0cm 17cm 0cm},clip,width=\textwidth]{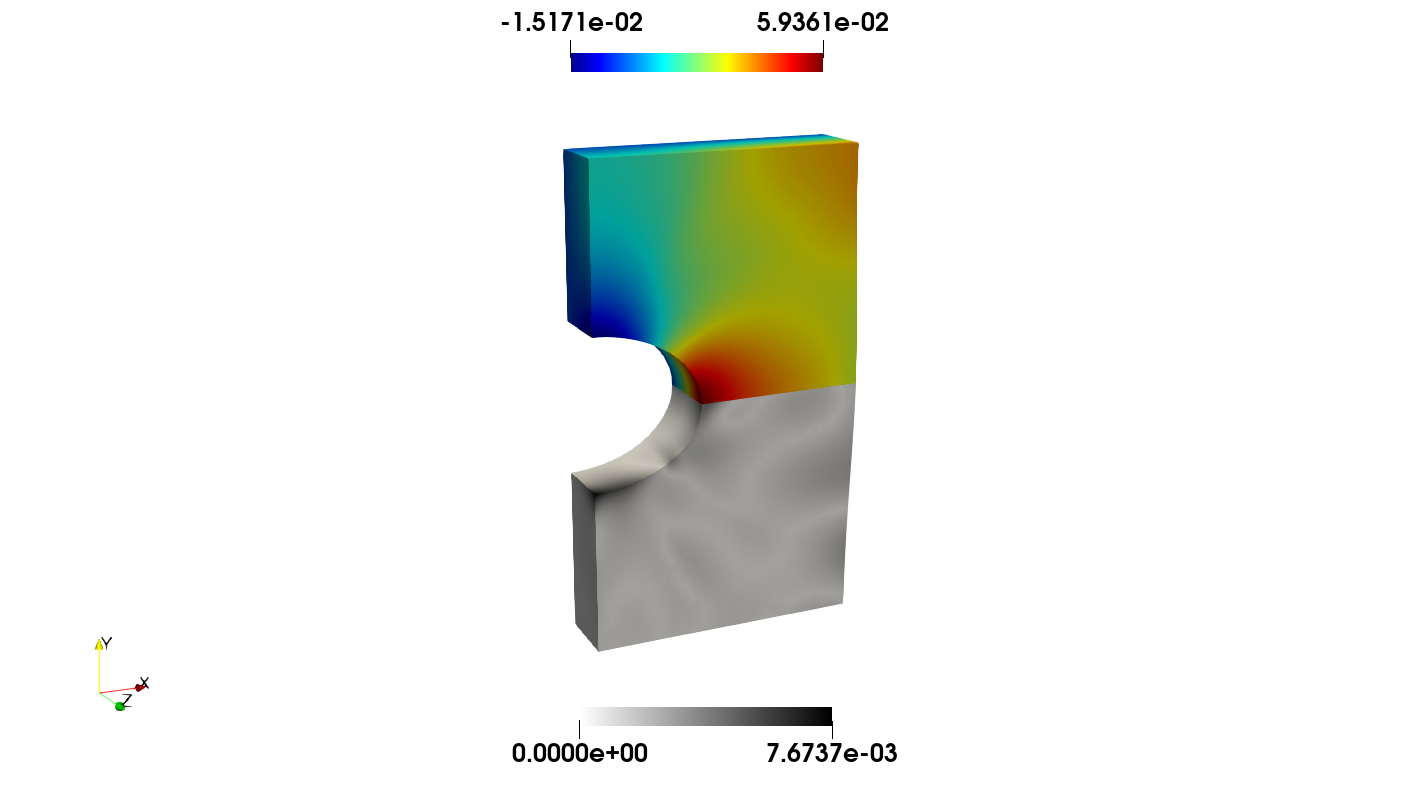}
        }
    \end{minipage}
    &
    \begin{minipage}[c]{\x\textwidth}
       \centering 
        \subfloat[Mesh 1, $\Bar{\sigma}$]{\includegraphics[trim={17cm 0cm 17cm 0cm},clip,width=\textwidth]{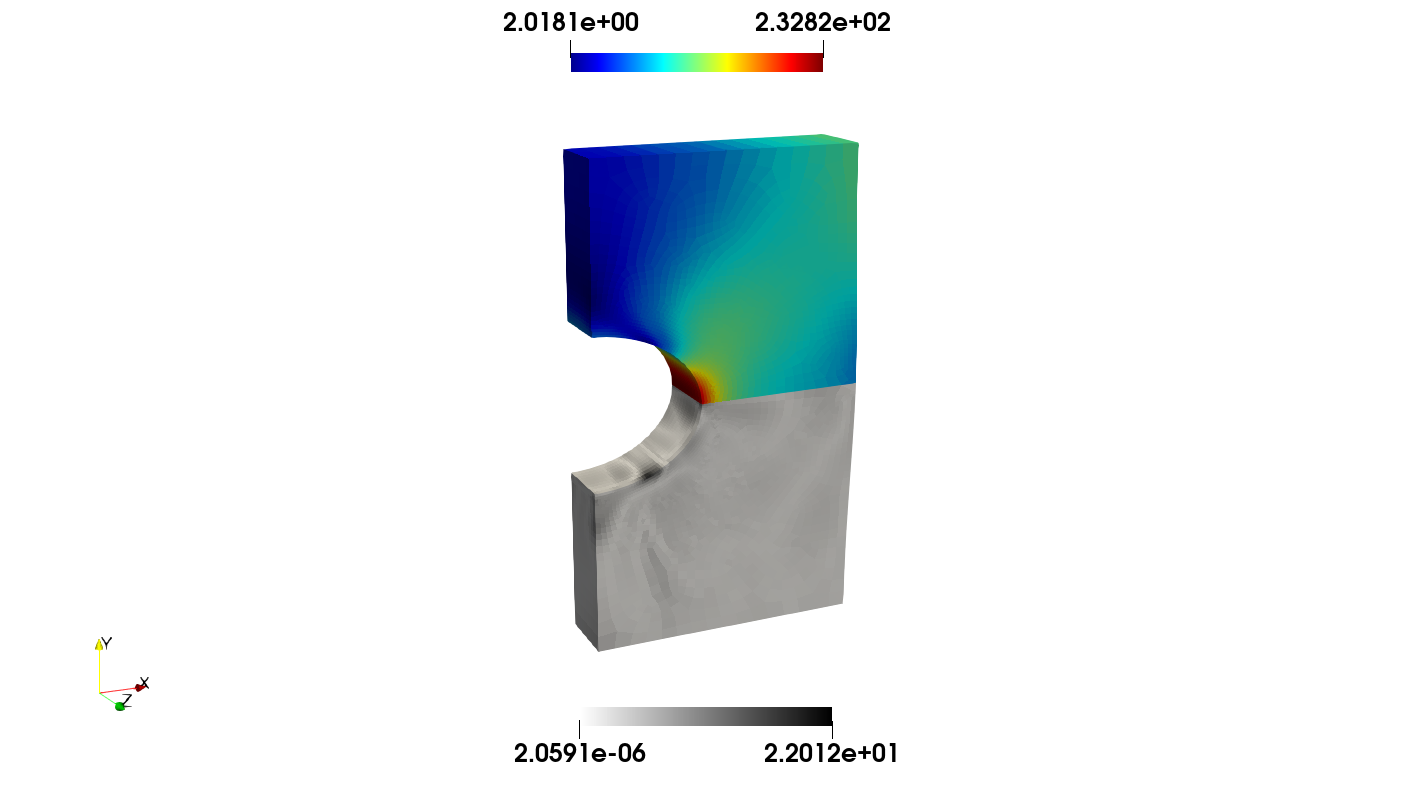}
        }
    \end{minipage}
    &
    \begin{minipage}[c]{\x\textwidth}
       \centering 
        \subfloat[Mesh 1, $\Bar{\epsilon}^p$]{\includegraphics[trim={17cm 0cm 17cm 0cm},clip,width=\textwidth]{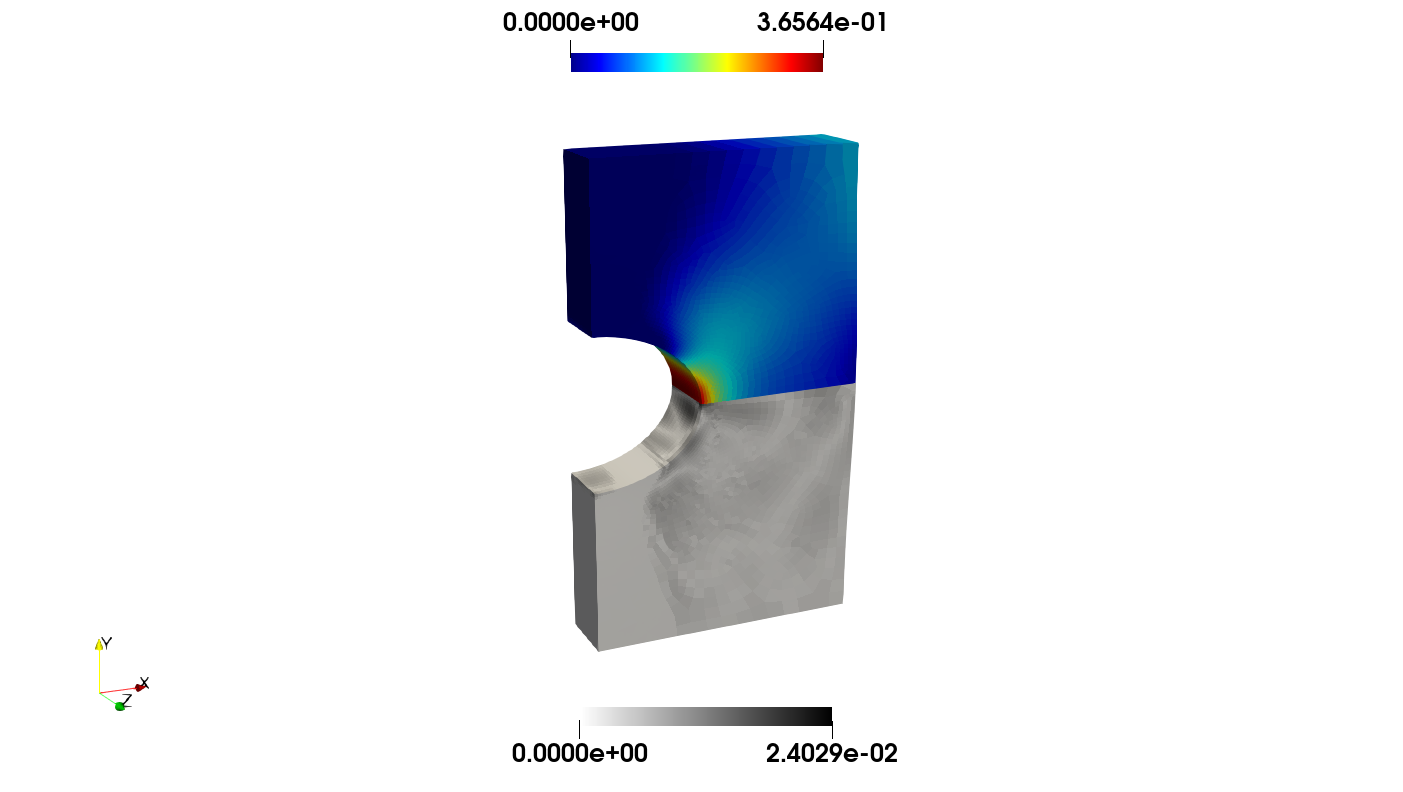}
        }
    \end{minipage}
    \\

     \begin{minipage}[c]{\x\textwidth}
       \centering 
        \subfloat[Mesh 2, $U_x$]{\includegraphics[trim={17cm 0cm 17cm 0cm},clip,width=\textwidth]{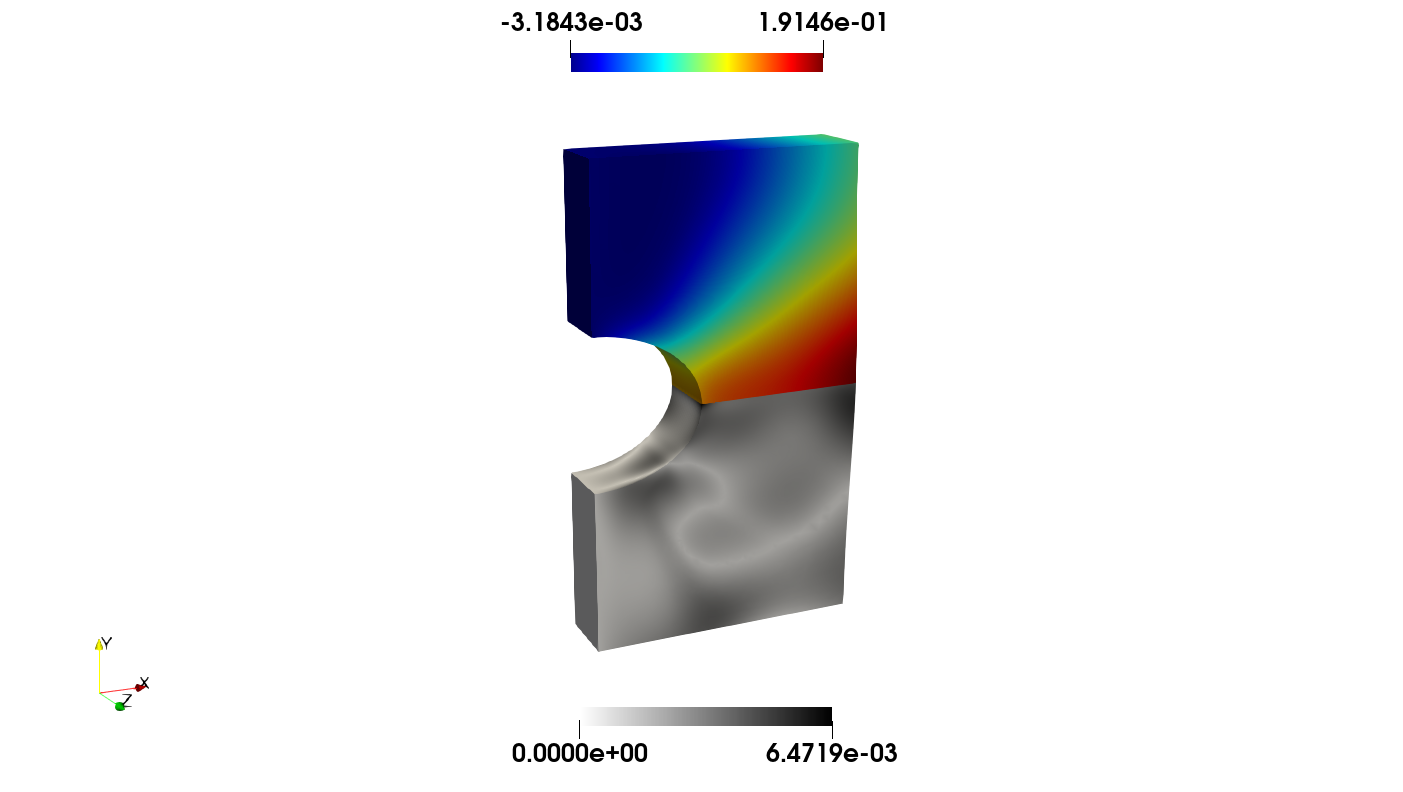}
        }
    \end{minipage}
    &
    \begin{minipage}[c]{\x\textwidth}
       \centering 
        \subfloat[Mesh 2, $U_y$]{\includegraphics[trim={17cm 0cm 17cm 0cm},clip,width=\textwidth]{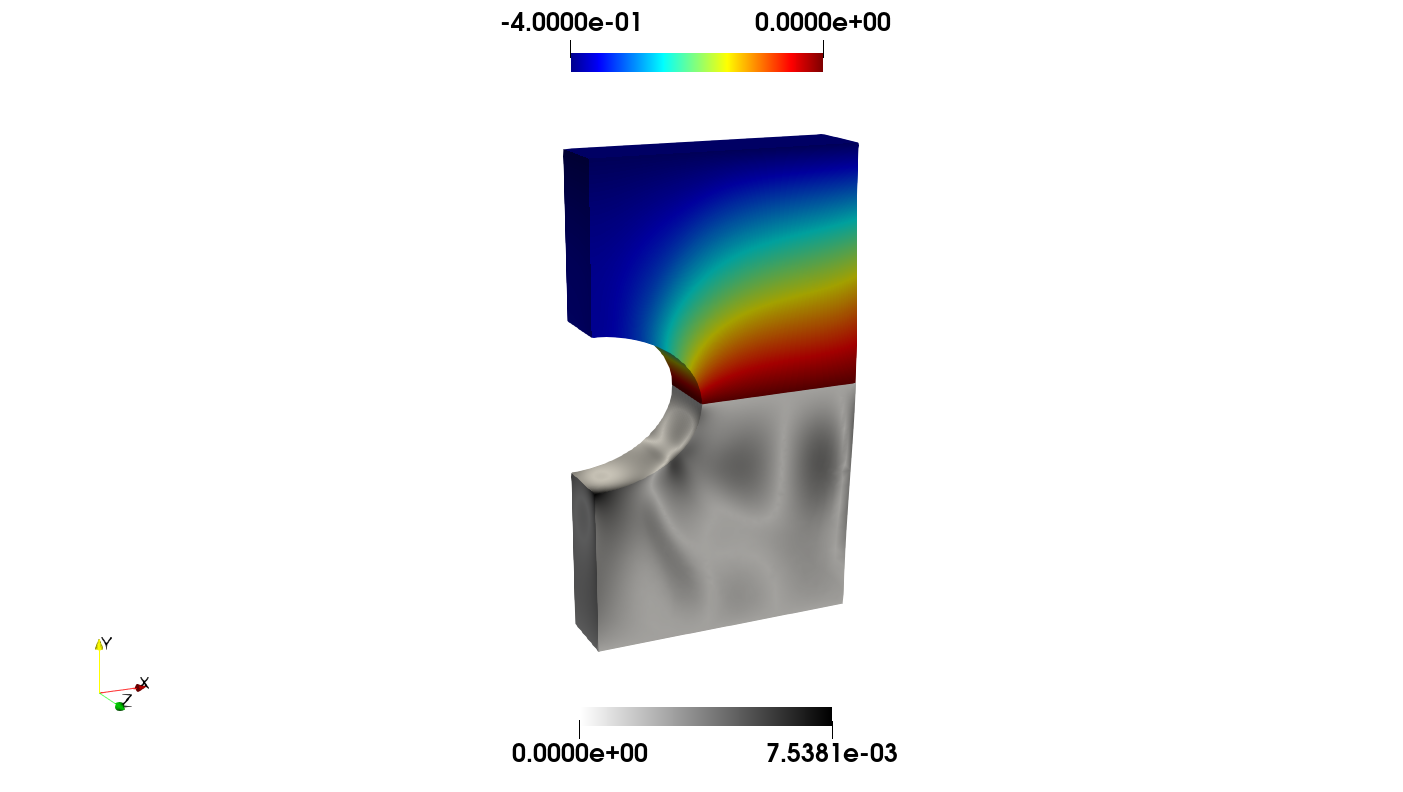}
        }
    \end{minipage}
    &
    \begin{minipage}[c]{\x\textwidth}
       \centering 
        \subfloat[Mesh 2, $U_z$]{\includegraphics[trim={17cm 0cm 17cm 0cm},clip,width=\textwidth]{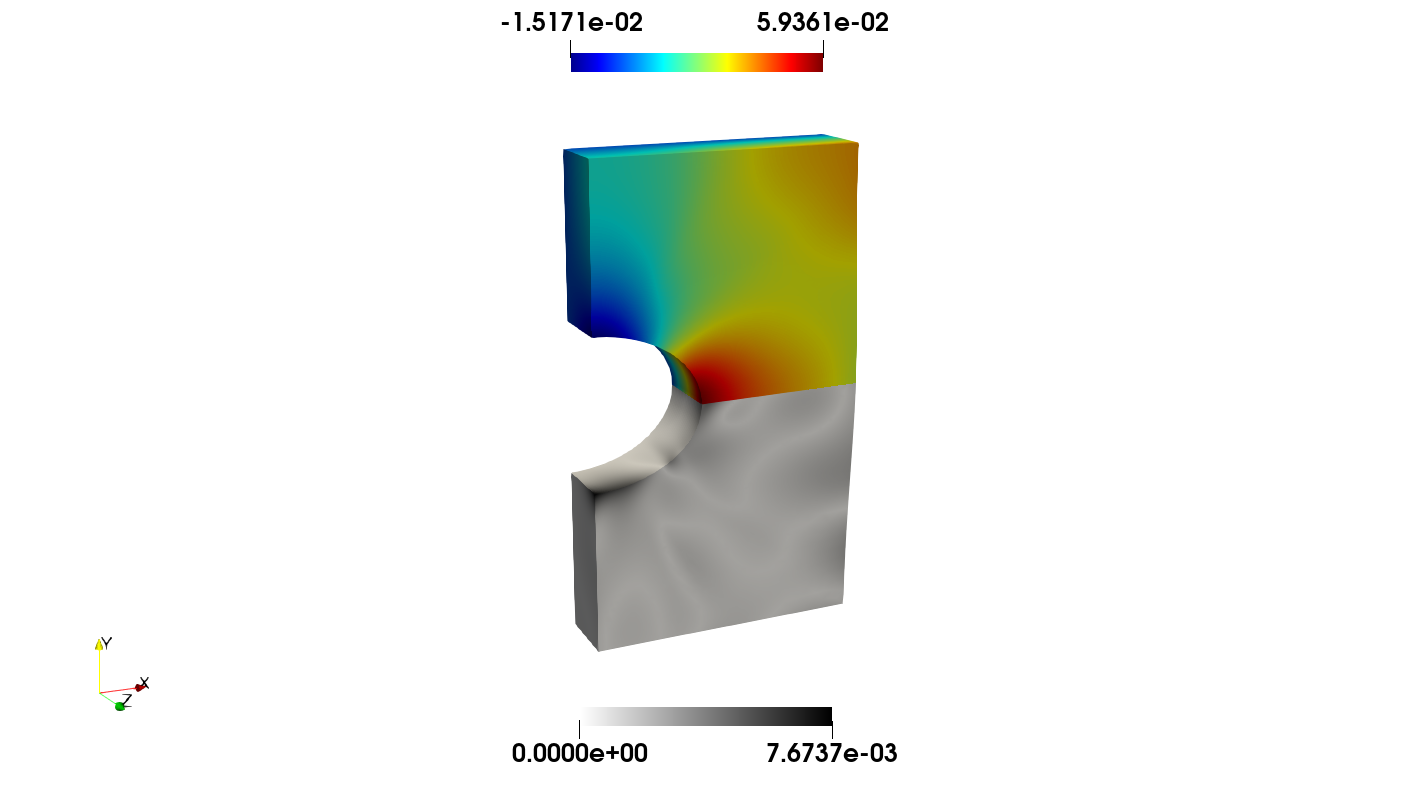}
        }
    \end{minipage}
    &
    \begin{minipage}[c]{\x\textwidth}
       \centering 
        \subfloat[Mesh 2, $\Bar{\sigma}$]{\includegraphics[trim={17cm 0cm 17cm 0cm},clip,width=\textwidth]{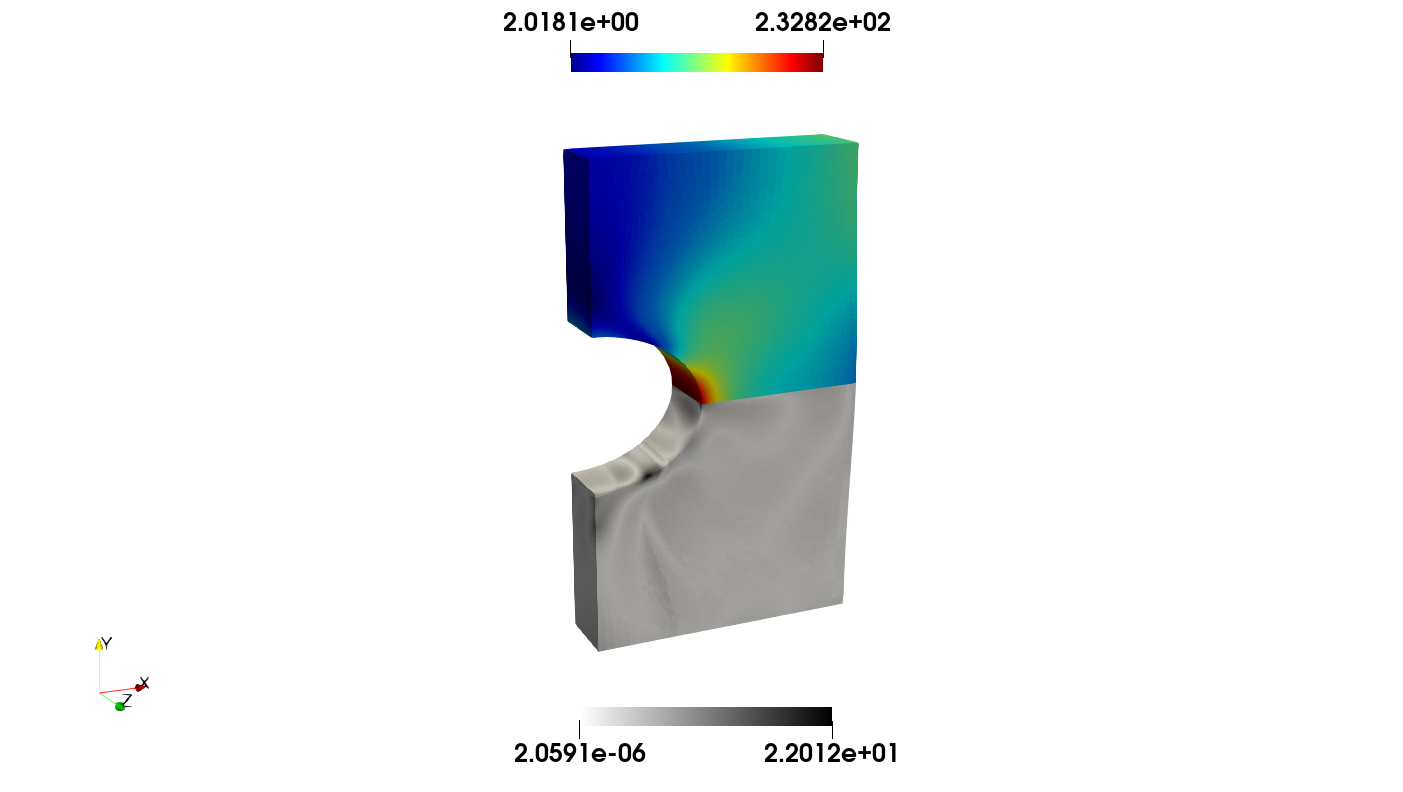}
        }
    \end{minipage}
    &
    \begin{minipage}[c]{\x\textwidth}
       \centering 
        \subfloat[Mesh 2, $\Bar{\epsilon}^p$]{\includegraphics[trim={17cm 0cm 17cm 0cm},clip,width=\textwidth]{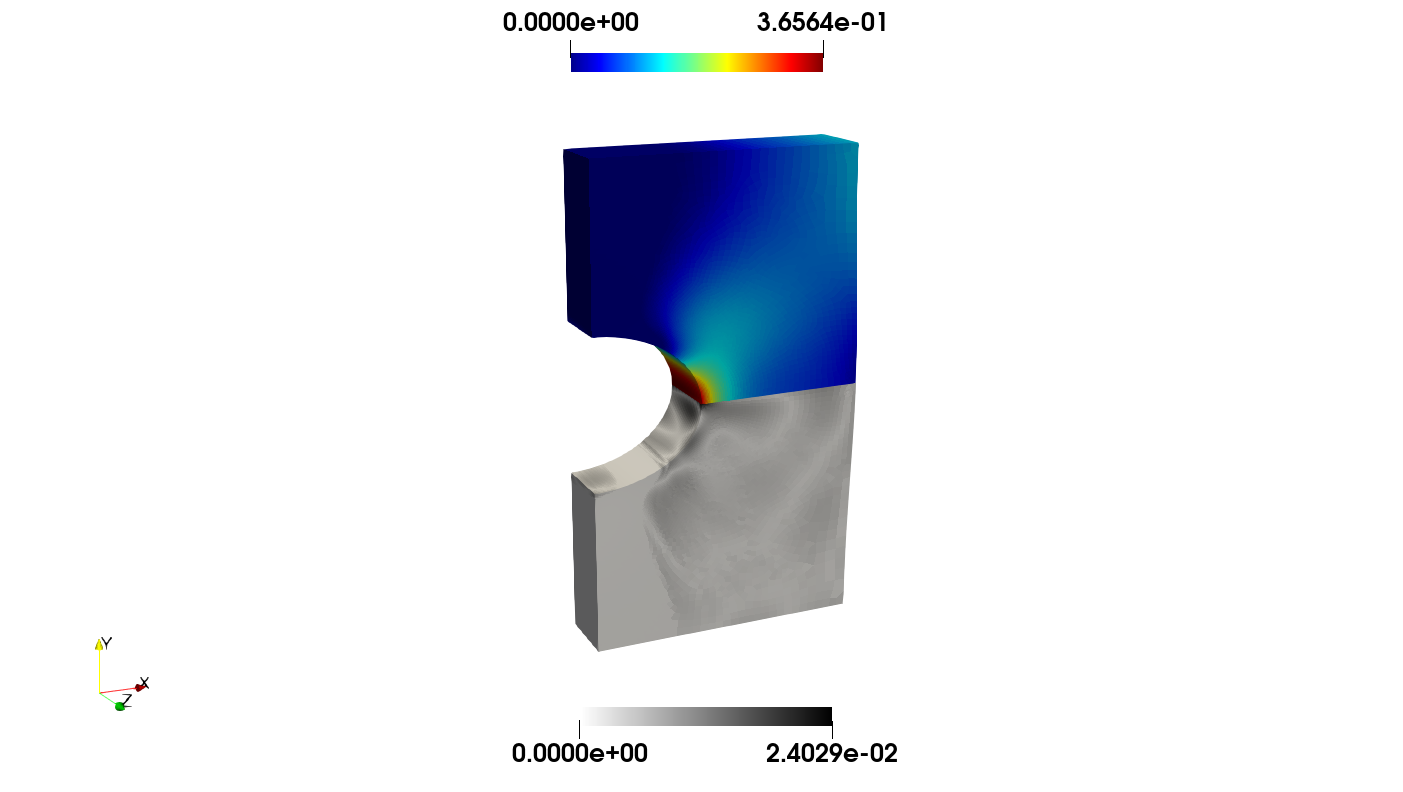}
        }
    \end{minipage}
    \\

    \begin{minipage}[c]{\x\textwidth}
       \centering 
        \subfloat[Mesh 3, $U_x$]{\includegraphics[trim={17cm 0cm 17cm 0cm},clip,width=\textwidth]{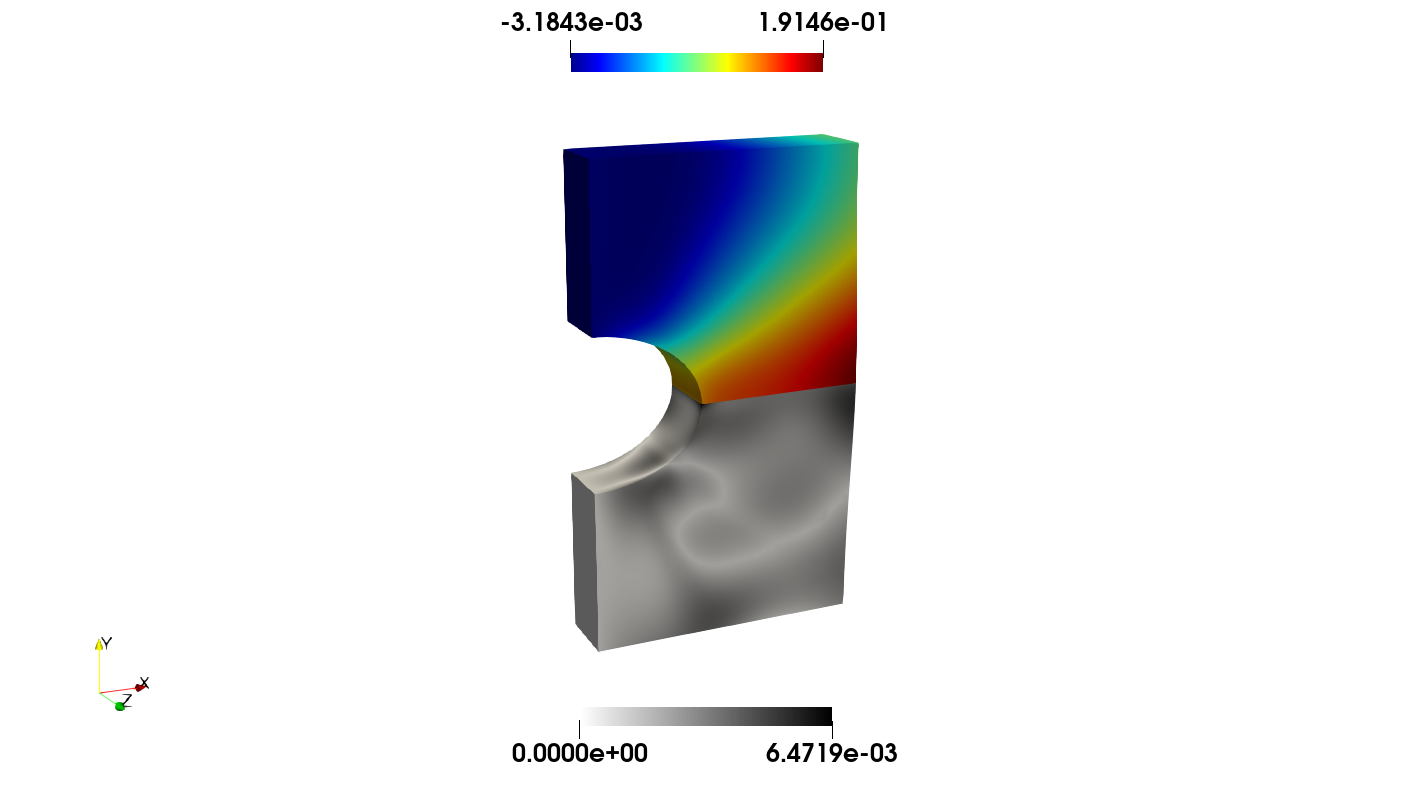}
        }
    \end{minipage}
    &
    \begin{minipage}[c]{\x\textwidth}
       \centering 
        \subfloat[Mesh 3, $U_y$]{\includegraphics[trim={17cm 0cm 17cm 0cm},clip,width=\textwidth]{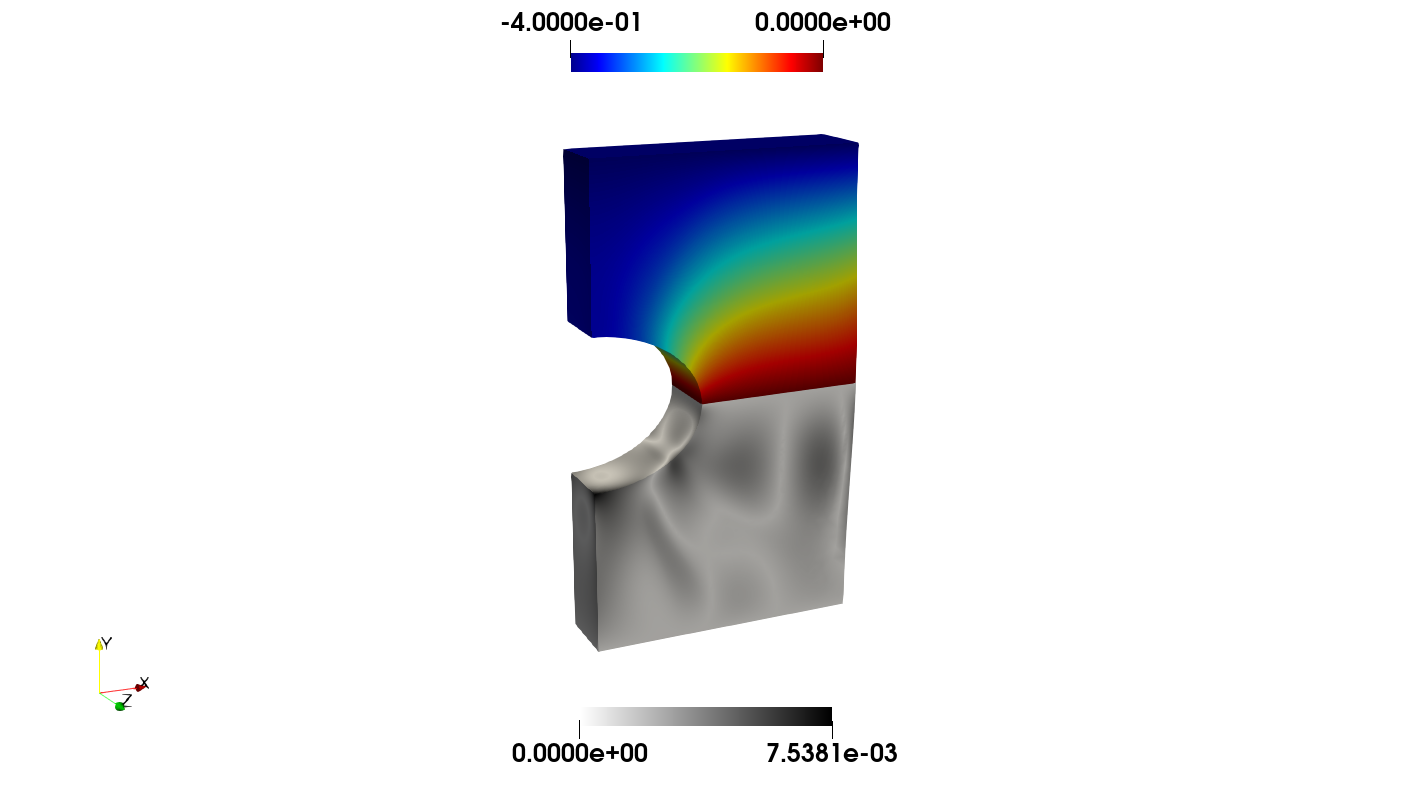}
        }
    \end{minipage}
    &
    \begin{minipage}[c]{\x\textwidth}
       \centering 
        \subfloat[Mesh 3, $U_z$]{\includegraphics[trim={17cm 0cm 17cm 0cm},clip,width=\textwidth]{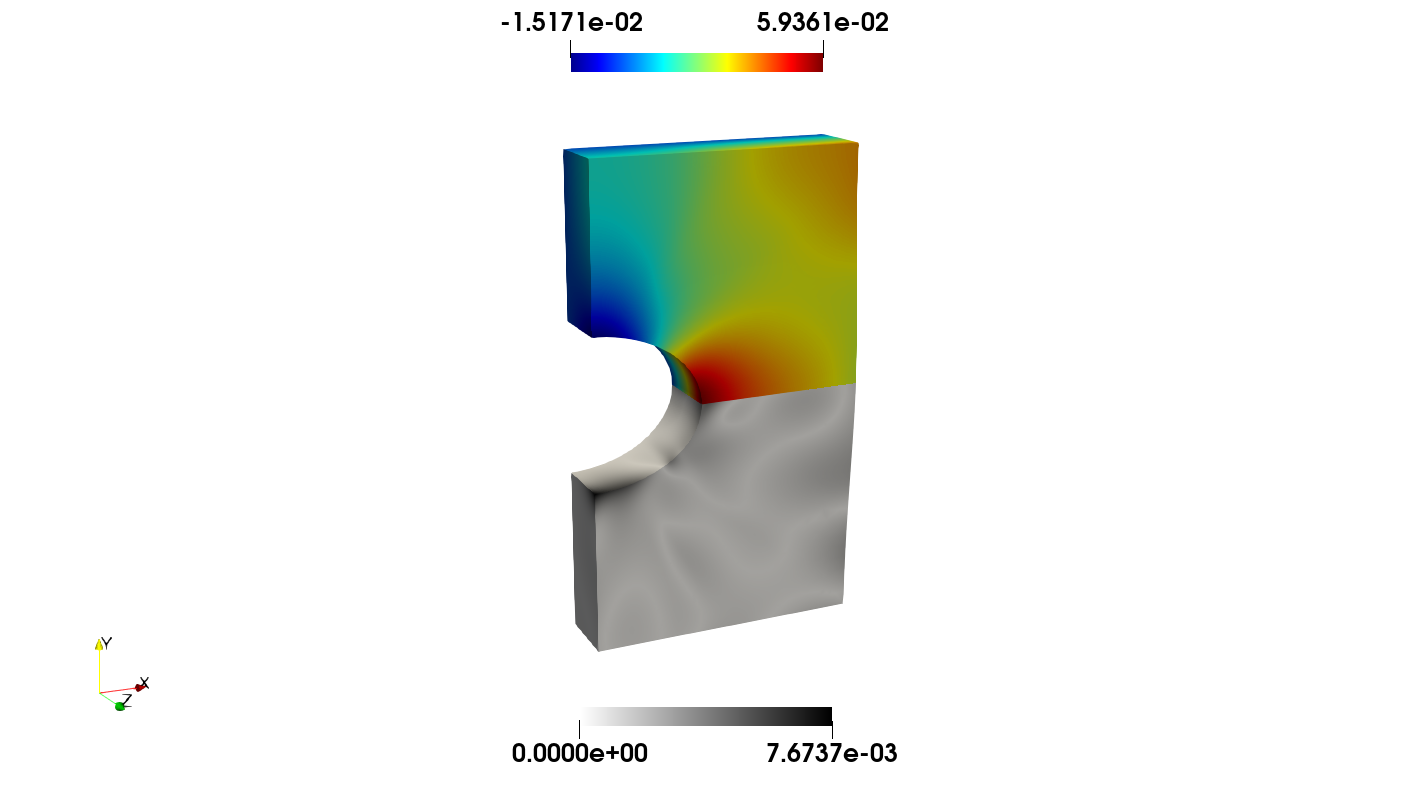}
        }
    \end{minipage}
    &
    \begin{minipage}[c]{\x\textwidth}
       \centering 
        \subfloat[Mesh 3, $\Bar{\sigma}$]{\includegraphics[trim={17cm 0cm 17cm 0cm},clip,width=\textwidth]{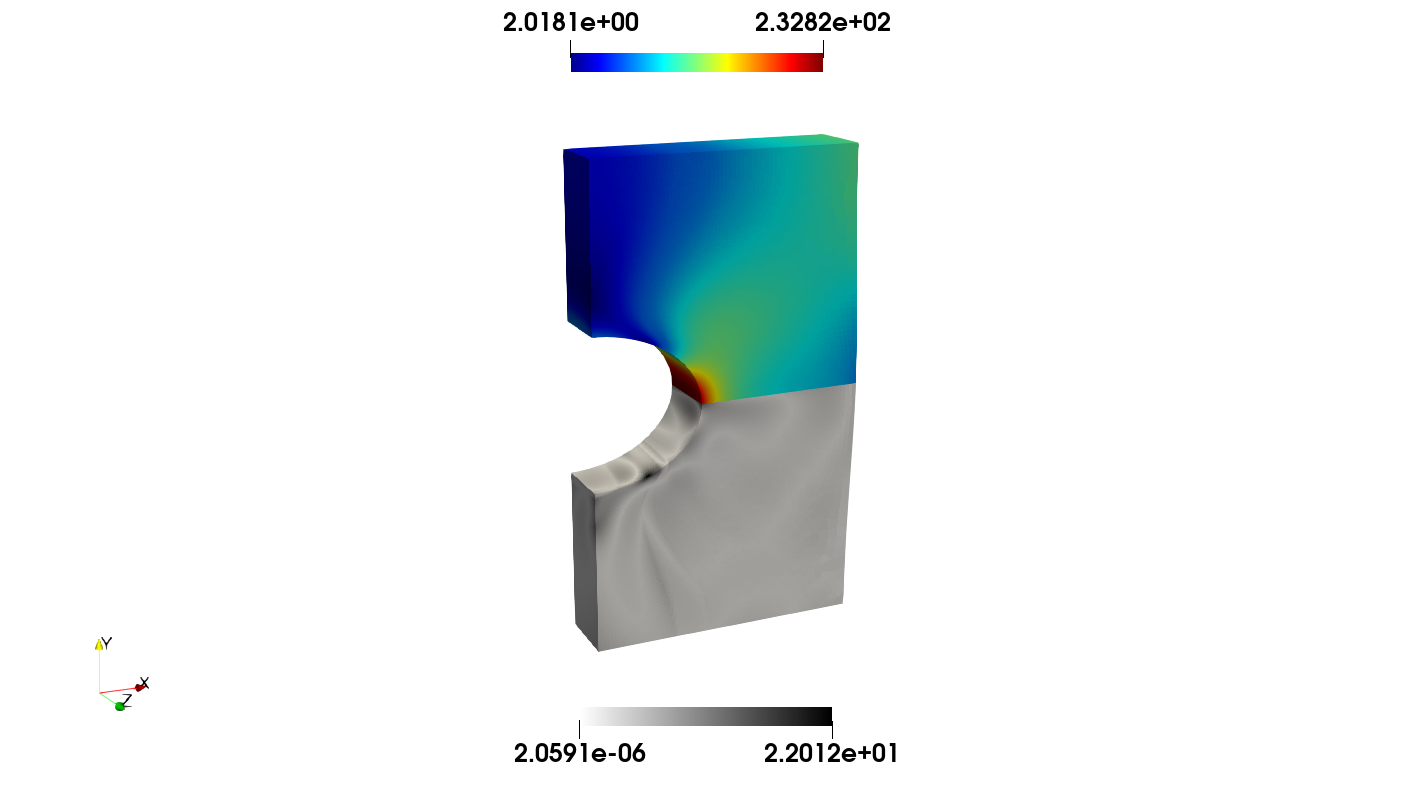}
        }
    \end{minipage}
    &
    \begin{minipage}[c]{\x\textwidth}
       \centering 
        \subfloat[Mesh 3, $\Bar{\epsilon}^p$]{\includegraphics[trim={17cm 0cm 17cm 0cm},clip,width=\textwidth]{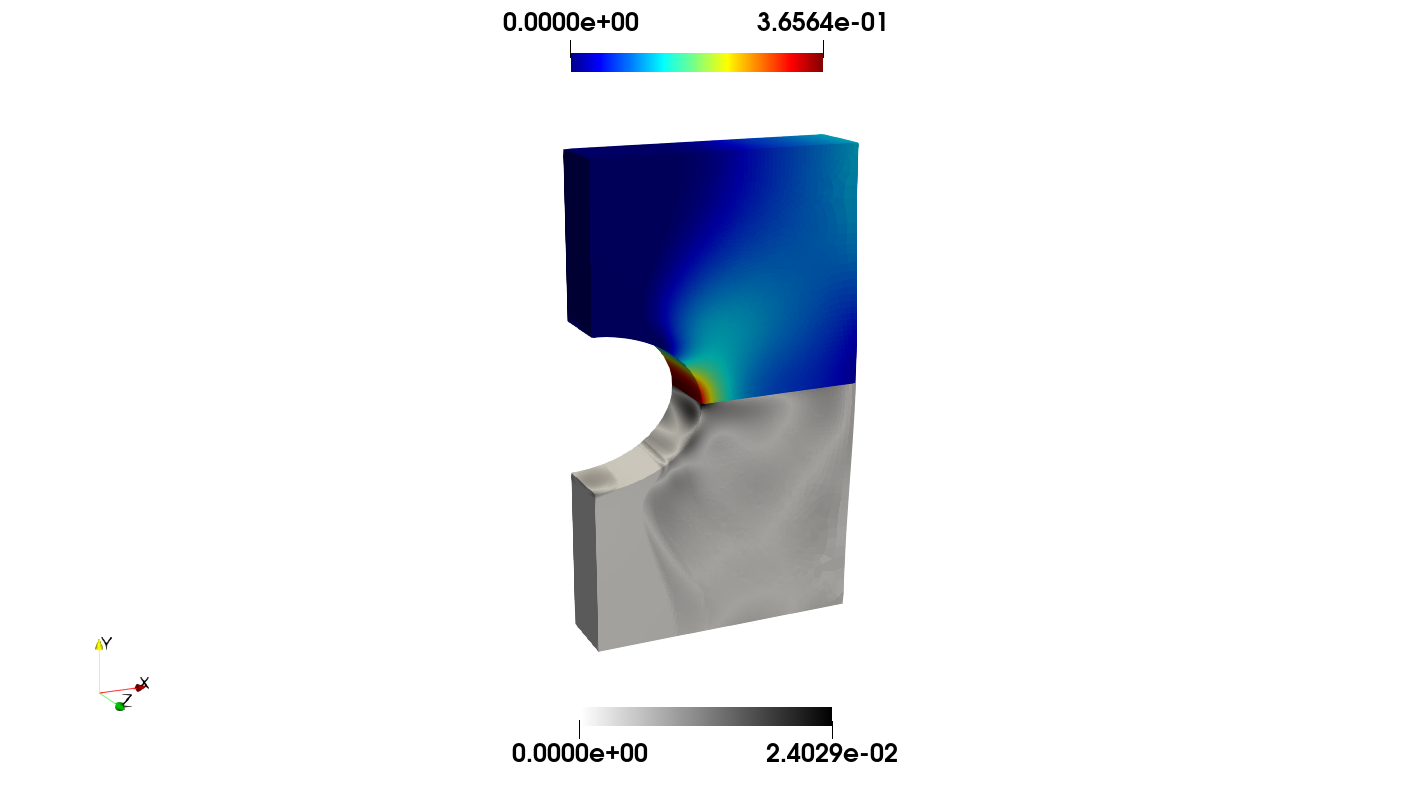}
        }
    \end{minipage}
    \\

    \end{tabular}
    \caption{Comparison between FEM and DEM inference in three progressively refined meshes. For each sub-figure, the top half, in the rainbow color map, shows the FEM reference solution. While the bottom half, in a white-to-black color map, shows the absolute difference in the DEM inference. The color scale for each field variable is identical for all three meshes.}
    \label{infer}
\end{figure}

Once the DEM model is trained on the coarsest mesh, we used it to infer the solution on Mesh 2 and Mesh 3. Due to the path-dependent nature of elastoplastic problems, the inference was also done in different load steps. A blank DEM model was first instantiated with randomly generated model parameters. For each load step, the previously saved parameters were loaded without any additional training, and a displacement field was predicted. The radial return algorithm was used to update internal state variables for the next load step, and the process was repeated until all 4 load steps had been inferred. Since the solution is path dependent, we see from \tref{train_accuracy} that the DEM errors gradually increase with load steps. As such, we present the solution contour plots at step 4 for all three meshes in \fref{infer} to show the worst-case error distributions. The accuracy metrics for all four load steps in all three meshes are listed in \tref{infer_accuracy}. A comparison of DEM training/inference time with FEM simulation time is shown in \tref{time_comparison}.

\begin{table}[h!]
    \caption{DEM performance metrics in 4 load steps}
    \small
    \centering
    \begin{tabular}{cccccccc}
     Case & \vline &  $AD_{U_{x}}$ [mm] & $AD_{U_{y}}$ [mm] & $AD_{U_{y}}$ [mm] & $AD_{\Bar{\sigma}}$ [MPa] & $AD_{\Bar{\epsilon}^p}$ [/] & $||\bm{e}||_{L^2}$ [\%] \\
    \hline
    Step1, Mesh1 & \vline & 3.61$\times 10^{-4}$ & 3.42$\times 10^{-4}$ & 1.51$\times 10^{-4}$ & 0.56 & 2.84$\times 10^{-4}$  & 0.56 \\
    Step1, Mesh2 & \vline & 3.48$\times 10^{-4}$ & 3.26$\times 10^{-4}$ & 1.62$\times 10^{-4}$ & 0.63 & 2.90$\times 10^{-4}$  & 0.53 \\
    Step1, Mesh3 & \vline & 3.48$\times 10^{-4}$ & 3.20$\times 10^{-4}$ & 1.73$\times 10^{-4}$ & 0.69 & 3.04$\times 10^{-4}$  & 0.52 \\
    \hline
    Step2, Mesh1 & \vline & 4.62$\times 10^{-4}$ & 4.29$\times 10^{-4}$ & 2.64$\times 10^{-4}$ & 0.72 & 4.21$\times 10^{-4}$  & 0.72 \\
    Step2, Mesh2 & \vline & 4.44$\times 10^{-4}$ & 4.18$\times 10^{-4}$ & 2.95$\times 10^{-4}$ & 0.81 & 4.33$\times 10^{-4}$  & 0.70 \\
    Step2, Mesh3 & \vline & 4.35$\times 10^{-4}$ & 4.09$\times 10^{-4}$ & 3.21$\times 10^{-4}$ & 0.88 & 4.57$\times 10^{-4}$  & 0.68 \\
    \hline
    Step3, Mesh1 & \vline & 6.17$\times 10^{-4}$ & 6.25$\times 10^{-4}$ & 4.09$\times 10^{-4}$ & 0.69 & 8.42$\times 10^{-4}$  & 0.51 \\
    Step3, Mesh2 & \vline & 6.00$\times 10^{-4}$ & 6.12$\times 10^{-4}$ & 4.56$\times 10^{-4}$ & 0.80 & 8.93$\times 10^{-4}$  & 0.49 \\
    Step3, Mesh3 & \vline & 6.06$\times 10^{-4}$ & 6.15$\times 10^{-4}$ & 5.00$\times 10^{-4}$ & 0.88 & 9.32$\times 10^{-4}$  & 0.49 \\
    \hline
    Step4, Mesh1 & \vline & 1.31$\times 10^{-3}$ & 1.00$\times 10^{-3}$ & 5.20$\times 10^{-4}$ & 1.03 & 1.46$\times 10^{-3}$  & 0.89 \\
    Step4, Mesh2 & \vline & 1.23$\times 10^{-3}$ & 9.11$\times 10^{-4}$ & 5.56$\times 10^{-4}$ & 1.17 & 1.52$\times 10^{-3}$  & 0.80 \\
    Step4, Mesh3 & \vline & 1.21$\times 10^{-3}$ & 8.67$\times 10^{-4}$ & 5.92$\times 10^{-4}$ & 1.28 & 1.58$\times 10^{-3}$  & 0.76 \\
    \end{tabular}
    \label{infer_accuracy}
\end{table}

\begin{table}[h!]
    \caption{Comparison of DEM training/inference and FEM simulation on different meshes}
    \small
    \centering
    \begin{tabular}{ccc}
     Case & \vline & Time [s] \\
    \hline
    FEM, mesh 1 & \vline  & 13 \\
    FEM, mesh 2 & \vline  & 55 \\
    FEM, mesh 3 & \vline  & 447 \\
    \hline
    DEM, mesh 1 & \vline  &  220 (Training) \\
    DEM, mesh 2 & \vline  &  0.78 (Inference) \\
    DEM, mesh 3 & \vline  &  1.05 (Inference) \\
    \end{tabular}
    \label{time_comparison}
\end{table}

Results in \tref{train_accuracy} show that the DEM prediction on Mesh 1 is accurate, having a $L^2$ displacement error norm of less than 1\%, and the mean difference in Mises stress is less than 1 MPa for the first three load steps. \fref{Contour3} shows error concentrations near the circular hole, where localized stress concentration is expected. The prediction errors generally increase with the load step, which is reasonable since elastoplasticity problems are path-dependent, and previous errors in plastic state variables accumulate in subsequent load steps. In this example, FEM is more efficient than DEM training and finishes about 16 times faster.

\fref{infer} shows that DEM inference on finer meshes does not significantly affect the spatial error distribution. This observation is further supported by the metrics in \tref{infer_accuracy}, where the inference accuracy on Mesh 2 and Mesh 3 remains similar to that obtained during training on Mesh 1. All inference results remain accurate with $L^2$ error norms less than 1$\%$. Finally, \tref{time_comparison} shows that once trained, DEM inference is highly efficient, thanks to GPU acceleration. Even on the finest mesh (Mesh 3) with 662120 elements, the total inference time is just over 1 second, which is about 0.25 seconds per load step. While for FEM, solution time increases dramatically as the number of elements increases due to the huge increase in the global stiffness matrix size. And in particular, for Mesh 3, the total DEM time (training on Mesh 1 + inference on Mesh 3) was about 221 seconds, which is half of the time required for FEM to solve on the finest mesh.

\section{Conclusions and future work}
\label{sec:conc}
In this work, we developed a novel DEM framework to solve elastoplasticity problems with $J_2$ plasticity. To the best of the authors' knowledge, this is the first time in the literature that an energy-based deep neural network solves plasticity problems. Contrary to the DCM for plasticity \citep{abueidda2021meshless}, which is based on the strong form and requires second-order spatial gradients of the displacement in the loss function, the current method only requires first-order spatial gradients in the loss function, making the gradient calculation more efficient. The loss function is based on the discrete variational formulation of the elastoplasticity problem. Radial return is used to update the plastic internal state variables given the DEM-predicted displacements, but other data-driven or physics-inform surrogate models can also be used. Spatial gradients of the displacement field are computed by finite element shape function gradients. 

Four numerical examples are presented to demonstrate the capability and application of the current DEM framework. Stress-strain curves under cyclic loading were generated for two hardening models using DEM. The ability to capture material heterogeneity was tested by simulating a bi-material plate under shear. A comparison with DCM was made using an elastoplasticity problem to compare the accuracy and computational cost of the two methods. The ability of the current model to be used on unstructured meshes was demonstrated by simulating the cyclic loading of a plate with a circular hole. The inference capability of the trained DEM model onto finer meshes was also tested. In all examples, the DEM predictions were accurate compared to FEM, while larger errors were commonly found near stress concentration points. As it currently stands, DEM is less computationally efficient than FEM, taking longer time to train than the FEM solution. However, once trained, DEM inference is efficient, and training on a coarse mesh and then inferring on a fine mesh can be faster than the FEM solution on the fine mesh.

The introduction of the total free energy of the system as a loss function for DEM in elastoplasticity problems is a large step forward from PMPE applied to elastic deformations. This opens a brand new door for the application of DEM to problems beyond the elastic range of the material. Compared to FEM, DEM is easier to implement, and extension to geometric and/or material nonlinearity is straightforward. The robust minimization structure of DEM also allows for the solution of large deformation problems in a single load step \citep{he2022use}, where FEM will typically require multiple load steps. All examples in this work were done while using the radial return method as the plastic state update kernel. Although this limits the application to materials whose return algorithm and direction are known, this can be easily remedied using a surrogate model (either data-driven or physics-informed) trained independently of the rest of the DEM framework. The rest of the DEM framework is designed to be agnostic to the plastic state update algorithm used. An important limitation to recognize is that, as it currently stands, training the elastoplastic DEM framework from scratch takes longer than the traditional FEM solution time. We emphasize that we are comparing our in-house DEM implementation, which was not specifically optimized for performance, to the highly optimized commercial FEM code Abaqus. If fewer CPU cores were used in FEM, or if some less-optimized open-source FE code was used, the benchmark performance results would have been much more favorable for DEM. The slower run time is also not unique to the current framework, as many PINNs that solve a PDE from scratch can take longer than FEM \citep{samaniego2020energy,nguyen2021parametric} due to the nonlinearity and non-convexity of the neural network training \citep{ew2018deep,samaniego2020energy} even when the original problem is linear. Despite the slower training time, the primary contribution that differentiates the current work from previous research efforts of using neural networks to solve elastoplasticity problems is that it provides a truly label-free framework to solve an elastoplasticity problem from scratch and does not need any assistance from the classical nonlinear numerical methods, such as FEM, except for the validation of results. Previous works rely on FEM solvers either to generate training data sets (which is typically the most time-consuming part of training a data-driven neural network \citep{abueidda2021meshless}) or use them to solve the discrete FE problem with the consistent stiffness matrix information provided by a trained neural network. Prior to the current work, the only other PINN with similar capabilities, to the best of the authors' knowledge, is the DCM model developed by \cite{abueidda2021meshless}. Compared to this previous work, the DEM loss function is more efficient to evaluate than the DCM loss function and requires less memory usage. The DEM model trains faster than a DCM model when the number of trainable parameters is identical, and the solution is more accurate. This work also marks the first time in the literature that a comparison between physics-informed models based on the PDE form (DCM) and energy form (DEM) is made on an elastoplasticity problem. Therefore, it will likely be highly appreciated by the entire engineering and scientific communities that are working in the confluence of artificial intelligence and classical numerical methods. In addition, PINNs and other neural network-based PDE solution methods can greatly benefit from GPUs. High-performance GPUs or even GPU clusters enable neural networks to train much faster than on a traditional CPU and significantly reduce the total computational time. In addition, once a DEM model has been trained, the solution field can be inferred for arbitrary points within the domain at speed much faster than FEM. 

Using deep neural networks to solve PDEs from scratch is a rising area of research and many open questions regarding their applicability, performance and convergence still exist. This work takes a step in expanding the applicability of energy-based neural network models to solve path-dependent elastoplasticity problems, but many future extensions are possible. In the future, we aim to extend the current framework to nonlinear hardening models such as the Voce \citep{follansbee1988constitutive} and Frederick-Armstrong \citep{armstrong1966mathematical} models. In addition, recent developments in data-driven \citep{lu2021learning} and physics-informed \citep{wang2021learning} deep operator networks (DeepONets) have allowed the approximation of nonlinear operators in parameterized partial differential equations. Once trained, DeepONets have exceptional generalizability and can infer solutions with different loading, boundary conditions, and material properties without any additional training. Investigating how DeepONets can be combined with the current DEM framework to enhance generalizability will be our future work.

\section*{Replication of results}
The data and source code that support the findings of this study can be found at \url{https://github.com/Jasiuk-Research-Group}. \textcolor{red}{Note to editor and reviewers: the link above will be made public upon the publication of this manuscript. During the review period, the data and source code can be made available upon request to the corresponding author.}

\section*{Acknowledgement}
This research is part of the Delta research computing project, which is supported by the National Science Foundation (award OCI 2005572), and the State of Illinois. Delta is a joint effort of the University of Illinois at Urbana-Champaign and its National Center for Supercomputing Applications.

\section*{Declaration of competing interest}
The authors declare that they have no conflict of interest.

\section*{CRediT authorship contribution statement}
\textbf{Junyan He}: Conceptualization, Methodology, Software, Formal analysis, Investigation, Data Curation, Writing - Original Draft.
\textbf{Diab Abueidda}: Conceptualization, Supervision, Writing - Review \& Editing.
\textbf{Rashid Abu Al-Rub}: Supervision, Writing - Review \& Editing.
\textbf{Seid Koric}: Supervision, Writing - Review \& Editing.
 \textbf{Iwona Jasiuk}: Supervision, Resources, Writing - Review \& Editing.

\bibliographystyle{plainnat}
\setlength{\bibsep}{0.0pt}
{\scriptsize \bibliography{References.bib} }

\begin{thebibliography}{47}
\providecommand{\natexlab}[1]{#1}
\providecommand{\url}[1]{\texttt{#1}}
\expandafter\ifx\csname urlstyle\endcsname\relax
  \providecommand{\doi}[1]{doi: #1}\else
  \providecommand{\doi}{doi: \begingroup \urlstyle{rm}\Url}\fi

\bibitem[Abueidda et~al.(2021{\natexlab{a}})Abueidda, Koric, Sobh, and
  Sehitoglu]{abueidda2021deep}
Diab~W Abueidda, Seid Koric, Nahil~A Sobh, and Huseyin Sehitoglu.
\newblock Deep learning for plasticity and thermo-viscoplasticity.
\newblock \emph{International Journal of Plasticity}, 136:\penalty0 102852,
  2021{\natexlab{a}}.

\bibitem[Abueidda et~al.(2021{\natexlab{b}})Abueidda, Lu, and
  Koric]{abueidda2021meshless}
Diab~W Abueidda, Qiyue Lu, and Seid Koric.
\newblock Meshless physics-informed deep learning method for three-dimensional
  solid mechanics.
\newblock \emph{International Journal for Numerical Methods in Engineering},
  122\penalty0 (23):\penalty0 7182--7201, 2021{\natexlab{b}}.

\bibitem[Abueidda et~al.(2022{\natexlab{a}})Abueidda, Koric, Al-Rub, Parrott,
  James, and Sobh]{abueidda2022deep}
Diab~W Abueidda, Seid Koric, Rashid~Abu Al-Rub, Corey~M Parrott, Kai~A James,
  and Nahil~A Sobh.
\newblock A deep learning energy method for hyperelasticity and
  viscoelasticity.
\newblock \emph{European Journal of Mechanics-A/Solids}, 95:\penalty0 104639,
  2022{\natexlab{a}}.

\bibitem[Abueidda et~al.(2022{\natexlab{b}})Abueidda, Koric, Guleryuz, and
  Sobh]{abueidda2022enhanced}
Diab~W Abueidda, Seid Koric, Erman Guleryuz, and Nahil~A Sobh.
\newblock Enhanced physics-informed neural networks for hyperelasticity.
\newblock \emph{arXiv preprint arXiv:2205.14148}, 2022{\natexlab{b}}.

\bibitem[Al-Haik et~al.(2006)Al-Haik, Hussaini, and
  Garmestani]{al2006prediction}
MS~Al-Haik, MY~Hussaini, and H~Garmestani.
\newblock Prediction of nonlinear viscoelastic behavior of polymeric composites
  using an artificial neural network.
\newblock \emph{International journal of plasticity}, 22\penalty0 (7):\penalty0
  1367--1392, 2006.

\bibitem[Ali et~al.(2019)Ali, Muhammad, Brahme, Skiba, and
  Inal]{ali2019application}
Usman Ali, Waqas Muhammad, Abhijit Brahme, Oxana Skiba, and Kaan Inal.
\newblock Application of artificial neural networks in micromechanics for
  polycrystalline metals.
\newblock \emph{International Journal of Plasticity}, 120:\penalty0 205--219,
  2019.

\bibitem[Armstrong et~al.(1966)Armstrong, Frederick,
  et~al.]{armstrong1966mathematical}
Peter~J Armstrong, CO~Frederick, et~al.
\newblock \emph{A mathematical representation of the multiaxial Bauschinger
  effect}, volume 731.
\newblock Berkeley Nuclear Laboratories Berkeley, CA, 1966.

\bibitem[Bonatti and Mohr(2022)]{bonatti2022importance}
Colin Bonatti and Dirk Mohr.
\newblock On the importance of self-consistency in recurrent neural network
  models representing elasto-plastic solids.
\newblock \emph{Journal of the Mechanics and Physics of Solids}, 158:\penalty0
  104697, 2022.

\bibitem[Bottou(2010)]{bottou2010large}
L{\'e}on Bottou.
\newblock Large-scale machine learning with stochastic gradient descent.
\newblock In \emph{Proceedings of COMPSTAT'2010}, pages 177--186. Springer,
  2010.

\bibitem[E~W(2018)]{ew2018deep}
Yu~B E~W.
\newblock The deep ritz method: A deep learning-based numerical algorithm for
  solving variational problems.
\newblock \emph{Commun Math Stat}, 6\penalty0 (1):\penalty0 1--12, 2018.

\bibitem[Follansbee and Kocks(1988)]{follansbee1988constitutive}
PS~Follansbee and UF~Kocks.
\newblock A constitutive description of the deformation of copper based on the
  use of the mechanical threshold stress as an internal state variable.
\newblock \emph{Acta Metallurgica}, 36\penalty0 (1):\penalty0 81--93, 1988.

\bibitem[Fuhg and Bouklas(2022)]{fuhg2022mixed}
Jan~N Fuhg and Nikolaos Bouklas.
\newblock The mixed deep energy method for resolving concentration features in
  finite strain hyperelasticity.
\newblock \emph{Journal of Computational Physics}, 451:\penalty0 110839, 2022.

\bibitem[Fuhg et~al.(2021)Fuhg, B{\"o}hm, Bouklas, Fau, Wriggers, and
  Marino]{fuhg2021model}
Jan~Niklas Fuhg, Christoph B{\"o}hm, Nikolaos Bouklas, Amelie Fau, Peter
  Wriggers, and Michele Marino.
\newblock Model-data-driven constitutive responses: application to a multiscale
  computational framework.
\newblock \emph{International Journal of Engineering Science}, 167:\penalty0
  103522, 2021.

\bibitem[Guo et~al.(2021)Guo, Zhuang, and Rabczuk]{guo2021deep}
Hongwei Guo, Xiaoying Zhuang, and Timon Rabczuk.
\newblock A deep collocation method for the bending analysis of kirchhoff
  plate.
\newblock \emph{arXiv preprint arXiv:2102.02617}, 2021.

\bibitem[Haghighat et~al.(2021)Haghighat, Raissi, Moure, Gomez, and
  Juanes]{haghighat2021physics}
Ehsan Haghighat, Maziar Raissi, Adrian Moure, Hector Gomez, and Ruben Juanes.
\newblock A physics-informed deep learning framework for inversion and
  surrogate modeling in solid mechanics.
\newblock \emph{Computer Methods in Applied Mechanics and Engineering},
  379:\penalty0 113741, 2021.

\bibitem[He et~al.(2022{\natexlab{a}})He, Abueidda, Koric, and
  Jasiuk]{he2022use}
Junyan He, Diab Abueidda, Seid Koric, and Iwona Jasiuk.
\newblock On the use of graph neural networks and shape-function-based gradient
  computation in the deep energy method.
\newblock \emph{arXiv preprint arXiv:2207.07216}, 2022{\natexlab{a}}.

\bibitem[He et~al.(2022{\natexlab{b}})He, Kushwaha, Abueidda, and
  Jasiuk]{he2022exploring}
Junyan He, Shashank Kushwaha, Diab Abueidda, and Iwona Jasiuk.
\newblock Exploring the structure-property relations of thin-walled, 2d
  extruded lattices using neural networks.
\newblock \emph{arXiv preprint arXiv:2205.06761}, 2022{\natexlab{b}}.

\bibitem[He et~al.(2022{\natexlab{c}})He, Kushwaha, Chadha, Koric, Abueidda,
  and Jasiuk]{he2022deep}
Junyan He, Shashank Kushwaha, Charul Chadha, Seid Koric, Diab Abueidda, and
  Iwona Jasiuk.
\newblock Deep energy method in topology optimization applications.
\newblock \emph{arXiv preprint arXiv:2207.03072}, 2022{\natexlab{c}}.

\bibitem[Hornik et~al.(1989)Hornik, Stinchcombe, and
  White]{hornik1989multilayer}
Kurt Hornik, Maxwell Stinchcombe, and Halbert White.
\newblock Multilayer feedforward networks are universal approximators.
\newblock \emph{Neural networks}, 2\penalty0 (5):\penalty0 359--366, 1989.

\bibitem[Huang et~al.(2020)Huang, Fuhg, Wei{\ss}enfels, and
  Wriggers]{huang2020machine}
Dengpeng Huang, Jan~Niklas Fuhg, Christian Wei{\ss}enfels, and Peter Wriggers.
\newblock A machine learning based plasticity model using proper orthogonal
  decomposition.
\newblock \emph{Computer Methods in Applied Mechanics and Engineering},
  365:\penalty0 113008, 2020.

\bibitem[Ibragimova et~al.(2021)Ibragimova, Brahme, Muhammad, L{\'e}vesque, and
  Inal]{ibragimova2021new}
Olga Ibragimova, Abhijit Brahme, Waqas Muhammad, Julie L{\'e}vesque, and Kaan
  Inal.
\newblock A new ann based crystal plasticity model for fcc materials and its
  application to non-monotonic strain paths.
\newblock \emph{International Journal of Plasticity}, 144:\penalty0 103059,
  2021.

\bibitem[Jang et~al.(2021)Jang, Fazily, and Yoon]{jang2021machine}
Dong~Phill Jang, Piemaan Fazily, and Jeong~Whan Yoon.
\newblock Machine learning-based constitutive model for j2-plasticity.
\newblock \emph{International Journal of Plasticity}, 138:\penalty0 102919,
  2021.

\bibitem[Liu et~al.(2020)Liu, Tao, Du, Yu, and Xu]{liu2020learning}
Xin Liu, Fei Tao, Haodong Du, Wenbin Yu, and Kailai Xu.
\newblock Learning nonlinear constitutive laws using neural network models
  based on indirectly measurable data.
\newblock \emph{Journal of Applied Mechanics}, 87\penalty0 (8):\penalty0
  081003, 2020.

\bibitem[Lu et~al.(2021)Lu, Jin, Pang, Zhang, and Karniadakis]{lu2021learning}
Lu~Lu, Pengzhan Jin, Guofei Pang, Zhongqiang Zhang, and George~Em Karniadakis.
\newblock Learning nonlinear operators via deeponet based on the universal
  approximation theorem of operators.
\newblock \emph{Nature Machine Intelligence}, 3\penalty0 (3):\penalty0
  218--229, 2021.

\bibitem[Masi et~al.(2021)Masi, Stefanou, Vannucci, and
  Maffi-Berthier]{masi2021thermodynamics}
Filippo Masi, Ioannis Stefanou, Paolo Vannucci, and Victor Maffi-Berthier.
\newblock Thermodynamics-based artificial neural networks for constitutive
  modeling.
\newblock \emph{Journal of the Mechanics and Physics of Solids}, 147:\penalty0
  104277, 2021.

\bibitem[Mozaffar et~al.(2019)Mozaffar, Bostanabad, Chen, Ehmann, Cao, and
  Bessa]{mozaffar2019deep}
M~Mozaffar, R~Bostanabad, W~Chen, K~Ehmann, Jian Cao, and MA~Bessa.
\newblock Deep learning predicts path-dependent plasticity.
\newblock \emph{Proceedings of the National Academy of Sciences}, 116\penalty0
  (52):\penalty0 26414--26420, 2019.

\bibitem[Muhammad et~al.(2021)Muhammad, Brahme, Ibragimova, Kang, and
  Inal]{muhammad2021machine}
Waqas Muhammad, Abhijit~P Brahme, Olga Ibragimova, Jidong Kang, and Kaan Inal.
\newblock A machine learning framework to predict local strain distribution and
  the evolution of plastic anisotropy \& fracture in additively manufactured
  alloys.
\newblock \emph{International Journal of Plasticity}, 136:\penalty0 102867,
  2021.

\bibitem[Nguyen-Thanh et~al.(2020)Nguyen-Thanh, Zhuang, and
  Rabczuk]{nguyen2020deep}
Vien~Minh Nguyen-Thanh, Xiaoying Zhuang, and Timon Rabczuk.
\newblock A deep energy method for finite deformation hyperelasticity.
\newblock \emph{European Journal of Mechanics-A/Solids}, 80:\penalty0 103874,
  2020.

\bibitem[Nguyen-Thanh et~al.(2021)Nguyen-Thanh, Anitescu, Alajlan, Rabczuk, and
  Zhuang]{nguyen2021parametric}
Vien~Minh Nguyen-Thanh, Cosmin Anitescu, Naif Alajlan, Timon Rabczuk, and
  Xiaoying Zhuang.
\newblock Parametric deep energy approach for elasticity accounting for strain
  gradient effects.
\newblock \emph{Computer Methods in Applied Mechanics and Engineering},
  386:\penalty0 114096, 2021.

\bibitem[Paszke et~al.(2019)Paszke, Gross, Massa, Lerer, Bradbury, Chanan,
  Killeen, Lin, Gimelshein, Antiga, Desmaison, Kopf, Yang, DeVito, Raison,
  Tejani, Chilamkurthy, Steiner, Fang, Bai, and Chintala]{NEURIPS2019_9015}
Adam Paszke, Sam Gross, Francisco Massa, Adam Lerer, James Bradbury, Gregory
  Chanan, Trevor Killeen, Zeming Lin, Natalia Gimelshein, Luca Antiga, Alban
  Desmaison, Andreas Kopf, Edward Yang, Zachary DeVito, Martin Raison, Alykhan
  Tejani, Sasank Chilamkurthy, Benoit Steiner, Lu~Fang, Junjie Bai, and Soumith
  Chintala.
\newblock Pytorch: An imperative style, high-performance deep learning library.
\newblock In H.~Wallach, H.~Larochelle, A.~Beygelzimer, F.~d\textquotesingle
  Alch\'{e}-Buc, E.~Fox, and R.~Garnett, editors, \emph{Advances in Neural
  Information Processing Systems 32}, pages 8024--8035. Curran Associates,
  Inc., 2019.
\newblock URL
  \url{http://papers.neurips.cc/paper/9015-pytorch-an-imperative-style-high-performance-deep-learning-library.pdf}.

\bibitem[Pi~Savall et~al.(2021)Pi~Savall, Mielke, and Ricken]{pi2021data}
Berta Pi~Savall, Andr{\'e} Mielke, and Tim Ricken.
\newblock Data-driven stress prediction for thermoplastic materials.
\newblock \emph{PAMM}, 21\penalty0 (1):\penalty0 e202100225, 2021.

\bibitem[Qu et~al.(2021)Qu, Di, Feng, Wang, and Zhao]{qu2021towards}
Tongming Qu, Shaocheng Di, YT~Feng, Min Wang, and Tingting Zhao.
\newblock Towards data-driven constitutive modelling for granular materials via
  micromechanics-informed deep learning.
\newblock \emph{International Journal of Plasticity}, 144:\penalty0 103046,
  2021.

\bibitem[Raissi(2018)]{raissi2018deep}
Maziar Raissi.
\newblock Deep hidden physics models: Deep learning of nonlinear partial
  differential equations.
\newblock \emph{The Journal of Machine Learning Research}, 19\penalty0
  (1):\penalty0 932--955, 2018.

\bibitem[Rezaei et~al.(2022)Rezaei, Harandi, Moeineddin, Xu, and
  Reese]{rezaei2022mixed}
Shahed Rezaei, Ali Harandi, Ahmad Moeineddin, Bai-Xiang Xu, and Stefanie Reese.
\newblock A mixed formulation for physics-informed neural networks as a
  potential solver for engineering problems in heterogeneous domains:
  comparison with finite element method.
\newblock \emph{arXiv preprint arXiv:2206.13103}, 2022.

\bibitem[Samaniego et~al.(2020)Samaniego, Anitescu, Goswami, Nguyen-Thanh, Guo,
  Hamdia, Zhuang, and Rabczuk]{samaniego2020energy}
Esteban Samaniego, Cosmin Anitescu, Somdatta Goswami, Vien~Minh Nguyen-Thanh,
  Hongwei Guo, Khader Hamdia, X~Zhuang, and T~Rabczuk.
\newblock An energy approach to the solution of partial differential equations
  in computational mechanics via machine learning: Concepts, implementation and
  applications.
\newblock \emph{Computer Methods in Applied Mechanics and Engineering},
  362:\penalty0 112790, 2020.

\bibitem[Simo and Hughes(2006)]{simo2006computational}
Juan~C Simo and Thomas~JR Hughes.
\newblock \emph{Computational inelasticity}, volume~7.
\newblock Springer Science \& Business Media, 2006.

\bibitem[SIMULIA(2020)]{Abaqus2021}
SIMULIA.
\newblock Abaqus, 2020.

\bibitem[Tancogne-Dejean et~al.(2021)Tancogne-Dejean, Gorji, Zhu, and
  Mohr]{tancogne2021recurrent}
Thomas Tancogne-Dejean, Maysam~B Gorji, Juner Zhu, and Dirk Mohr.
\newblock Recurrent neural network modeling of the large deformation of
  lithium-ion battery cells.
\newblock \emph{International Journal of Plasticity}, 146:\penalty0 103072,
  2021.

\bibitem[Wang et~al.(2021)Wang, Wang, and Perdikaris]{wang2021learning}
Sifan Wang, Hanwen Wang, and Paris Perdikaris.
\newblock Learning the solution operator of parametric partial differential
  equations with physics-informed deeponets.
\newblock \emph{Science advances}, 7\penalty0 (40):\penalty0 eabi8605, 2021.

\bibitem[Wilkins(1964)]{wilkins1964methods}
Mark~L Wilkins.
\newblock Methods in computational physics.
\newblock \emph{Calculation of elastic--plastic flow}, pages 211--263, 1964.

\bibitem[Yang et~al.(2020)Yang, Xiang, Tang, and Guo]{yang2020learning}
Hang Yang, Qian Xiang, Shan Tang, and Xu~Guo.
\newblock Learning material law from displacement fields by artificial neural
  network.
\newblock \emph{Theoretical and Applied Mechanics Letters}, 10\penalty0
  (3):\penalty0 202--206, 2020.

\bibitem[Yu et~al.(2022)Yu, Han, Yang, Wang, Tang, and
  Guo]{yu2022elastoplastic}
Zefeng Yu, Chenghang Han, Hang Yang, Yu~Wang, Shan Tang, and Xu~Guo.
\newblock Elastoplastic constitutive modeling under the complex loading driven
  by gru and small-amount data.
\newblock \emph{Theoretical and Applied Mechanics Letters}, page 100363, 2022.

\bibitem[Zehnder et~al.(2021)Zehnder, Li, Coros, and
  Thomaszewski]{zehnder2021ntopo}
Jonas Zehnder, Yue Li, Stelian Coros, and Bernhard Thomaszewski.
\newblock Ntopo: Mesh-free topology optimization using implicit neural
  representations.
\newblock \emph{Advances in Neural Information Processing Systems},
  34:\penalty0 10368--10381, 2021.

\bibitem[Zhang and Mohr(2020)]{zhang2020using}
Annan Zhang and Dirk Mohr.
\newblock Using neural networks to represent von mises plasticity with
  isotropic hardening.
\newblock \emph{International Journal of Plasticity}, 132:\penalty0 102732,
  2020.

\bibitem[Zhang et~al.(2022)Zhang, Wang, Zhu, Diehl, Maldar, Shang, and
  Zeng]{zhang2022predicting}
Sheng Zhang, Leyun Wang, Gaoming Zhu, Martin Diehl, Alireza Maldar, Xiaoqing
  Shang, and Xiaoqin Zeng.
\newblock Predicting grain boundary damage by machine learning.
\newblock \emph{International Journal of Plasticity}, 150:\penalty0 103186,
  2022.

\bibitem[Zhu et~al.(1997)Zhu, Byrd, Lu, and Nocedal]{zhu1997algorithm}
Ciyou Zhu, Richard~H Byrd, Peihuang Lu, and Jorge Nocedal.
\newblock Algorithm 778: L-bfgs-b: Fortran subroutines for large-scale
  bound-constrained optimization.
\newblock \emph{ACM Transactions on mathematical software (TOMS)}, 23\penalty0
  (4):\penalty0 550--560, 1997.

\bibitem[Ziegler(1959)]{ziegler1959modification}
Hans Ziegler.
\newblock A modification of prager’s hardening rule.
\newblock \emph{Quarterly of Applied mathematics}, 17\penalty0 (1):\penalty0
  55--65, 1959.

\end{thebibliography}
\end{document}